# Designing diagnostic platforms for analysis of disease patterns and probing disease emergence

Ali Rohani

# Abstract

The emerging era of personalized medicine relies on medical decisions, practices, and products being tailored to the individual patient. Point-of-care systems, at the heart of this model, play two important roles. First, they are required for identifying subjects for optimal therapies based on their genetic make-up and epigenetic profile. Second, they will be used for assessing the progression of such therapies. Central to this vision is designing systems that, with minimal user-intervention, can transduce complex signals from biosystems in complement with clinical information to inform medical decision within point-of-care settings. To reach our ultimate goal of developing point-of-care systems and realizing personalized medicine, we are taking a multistep systems-level approach towards understanding cellular processes and biomolecular profiles, to quantify disease states and external interventions.

The first step is to generate data. In this step we develop necessary tools to enable reliable and repeatable set of experiments, to facilitate sensitive characterization of biosystems of interest. To this aim, we have developed AC electrokinetic methods within microfluidic systems to translate the bioparticles based on their unique electrophysiology. Innovative device structures and signaling schemes are essential for sensitive characterization of biosystems and deciphering the underlying mechanism responsible for emergent behavior of these systems. To accomplish our goals in this step, we have designed relevant device structures or signaling schemes to generate data. In the first chapter, we presented a nano-slit structure to create conductivity gradients. These patterned conductivity gradients are used to enhance dielectrophoretic trapping of nanometer sized proteins, antibodies and antigens to enable and enhance discovery of rare cancer biomarkers in high conductivity physiological media. Similarly in Chapter Two we employ AC dielectrophoresis inside an insulator based microfluidic structure to characterize and separate Human Embryonic Kidney (HEK) and Mouse Embryonic Fibroblast (MEF) cells, based on their unique mitochondrial morphology and subsequent dissimilar electrophysiology.

The second step is turning data into information. To this end, we develop tools to monitor bioparticles and perform accurate measurements on them. These measurements range from tracking bioparticles to quantifying their morphometric measures. Finding proper figures to measure and perform accurate measurements is the key factor for sensitive characterization of biosystems. To achieve this goal, in chapter one, we develop a sensitive methodology to quantify pre-concentration and depletion of different biomarkers in micro/nano fluidic systems. The unique characteristics of this methodology are: highly reduced measured noise and geometry free analysis capabilities. In Chapter Two, we developed MyMiA; an image analysis software to quantify morphometric properties of mitochondrial networks.

The third step is to turn information into insight. In this step, we use our gained information in conjunction with the underlying mechanistic knowledge to predict cellular/molecular behaviors leading to disease states and drug responses. To this end, we benefit from modeling, computation and simulations to better understand the systems of interest. To achieve this goal, in Chapter One, mathematical modeling of protein dielectric properties along with intense numerical simulations enabled us to design, enhance and optimize a biomarker-sensing system, and in chapter two, by using mathematical modeling we found the correlation between mitochondrial network properties (connectedness) and cells electrophysiological response, which enabled us to speared cells based on the level of connectivity in their mitochondrial network in a completely label free manner. Also by using morphometric measures from MyMiA and applying supervised learning algorithms, we showed cell line and certain proteomic activities in cells could be predicted based on their mitochondrial morphology which is of great importance for predicting patient outcomes and also developing new studies on cancer cells.

# Table of Contents

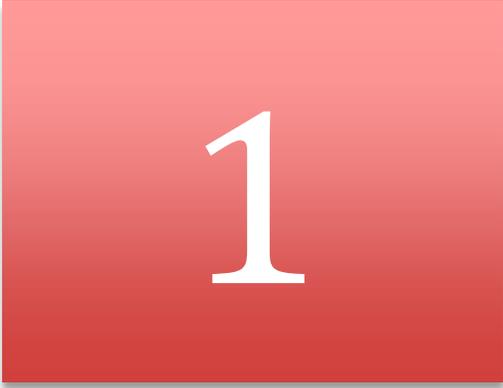

**Analyzing biomarker profiles in blood serum to enable early detection of cancer**

Selective and rapid enrichment of biomolecules is of great interest for biomarker discovery, protein crystallization and in biosensing for speeding assay kinetics and reducing signal interferences. The current state of the art is based on DC electrokinetics, wherein localized ion depletion at the anodic interface of the microchannel to perm-selective nanochannel is used to enhance electric fields and the resulting biomarker electromigration is balanced against electro-osmosis in the microchannel to cause high degrees of biomarker enrichment. However, biomarker enrichment is not selective and the levels fall off within physiological media of high conductivity, due to a reduction in both, ion concentration polarization and electro-osmosis effects. In the first part of this chapter we present an AC electrokinetic methodology for initiating frequency-selective and ultrafast enrichment of biomarkers in physiological media. Using AC fields at a critical frequency to generate negative dielectrophoresis of the biomarker of interest and an offset DC field to initiate ion conductivity gradients due to ion concentration polarization along the cathodic interface of the perm-selective region inside a nanoslit channel, we enhance the localized field due to ion depletion and the field gradient due to ion accumulation. Since the surface charge non-uniformity that causes perm-selectivity is placed inside the nanoslit, the spatial extent for biomarker electrophoresis, as well as its localized trapping by negative dielectrophoresis is enhanced, thereby enabling ultrafast biomarker enrichment within physiological media.

To maximize the sensitivity in our platform, we used quantitative information on the spatial and temporal dynamics of pre-concentration, such as position, extent and time towards sensor design for coupling pre-concentration to detection. Current quantification methodologies are based on the time evolution of fluorescence signals from biomarkers within a statically defined region of interest, which does not offer information on the spatial dynamics of pre-concentration and leads to significant errors when the pre-concentration zone is delocalized or exhibits wide variations in size, shape and position over time under the force field. We present a dynamic methodology for quantifying the region of interest by using a statistical description of particle distribution across the device geometry to determine the intensity thresholds for particle pre-concentration. We applied our method to study the delocalized pre-concentration dynamics under an electrokinetic force balance driven by negative dielectrophoresis, for aligning the pre-concentration and detection regions of neuropeptide Y, and for quantifying the polarizability dispersion of silica nano-colloids with frequency of the alternating .orce field.

Since proteomic biomarkers of interest to the early diagnosis of diseases and infections are present at trace levels versus interfering species, their selective enrichment is needed within bio-assays for speeding binding kinetics with receptors and reducing signal interferences. In the third part of this chapter we used our platform to enable frequency-selective enrichment of nanocolloidal biomolecules, based on the characteristic time constant for polarization of their electrical double-layer, since surface conduction in their ion cloud depends on colloidal size, shape and surface charge. In this manner, using DC-offset AC fields, differences in frequency dispersion for negative dielectrophoresis are balanced against electrophoresis in a nanoslit channel to enable the selective enrichment of prostate specific antigen (PSA) versus anti-mouse immunoglobulin antibodies that cause signal interferences to immunoassays. Through coupling enrichment to capture by receptors on graphene-modified surfaces, we demonstrate the elimination of false positives caused by anti-mouse immunoglobulin antibodies to the PSA immunoassay.

## 1.1. Surface conduction enhanced dielectrophoresis in nanoslits for rapid and selective biomarker enrichment in physiological media

**Introduction**

Biomarker discovery requires the identification of variations in rare sets of biomolecules within bio-fluids that also contain common interfering species, such as circulating antibodies, at million to billion-fold higher levels. Hence, there is a need for methods to cause rapid and selective enrichment of rare biomarkers versus the interfering species[1], preferably within physiological media, to maintain their ability to selectively bind with receptors, without needing buffer changes that lead to dilution. Since affinity methods based on antibody depletion can cause enrichment levels of no more than two to three orders of magnitude[2]; there is a need for complementary enrichment modalities. A selectively enriched plug of the target analyte is especially of interest within heterogeneous immunoassays for speeding binding kinetics to receptors due to the prevalence of mass transport limitations at detection limit target levels[3], while also reducing signal interferences from circulating antibodies that bind to the receptors to often cause false positives and false negatives[4]. In this manner, immunoassay signals at detection limit levels can reach their saturation signal at earlier time points[5], thereby leading to reduced errors for analyte quantification, due to the invariance of signals with incubation time and method[6].

A commonly investigated methodology for achieving highly enriched analyte plugs from dilute samples is through applying an electrokinetic force balance in conjunction with localized ion depletion to create a trap due to the sharp spatial profile in the field. Typically, ion concentration polarization (ICP) effects at the entrance of perm-selective nanochannels are used to cause sharp field gradients due to ion depletion at the anodic interface of microchannel to the nanochannel, and this is balanced against electro-osmosis within the microchannel[7] to cause high degrees of biomarker enrichment[8]. However, due to the abrupt field profile, the trapped biomarkers are co-localized within a tightly confined region, which limits the scope for spatially graded stacking towards selectivity. Furthermore, the degree of biomarker enrichment obtained by this method can fall sharply within physiological media of high conductivity, due to a reduction in electro-osmosis, as well as ICP effects. While a limited degree of biomarker enrichment can be obtained within conductive media through utilizing tighter nanochannels and/or higher applied fields to enhance ICP, these strategies are not practical due to fabrication challenges, disruptive electrothermal flow due to Joule heating and electrical-double layer induced field screening.

An alternate strategy is to carry out biomarker enrichment under the field profile due to ion conductivity gradients inside a nanoslit channel, where surface conduction effects are more pronounced than those at the microchannel interface to the nanoslit. Hence, the field gradient can be extended along the nanoslit length for enhancing the spatial extent of biomarker depletion, in parallel to enabling spatially graded biomarker stacking. In prior work, this ion conductivity gradient was created inside the nanochannel by connecting it to microchannels loaded with buffers with a high and low conductivity media, so that the field profile can be modulated due to electro-osmosis in the nanochannel[9], to achieve biomarker enrichment[10]. In this current work, we implement a similar electrokinetic force balance in conjunction with ion conductivity gradients generated inside a nanoslit channel, rather than in a microchannel. However, the ion conductivity gradients inside the nanoslit are initiated by ICP effects created using lateral constrictions with large surface charge non-uniformities, so that DC field-induced surface conduction continues to be significant versus bulk conduction, even in physiological media. Furthermore, instead of electroosmosis as the opposing force field, we utilize AC fields for frequency-selective negative dielectrophoresis (nDEP) at the constriction tips, since nDEP continues to be a significant force field within media of high conductivity. In this manner, we utilize ICP effects to enhance biomarker electrophoresis (EP) and nDEP trapping, thereby enabling the net force balance to cause ultrafast biomarker enrichment within physiological media.

Dielectrophoresis (DEP) enables highly selective trapping of bio-particles based on the characteristic frequency response of dielectric permittivity of the particle versus the medium[11],[12],[13],[14] for causing frequency-selective enrichment of target DNA[15],[16],[17] and protein molecules in physiological media[18],[19], [20], [21],[22].

However, the highly localized nature of DEP behavior, due to its dependence on $\nabla E^2$, limits its spatial extent[14],[23]. Herein, utilizing ICP effects at lateral constrictions to generate conductivity differences along the cathodic region of the perm-selective interface, the localized field ($E$) is enhanced due to ion depletion in the vicinity of the constriction tips, whereas ion accumulation along the constriction sidewalls increases the field gradient ($\nabla E$), thereby dramatically enhancing the magnitude and spatial extent of nDEP, due to its $\nabla E^2$ dependence. While our prior work attributed the enhanced biomarker enrichment in this device to tilting of the nDEP potential barrier under DC bias; herein, we elucidate the role of surface conduction in the nanoslit towards creating ion conductivity gradients that enhance the height and depth of the potential barrier for trapping, thereby exponentially enhancing localized biomarker numbers due to its Boltzmann distribution profile. Furthermore, while ICP effects have been extensively applied within prior work to cause ion gradients in the microchannel region interfacing the nanochannel, this is the first report on using ICP to initiate ion conductivity gradients inside the nanochannel, where surface charge-induced conduction effects can cause more pronounced ICP effects than within microchannels. As a result, the presented AC electrokinetic method can cause biomarker enrichment within physiological media of high conductivity and with the necessary frequency-selectivity to reduce signal interferences during biomarker detection, whereas these are not possible with DC electrokinetic methods coupled to ICP.

**Results and Discussion**

**A. Device Geometry for ICP-enhanced nDEP enrichment**

We begin with schematically illustrating the mechanism for generating ion conductivity gradients inside the nanoslit and illustrate its application towards enhancing the electrokinetic enrichment of biomarkers. The device geometry is subsequently optimized using the Poisson-Nernst-Plank (PNP) model to simulate the conductivity gradients and its application towards frequency-selective biomarker enrichment is validated using spatio-temporal profiles of fluorescently-labeled streptavidin and prostate specific antigen (PSA). Our prior work has implemented this methodology within various sensing paradigms for improving biomarker sensitivity [20],[24]. A cross-section view of the device geometry used within this work is schematically presented in Figure 1a, wherein glassy carbon modified Pt electrodes within the reservoirs leading to a microchannel on each side are used to initiate the discussed electrokinetic effects under a DC-offset (2 V/cm) AC field (70 $V_{rms}$/cm), that is set at the critical frequency required to cause optimal biomarker nDEP, which is 1 MHz for streptavidin, 3 MHz for Neuropeptide Y (NPY) [20] and 4-6 MHz for Prostate Specific Antigen (PSA)[25]. The microchannels (5 µm depth) are connected by slit-shaped channels of 200 nm depth (henceforth called nanoslit). Within each nanoslit (300 µm length / 200 nm depth), a lateral constriction (optimized at 300 nm length and 40 nm gap) with a sharp surface charge non-uniformity is used to initiate ICP (see magnified top-view in Figure 1b), whereas no significant ICP occurs at the micro/ nanochannel interface due to the wide nanoslit entry points (30 µm width). The geometry of this constricted perm-selective region is designed so as to enable ion conductivity gradients over a wide extent across its cathodic interface, due to interaction of ion depletion and ion accumulation profiles under field-induced ICP conditions. First, the spatial extent of the constriction is designed to be short ($l\sim300$ nm) to reduce the length over which the voltage drop occurs inside the perm-selective region, thereby enhancing the localized field and extending ion depletion through a significant portion of the nanoslit, beyond the immediate perm-selective region. In this manner, due to enhanced surface conduction of counter-ions and the exclusion of co-ions in the perm-selective region, the ion depletion region extends from across the anodic interface through inside of the perm-selective region, up to portions across the cathodic interface inside the nanoslit, as reported within other similar geometries[26]. Second, the sloping constriction geometry enables a sharp drop in the field profile along its sidewall directions versus a more gradual drop along the device centerline. As a result, ion accumulation at its cathodic interface is higher along the constriction sidewall directions and occurs within a short distance of the constriction tip, whereas the accumulation is lower along the device centerline direction and is pushed further away from the constriction tip. Due to current conservation, these ion conductivity gradients at the cathodic interface modify the localized field and nDEP trapping potential profiles. On one hand, the reduction in localized ion conductivity at the constriction tip due to the extension of ion depletion across from the anodic to the cathodic interface ($\sigma_{depletion}$), increases the field at the constriction tip: $E$. On the other hand, the sharp rise in localized ion conductivity along the sidewall directions due to the ion accumulation ($\sigma_{acc}$), creates a localized reduction in the field away from the constriction tip, thereby increasing the field gradient: $\nabla E$. In Figure 1c, we represent these field profile changes in terms of alterations in the nDEP trapping potential profile, from the green solid line (nDEP with no ICP due to low DC field) to the blue dashed line (nDEP with ICP due to a critical DC field). Since nDEP depends on the product of the field and its gradient, its enhancement under these ICP conditions can be represented as a higher potential barrier to indicate the higher field at the constriction tips and a deeper potential well to indicate the steeper field gradient along the constriction sidewall. Hence, biomarker numbers ($n$) rise with the potential ($U$) as per the Boltzmann distribution: $n=n_o exp(-U/k_B T)$; thereby causing exponential biomarker enrichment with an increase in $U$.

Furthermore, as a consequence of the extension in ion depletion along the cathodic interface to well beyond the immediate vicinity of the perm-selective region, the local ion conductivity along the centerline direction ($\sigma_{centerline}$) remains low over an extended length away from the constriction tip. Hence, due to current conservation, the field enhancement arising from ion depletion at the constriction tips is extended along the centerline direction at the cathodic interface, whereas this field enhancement is abruptly lowered along the sidewall directions, due to ion accumulation within a short distance of the constriction tip. As a result, electrophoresis (EP) is enhanced along the centerline, whereas the nDEP trapping is enhanced along the sidewall in the vicinity of the constriction tip, thereby routing biomarkers from the off-sidewall directions towards localized nDEP traps at the constriction sidewall. In the absence of ICP effects, such as by using AC fields with a DC field that is not sufficient to create ICP, we would still have biomarker enrichment along the cathodic interface due to the balance of nDEP versus electrophoresis. However, in the presence of a critical DC field to cause ICP, the enhanced EP velocities and nDEP trapping alters the net potential profile to exponentially enhance biomarker enrichment due to its Boltzmann distribution.

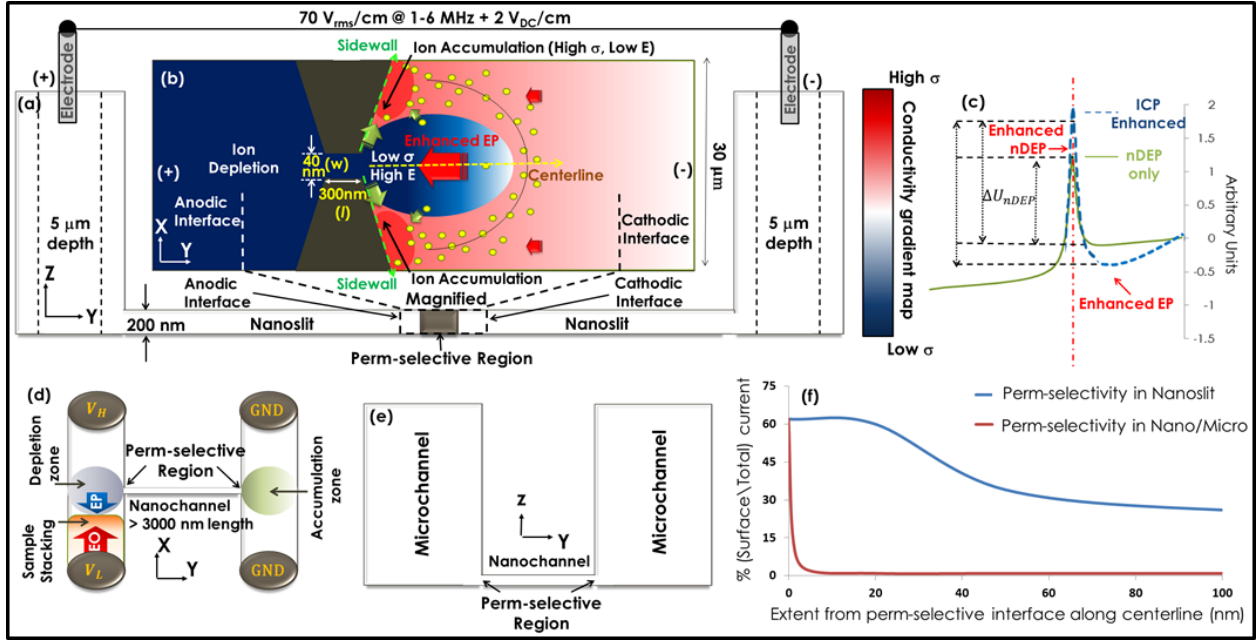

FIG. 1. (a) Cross-section view of the nanoslit device. (b) Top-view of the lateral constriction used to initiate ICP in the nanoslit (biomarkers shown as green particles). (c) alteration of electrokinetic trapping potential profile due to ICP enables exponential enhancement in biomarker numbers. The H-shaped ICP device from prior work: (d) top-view (e) cross-section. (f) Comparison of calculated surface to total ion current away from the perm-selective interface for the different geometries.

## B. Comparison to DC electrokinetics at micro/nano interfaces

It is noteworthy to compare the current device geometry (Figure 1a), which enables the initiation of ICP effects inside the nanoslit channel, to the commonly used H-shaped device geometry (Figure 1d and 1e), where ion depletion due to ICP at the microchannel interface to nanochannel enhances the local field (by factor $\alpha_1$) for biomarker electrophoresis ($F_{EP}$) versus electro-osmosis ($F_{EO}$), as given by the force balance:

$$F_{net} = \alpha_1 F_{EP} - F_{EO} \qquad (1)$$

With increasing media conductivity ($\sigma_m$), the increasing competition of bulk conduction with surface conduction in the nanochannel causes the level of ICP enhancement ($\alpha_1$) to drop, whereas the less thick electrical double-layer causes $F_{EO}$ to drop, thereby reducing the net level of biomarker enrichment ($F_{net}$). On the other hand, in the geometry of Figure 1a, by initiating ICP inside the nanoslit rather within the interfacing microchannel, we seek to maintain a significant level of ICP (which enhances $F_{EP}$ and $F_{nDEP}$ by factors of by $\alpha_2$ and $h$, respectively. Even within conductive physiological media, due to the strong influence of the surface charge non-uniformity ($\rho_c$) in the nanoslit on enhancing surface conduction over bulk conduction. With the H-shaped device geometry, on the other hand, ICP effects in the microchannel are less influenced by the surface charge non-uniformity created by the nanochannel / microchannel interface. This comparison of the influence of $\alpha_1$ versus that of $\alpha_2$ is shown in Figure 1f, in terms of the calculated surface to total ion current away from the perm-selective interface. For the device geometry of Fig. 1a, ICP due to the constriction-induced surface charge non-uniformity in the nanoslit enables an enhancement (due to $\alpha_2$) in surface conduction over an extent of 20 nm from the perm-selective interface, with only a gradual fall off over the next 80 nm to a steady enhanced level through the rest of the nanoslit channel length. In comparison, at the H-shaped device geometry, while the surface conduction effects are strong at the microchannel interface to the nanochannel due to the sharp field profile, this surface conduction drops off sharply within a few nanometers from the perm-selective interface into the microchannel, since the nanochannel surface charge presents a less weak influence on ion conduction within the microchannel. As a result, in order to maintain a significant degree of ion depletion within conductive physiological media, due to ICP at the microchannel to nanochannel interface, there is a need for tighter nanochannels and/or higher applied fields to enhance ICP conditions (or $\alpha_1$). These strategies are not practical due to fabrication challenges, disruptions from Joule heating induced electrothermal flow and voltage drop due to field screening by the electrical-double layer. A second major difference between these two ICP-based electrokinetic biomarker enrichment methods arises from how the ICP is utilized. Within the H-shaped device geometry, while ICP at the sharp microchannel to nanochannel interface creates a region with ion depletion at the anodic interface and ion accumulation at the cathodic interface, these ion depletion and accumulation regions do not interact due to their spatial separation. In fact, a spatial separation of at least 300 μm is required, especially within media of successively higher levels of conductivity, to prevent co-ion leakage from the cathodic to anodic portion of the microchannel / nanochannel interface, since this leakage will reduce the level of depletion along the anodic interface. Due to this spatial separation, only the localized ion depletion at the anodic interface is used to enhance the fields for DC-based electrokinetic enrichment, thereby creating sharp field gradients, with no ability to spatially modulate the field profile. Within our constriction-based nanoslit device geometry, on the other hand, the proximity of the ion depletion and ion accumulation regions cause their interaction, so that both regions act together to modulate the field profiles and create localized traps for enhancing the spatial extent of biomarker trapping. In fact, to maximize the voltage drop across the perm-selective region for enabling higher localized fields, we prefer designs with a short perm-selective spatial extent (usually 300 nm), so that it is just sufficient to maintain ion depletion along the cathodic interface. The third distinguishing feature of our approach is that instead of an opposing electroosmotic force field, we utilize AC fields to initiate frequency-selective nDEP away from the constriction tips, since nDEP continues to be a significant force field even within media of high conductivity. This is apparent considering the force balance at the cathodic interface in Figure 1a:

$$F_{net} = \alpha_2 F_{EP} - \beta F_{nDEP} - F_{EO} \qquad (2)$$

Hence, at higher $\sigma_m$, the drop in ICP-induced enhancement ($\alpha_2$ & $\beta$) can be arrested by using a sharper surface charge non-uniformity ($\rho_c \sim 0.08$-$0.3$ C/m$^2$) in the nanoslit, for enabling sufficient surface conduction versus increasing bulk conduction (Figure 1f). Furthermore, by using an opposing nDEP force field, a higher level of $F_{nDEP}$ in Eq. 2 is maintained at high $\sigma_m$, due to its dependence on the difference of particle versus medium conductivity: ($\sigma_p$-$\sigma_m$). In this manner, we continue to enhance the net biomarker transport towards the cathodic interface for biomarker enrichment within physiological media, since: (a) ICP effects that drop off within conductive media (lower $\alpha$ & $\beta$) continue to be significant within our device geometry, due to the significant role of surface conduction inside the nanoslit arising from the large surface charge non-uniformity of the lateral constriction ($\rho_c$), and: (b) nDEP effects rise rather than fall within conductive media, thereby maintaining a localized opposing force balance. In the subsequent sections, we compute the ion conductivity profiles to optimize the device geometry for enhanced biomarker trapping, and experimentally validating these ion conductivity gradients by mapping spatio-temporal biomarker trapping profiles under various applied field magnitude, frequency and bulk media conductivity conditions.

## C. Optimizing device geometry for enhancing nDEP

We consider optimization of the geometry of the perm-selective region within the nanoslit for enhancing electrokinetic biomarker enrichment under ICP conditions. Specifically, we compute ICP-induced ion conductivity profiles along the constriction sidewall and centerline directions, for varying lengths ($l$) and widths ($d$) of the perm-selective region, thereby enabling an assessment of the ensuing enhancement in nDEP. The angle of lateral constriction in the nanoslit is held constant at 30° to enable the highest possible field gradient ($\nabla|E|$), under the fabrication limitations. Due to the $\nabla|E|^2$ dependence of nDEP, we seek device geometries that promote ion depletion at the constriction tip to enhance the field ($E$) and initiate ion accumulation along the constriction sidewall to enhance the field gradient ($\nabla|E|$). Since the ion conductivity profiles in Figure 2 are computed in media with bulk conductivity of 1 S/m, the regions with localized conductivity less than 1 S/m exhibit ion depletion and those at greater than 1 S/m exhibit ion accumulation, with the insets (i-iv) indicating the respective region of counter-ion accumulation (solid orange circles). For the case of a perm-selective region with a large length (3000 nm), the surface charge non-uniformity has a greater electrostatic interaction length with counter-ions undergoing surface conduction. This will limit the level of surface conduction of counter-ions out of the perm-selective region, thereby causing counter-ion accumulation inside the perm-selective region, including at the constriction tips. As a result, the degree of ion depletion inside the perm-selective region is reduced, thereby limiting the degree of field enhancement at the constriction tip along the cathodic interface. The computed ion conductivity profile in Figure 2a for this case (3000 nm length of perm-selective region) shows no apparent ion depletion, with only a maximum in ion accumulation that occurs right at the constriction tips, as per inset (iv). For successively smaller length extents of the perm-selective region, the higher localized field and lower electrostatic interaction length with counter-ions enhances the net surface conduction. As a result, ion depletion at the constriction tips is enhanced (Figure 2a, inset v) to cause further increases in the localized field ($E$). However, the level of ion accumulation along the sidewall direction is seen to gradually decrease with increasing perm-selective extent, thereby lowering the field gradient ($\nabla|E|$), with the peak region for ion accumulation being pushed further away from the constriction tip along the sidewall. This suggests that an optimal length extent of the perm-selective region is needed for maximum electrokinetic enhancement, due to its $\nabla |E|^2$ dependence. Hence, we compute an optimal perm-selective length for surface conduction enhanced nDEP (factor "$h$" in Eq. 2), based on the product of the rise in potential energy barrier height due to field enhancement ($E$) under ion depletion (by factor: β) and the deeper potential well due to field gradient ($\nabla |E|$) under ion accumulation (by factor: κ).

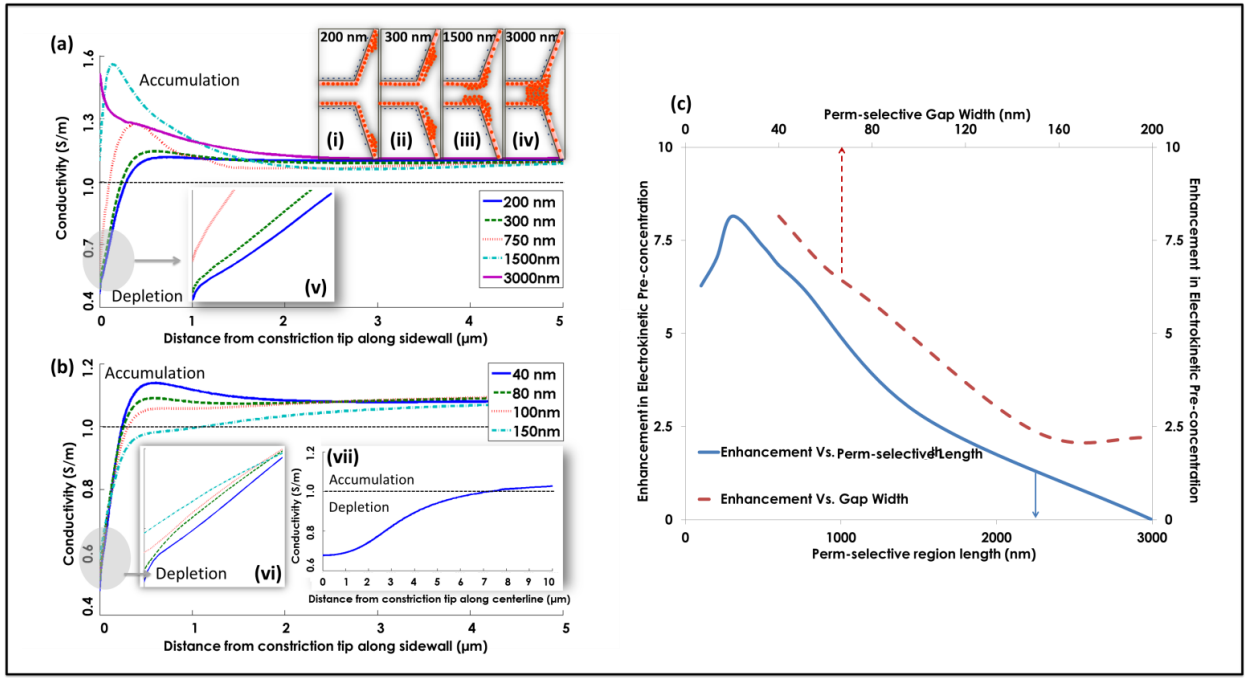

FIG. 2. (a) Ion conductivity profile along the constriction sidewall for different perm-selective lengths (gap width: 40nm) within media of bulk conductivity of 1 S/m. Insets (i-iv) show schematic of counter-ion distribution for different perm-selective lengths and (v) shows a magnified view of the depletion for each perm-selective length. (b) Ion conductivity profile along the constriction sidewall for different perm-selective gap widths (length: 300nm) at medium conductivity of 1 S/m. Inset (vi) shows a magnified view of the depletion at each constriction gap and inset (vii) shows ion conductivity profile along the constriction centerline for perm-selective region with 40nm width and 300 nm length with media of bulk conductivity of 1 S/m. (c) Calculated electrokinetic enhancement due to ICP for different perm-selective lengths (solid blue line) and gap sizes (dashed red line).

It is apparent from the plot in Figure 2c that a perm-selective region with a length of 300 nm is optimal, since it is capable of causing ~8-fold level of electrokinetic trapping enhancement (factor "$h$" in Eq. 2) at bulk media conductivity of 1 S/m to cause exponentially higher biomarker numbers (e$^{hU}$) versus that for nDEP with no ICP. At a smaller length of the perm-selective region (200 nm), while the degree of ion depletion is higher at the constriction tips (increasing $E$), the ion accumulation is lower and occurs too far away from the constriction tip to cause the necessary rise in $\nabla E^2$, for enhancing ICP-induced trapping. On the other hand, with gradual lowering of the constricting perm-selective gap, the enhanced exclusion of co-ions and surface conduction of counter-ions causes a higher level of ion depletion (Figure 2b, inset vi), as well as ion accumulation.

This explains the monotonic rise in the net electrokinetic enhancement in Figure 2c, for gaps varying from 160 nm to 40 nm. Finally, we compare the ion conductivity profile along sidewall (Figure 2b) versus centerline directions (Figure 2b, inset vii). For the optimal perm-selective device geometry to enhance electrokinetic enrichment (40 nm constriction gap with a 300 nm length extent), the centerline direction undergoes ion depletion over a wide extent (~7.5 μm from constriction tip), with more modest levels of ion accumulation thereafter, whereas the ion depletion region along the sidewall direction is localized within 0.5 μm from the constriction tip, followed by strong levels of ion accumulation. This quantitative picture is consistent with the schematic ion conductivity profiles, shown earlier in Figure 1b to explain the enrichment mechanism.

## C. Dimensionless perm-selectivity factor for AC electrokinetic enhancement

A critical level of perm-selectivity is required to initiate the necessary ion conductivity gradients that arise due to spatially proximal ion depletion and ion accumulation regions, for enhancing the net electrokinetic enrichment. We describe the necessary perm-selective geometry by the following dimensionless analysis. The necessary perm-selectivity ($PS$) depends on the perm-selective gap length ($l$), its cross-sectional area ($A$: product of gap width to nanoslit depth), its surface charge distribution ($\rho_c$), the ion mobility (μ) and media conductivity ($\sigma_m$). We write this in terms of the following dimensionless parameter:

$$PS = \frac{\rho_c \mu L^2}{\sigma_m A} \qquad (3)$$

Based on this, critical $PS$ values are in the range ($6 \times 10^{-9}$ to $6 \times 10^{-6}$). Unlike the case of DC electrokinetics, where maximum levels of perm-selectivity are needed, as achieved through reducing perm-selective gap (to reduce $A$) or increase length ($L$) to exponentially increase *PS*, we prefer *PS* values inside this range. For *PS* values above: $6 \times 10^{-6}$; ion accumulation and depletion are too far spatially separated, so that accumulation occurs only on the cathodic side and depletion occurs only on the anodic side. Such wide spatial separation of the ion accumulation and depletion regions coupled to ion accumulation at the cathodic side where trapping occurs under nDEP, reduces the electric field and their gradients, thereby reducing the net nDEP trapping efficiency. On the other hand, at *PS* values less than $6 \times 10^{-9}$, ICP effects are not present, thereby causing no enhancement in nDEP trapping due to ICP.

**D. Influence of bulk media conductivity**

Conductivity of the bulk media ($\sigma_m$) determines the electrical double-layer (EDL) thickness within the perm-selective region (schematically shown within insets (i-iii) in Figure 3a), so that the ensuing differences in surface conduction strongly influence the ion conductivity profiles along the cathodic interface of the constriction. At lower bulk media conductivities ($\sigma_m$=0.2 S/m), the thicker EDL causes enhanced levels of co-ion exclusion and counter-ion surface conduction within the perm-selective region, which in turn enhances the levels of ion depletion and ion accumulation versus those achieved within media of higher conductivities ($\sigma_m$=1 and 2 S/m). Figure 3 presents this information in terms of the normalized conductivity; i.e. as a ratio of the local conductivity to the bulk conductivity, with ion depletion in regions below unity and ion accumulation in regions above unity. Comparing Figure 3a and Figure 3b, it is clear that the degree of ion accumulation is significantly greater along the sidewall direction versus the centerline, and it peaks at well less than 0.5 µm from the constriction tip. Along the centerline direction, on the other hand, the effects of ion depletion from the perm-selective region continue to be apparent over several microns from the constriction tip, with the peak in ion accumulation occurring much further along, thereby creating a wide region for the localized ion conductivity gradient along the cathodic interface. These results are consistent with the schematic in Figure 1b, showing ion depletion over a broad elliptical region along the off-sidewall directions and ion accumulation close to the sidewall directions.

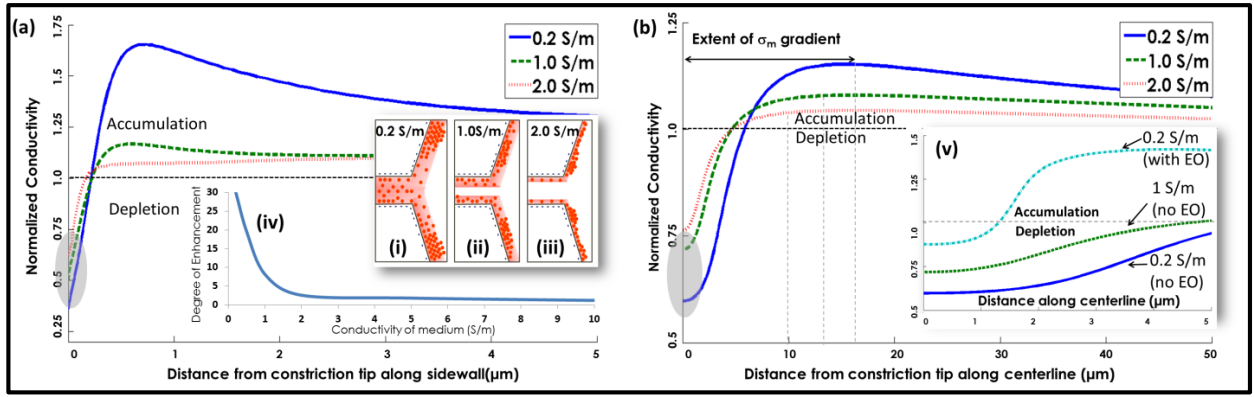

FIG. 3. Normalized ion conductivity profile for different bulk media conductivities (Perm-selective length of 300 nm and gap width of 40nm): (a) along the constriction sidewall, and (b) along the constriction centerline. Insets (i-iii) show a schematic of how the electrical double layer affects the counter-ion distribution, inset (iv) shows the net electrokinetic enhancement due to ICP effects within media of different bulk conductivities (neglecting electro-osmosis), and inset (v) shows the alteration of the extent of ion depletion when electro-osmosis effects are considered at $\sigma_m$ of 0.2 S/m.

At $\sigma_m$ of 1 S/m, there is no ion depletion beyond the constriction tip along the sidewall directions and ion accumulation peaks within 0.4 μm from the constriction tip, whereas along the centerline direction, the ion depletion extends to ~5 μm from the constriction tip and the localized conductivity gradient due to differential ion accumulation extends ~13 μm. As a result, the computed enhancement in potential barrier for trapping due to ion conductivity gradients at $\sigma_m$ of 1 S/m, continues to be significant in Figure 3c (~8-fold enhancement, with a conservative $\rho_c$: 0.08 C/m$^2$), even though the EDL is not thick enough to ensure maximum co-ion exclusion and counter-ion surface conduction within the perm-selective region versus the respective level at $\sigma_m$ of 0.2 S/m. However, while electro-osmotic effects on ion migration can be neglected at higher $\sigma_m$ (≥ 1 S/m), since electrophoretic ion mobilities are significantly greater than electro-osmotic mobilities in nanochannels[27],[28],[29] (see Supporting Information), this is not the case at $\sigma_m$ of 0.2 S/m, where the two mobilities are comparable[30]. The inset (v) of Fig. 3b shows the effect of including electro-osmosis at $\sigma_m$ of 0.2 S/m, on the localized ion conductivity profiles. While the ion depletion region extends over a significant portion of the device centerline in the absence of disruptions from electro-osmotic flow (~5 μm and ~8 μm from constriction tip at 1 S/m and 0.2 S/m, respectively), it is significantly reduced due to disruptions from electro-osmotic flow at $\sigma_m$ of 0.2 S/m (down to ~1.5 μm). We attribute this lowering of ion depletion along the cathodic interface at $\sigma_m$ of 0.2 S/m to the reduced co-ion leakage from the cathodic to anodic interface due to greater levels of co-ion exclusion at lower $\sigma_m$ and the opposing influence of electro-osmotic flow. Hence, the degree of biomarker enrichment at lower $\sigma_m$ is likely to be less optimal due to greater opposing electro-osmotic flow and lower localized nDEP trapping (see Eq. 2), which suggests a more dispersed biomarker trapping profile with lowering $\sigma_m$, as validated subsequently. Assuming a Boltzmann distribution for biomarkers, the ~5-fold enhanced potential barrier for electrokinetic trapping under ICP at $\sigma_m$ of 1.6 S/m using a conservative estimate for the surface charge non-uniformity ($\rho_c$: 0.08 C/m$^2$) in Fig. 3a (iv), suggests >100x higher levels of preconcentration ($e^{hU}$) than obtained solely under nDEP and up to 25-fold reduction in biomarker accumulation times.

**E. Validation using spatio-temporal biomarker profiles**

In order to explain how these conductivity gradients along the cathodic interface of the constriction region influence the profile and level of electrokinetic biomarker enrichment, we present spatio-temporal images and temporal plots of enrichment of labeled streptavidin at 1 MHz in Figure 4 (a-k), from a starting level of 100 ng/mL. Furthermore, to illustrate the frequency selectivity offered by this method, we identify conditions for enrichment of PSA versus interfering antibodies (Figure 4l).

The fluorescence intensity versus time plots for various conditions of nDEP induced electrokinetic trapping are determined by averaging the maximum intensities from 20 pixels along the sidewall direction and normalizing these fluorescence levels to those obtained under purely DC fields, where no perceptible rise in fluorescence is apparent. For comparison, we choose the following: (1) trapping within constriction gaps of varying size: ~ 40 nm in Figure 4a-4c versus ~130 nm in Figure 4f-4h; (2) trapping under varying degrees of surface conduction for inducing ICP conditions, using a critical DC offset field (1.5 $V_{DC}$/cm) to enhance ICP in Figure 4c versus a sub-critical DC offset field (0.2 $V_{DC}$/cm) to obviate ICP conditions in Figure 4e; and (3) trapping within physiological media, using PBS buffers with $\sigma_m$ of 1.6 S/m versus 10-fold diluted PBS buffers with $\sigma_m$ of 0.2 S/m (Figure 4b versus Figure 4j, after 1s trapping). For nanoslits with 40 nm constrictions (Figure 4a-4c), within 0.3s of field application, a significant level of biomarker preconcentration is apparent along the constriction sidewall within 0.3 µm of the constriction tip and biomarker depletion over a broad elliptical region that extends ~20 µm along the centerline is also clear. Over time, the biomarker depletion region remains more-or-less unchanged, while biomarker preconcentration gradually extends from the constriction sidewall directions to cover the off-sidewall directions. For the case of nanoslits with 130 nm constrictions, the profiles are somewhat similar (Figure 4f-4h), with initial preconcentration along constriction sidewalls and a biomarker depletion region along the centerline. However, the preconcentration is considerably slower[31], as apparent from the longer time intervals, and the biomarker depletion region extends no more than ~3 µm along the centerline. For the case of trapping under nDEP conditions with a sub-critical DC field for initiating ICP, the biomarker enrichment rate is gradual (Figure 4k) and the trapping profile shows only a very small biomarker depletion region (~0.3 µm along centerline in Figure 4e).

Within 10-fold diluted PBS media ($\sigma_m$~0.2 S/m), the trapping is dispersed, with no biomarker depletion region, even with a critical DC field for initiating ICP alongside nDEP. These enrichment profiles under various field, media conductivity and device geometries validate the role of $\sigma_m$ gradients generated under ICP conditions along the cathodic interface of the constriction region inside the nanoslit (as described in Figure 2 and Figure 3). As per the schematic in Figure 4d depicting biomarker and ion transport in nanoslits with constrictions of 40 nm gap, the high field due to ion depletion near the constriction tip and the high field gradient due to ion accumulation along the constriction sidewall cause a localized nDEP trap (green arrows) within ~0.3 µm from the constriction tip along the sidewall, which correlates with region of maximum biomarker in Figure 4a-4c.

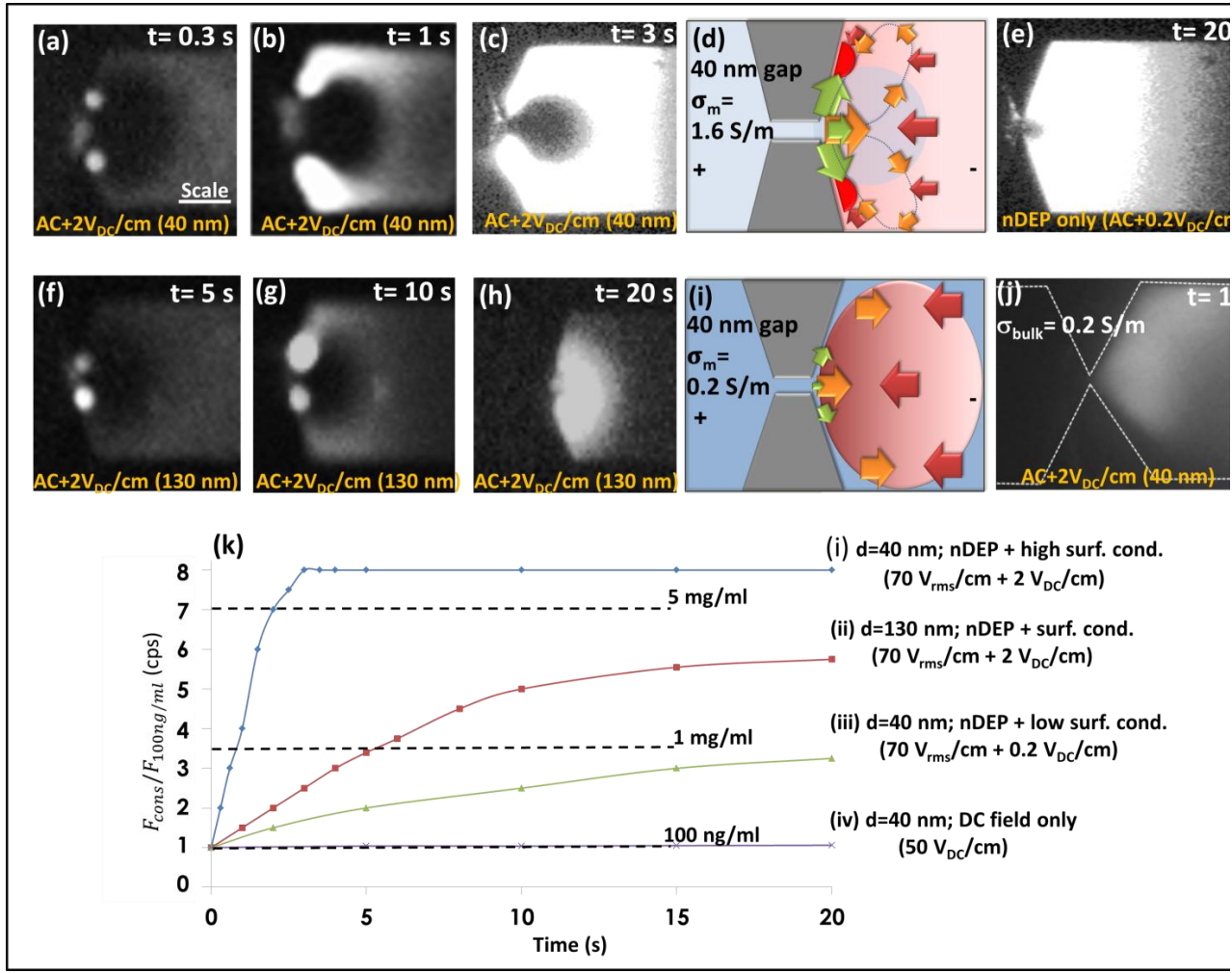

FIG. 4. Fluorescence images of streptavidin preconcentration in physiological media ($\sigma_m \sim 1.6$ S/m) from a starting level of 100 ng/mL, under an electrokinetic force balance due to 70 V$_{rms}$/cm (AC) plus 1.5 V$_{DC}$/cm offset for: (a) t=0.3s, (b) t=1s, (c) t=3s of enrichment at nanoslits with 40nm constriction gap widths and (f) t=5s, (g) t=10s, (h) t=20s of enrichment at nanoslits with 130nm constriction gap widths. (d) Preconcentration under predominantly nDEP conditions (no ICP enhancement) after 20 s of enrichment; and (j) preconcentration under conditions similar to (a-c), but in media of 10-fold lower conductivity ($\sigma_m \sim 0.2$ S/m). A schematic explanation of the biomarker preconcentration profile under the force balance of electrophoresis (red) vs. nDEP (Green) and enhanced second order electroosmosis (orange) is shown for: (d) physiological media, and (i) 10-fold diluted media. (k) Fluorescence intensity versus time averaged for 20 pixels along the constriction sidewall with highest intensity and normalized to fluorescence under purely DC fields. (l) Biomarker enrichment versus frequency of applied field (0.8-6 MHz) for streptavidin (light blue), PSA (orange) and anti-mouse IgG (green). Blue scale bar = 10 μm.

On the other hand, the high field along the device centerline arising from extension of ion depletion onward from the constriction tip drives biomarkers by electrophoresis (red arrows) away from the centerline towards the nDEP trapping region along the sidewall, correlating to the broad biomarker depletion region away from the off-sidewall directions up towards the device centerline. The localized $\sigma_m$ gradients due to ICP at bulk $\sigma_m$ of 1.6 S/m (Figure 3a (iv)) that cause at least 5-fold enhancement in potential barrier for trapping (assuming a conservative $\rho_c$: 0.08 C/m$^2$) suggest >100x higher levels of enrichment than obtained solely under nDEP. As per the fluorescence levels in Figure 5k, this causes localized enhancement of >5000x in streptavidin concentration. In fact, our prior work quantified this at $10^4$-$10^5$-fold concentration enhancement in the preconcentration region and ~$10^3$-fold enhancement in analyte binding at capture probe surface[20]. This is apparent from the sharp rise in fluorescence signal to saturation levels in Figure 5k, within just 3 seconds (curve (i)). For the nanoslit with a constriction of 130 nm gap (Figure 4f-4h), the biomarker enrichment profile is quite similar to that for the equivalent device at 40 nm, but the lower level of nDEP trapping, as well as the ICP enhancement of electrokinetic trapping due to lower ion depletion levels cause a less sharp rise in fluorescence signal versus time (Figure 4k, curve (ii)) and smaller region of biomarker depletion. It is noteworthy that ICP-enhanced nDEP trapping with a 130 nm constriction gap is greater than that obtained under nDEP trapping conditions at a 40 nm constriction gap, with no significant ICP enhancement (Figure 4k, curve (ii) vs. (iii)). This suggests that nDEP trapping within nanoslits containing larger constriction gaps (~130 nm) can be enhanced over that obtained utilizing smaller constriction gaps (~40 nm), by sufficiently increasing DC offset field to enhance surface conduction for inducing ICP. Within media of lower bulk conductivity (10-fold diluted PBS media at $\sigma_m$~0.2 S/m), the highly dispersed biomarker enrichment along the entire cathodic interface (Figure 4j) is explained by the schematic in Figure 4i. Herein, the force balance is dominated solely by biomarker electrophoresis (red) versus electro-osmosis in the nanoslit (orange), with a minimal role for nDEP trapping (green), thereby explaining the absence of the biomarker depletion region along the centerline and the highly focused biomarker trapping region along the sidewall directions that are seen within PBS media.

# Methods

## A. Theoretical Formulation

Ion conductivity profiles were computed for perm-selective geometries within the nanoslit, assuming constant surface charge non-uniformity ($\rho_c$) through applying Neumann flux/source boundary condition as shown in equation 5. We use a conservative estimate of: $\rho_c = -0.08\ Cm^{-2}$, [32], which is well lower than possible higher end estimates ($\rho_c = 0.3\ Cm^{-2}$).[33] At equilibrium the electric potential, $\psi$, distribution is governed by the Poisson equation[34]:

$$\nabla^2 \psi = -\frac{1}{\varepsilon_r \varepsilon_0} \sum_{i=1}^{N} z_i e n_i \qquad (4)$$

To solve Poisson equation, Dirichlet boundary condition was applied to inlet/outlet walls in reservoirs, where electrodes are inserted, to simulate the voltage source. All other boundaries are set to zero flux. In $NaCl$ solution, $N = 2$ is the number of ionic species, where $n_i$ ($i = Na, Cl$) and $z_i$ ($i = Na, Cl$) are the number density and valence of $i^{th}$ ionic species respectively. $\varepsilon_0$ and $\varepsilon_r = 80$, represent permittivity of vacuum and relative permittivity of the medium, respectively. Using a continuum treatment of the problem, in the absence of chemical reactions, the flux of each of these ionic species is described by Nernst-Planck equation, which represents the balance between convection, diffusion and ionic drift due to presence electric field[32],[35].

$$\frac{\partial c_i}{\partial t} = \nabla \cdot \left( D_i \nabla c_i + \frac{D_i e z_i \nabla \psi}{K_B T} c_i - \vec{u} c_i \right) \qquad (5)$$

With ($i = Na\ or\ Cl$), $D_{Na} = 1.33 \times 10^{-9}\ \frac{m^2}{s}$ and $D_{Cl} = 2.03 \times 10^{-9}\ \frac{m^2}{s}$ are the diffusion coefficient of ion species, $K_B$ is Boltzmann constant, $T$ is the temperature, $c_i$ is the molar concentration of species at each point of the device and $u$ is the fluid velocity vector. To solve this equation, all boundaries except inlet and outlet walls were assigned zero flux boundaries for both ionic species. Dirichlet boundary condition equal to bulk concentration is assigned to both Inlet/outlet for both ionic species. In steady state, continuity equation requires: $\nabla \cdot J_i = 0$. A closer examination of equations (3) and (4) reveals their interdependence, thereby requiring them to be solved simultaneously. Together they are known as Poisson-Nernst-Plank (PNP) equations, which should be solved numerically, since analytical solutions are not available for their nonlinear form[32]. Using standard boundary conditions for continuity of: $\psi$ and $C_i$; and of their first derivatives, we used COMSOL Multiphysics to solve these equations over 2D space to find the concentration distribution of ionic species at any point in the device. The only exceptions are at the constriction tip and within the walls of the perm-selective region, wherein the electrical boundary condition is set to account for the surface charge density ($\rho_c$):

$$\frac{d\psi}{dr} = \frac{\rho_c}{\varepsilon_0 \varepsilon_r} \tag{6}$$

Assuming incompressibility of the fluid, $\nabla \cdot u = 0$, Navier-Stockes equation, which governs the fluid momentum completes these set of equations:

$$\rho_m \frac{\partial u}{\partial t} = -\nabla p + \eta \nabla^2 u + f \tag{7}$$

In this equation $\rho_m = 1000 \ kg/m^3$ is the water density, $p$ is the pressure, $\eta = 1.002 \ mPa.s$ is the water dynamic viscosity and $f$ is the body force. In the nanoslit design, due to the high surface to volume ratio, we can neglect dielectric forces [36], and electric body force can be defined as follows, with $\rho_f$, as the free charge density:

$$f = -\rho_f \nabla \psi \tag{8}$$

To solve for fluid flow, the no-slip Dirichlet boundary condition was applied to walls of the microchannel, constriction and perm-selective regions. Along these boundaries, a layered quadrilateral mesh with dense element distribution in the normal direction was used to resolve the thin boundary layers along the no-slip boundaries. However, for simulations in media of bulk conductivity of ≥1 S/m, the double layer thickness is within 1% of nano slit depth. Hence, in order to significantly lower computational requirements in our set of simulations within the microchannel region, which is very far from perm-selective zone, Dirichlet electroosmotic velocity boundary condition can be applied to the boundaries (walls) through choosing proper electroosmotic mobility for each bulk conductivity value.

After solving equations (4, 5 & 7), the current through any particular cross sectional area of the channel can be calculated [37]:

$$I = \iint (-\Lambda\nabla(\psi + \phi) - F\sum z_i D_i \nabla c_i + Fu\sum z_i c_i) \tag{9}$$

Here: $\Lambda = F^2 \sum z_i^2 m_i c_i$ is the electrolyte electrical conductivity, $m_i$ is the electrical mobility, $c_i$ is the molar concentration, $\phi$ is the electrical potential distribution due to the external applied voltage, and finally $F$, is the Faraday's constant. At steady state, non-uniform distribution of electric field, $E$, will cause the particles in the medium to experience dielectrophoretic force ($F_{DEP}$). The average $F_{DEP}$ on a homogeneous spherical particle with electrical permittivity $\varepsilon_p$, conductivity $\sigma_p$, and radius, $a$, suspended in a fluid with electrical permittivity $\varepsilon_m$, and conductivity $\sigma_m$, is given by the following:

$$\vec{F}_{DEP} = 2\pi a^3 \varepsilon_m Re \underbrace{\left(\frac{\varepsilon_p^* - \varepsilon_m^*}{\varepsilon_p^* + 2\varepsilon_m^*}\right)}_{K_{CM}} E.\nabla E = \xi |E|.\nabla|E| \tag{10}$$

$K_{CM}$ is the Clausius–Mossotti factor, which depends on the frequency ($\omega$) of the applied field as given by: $\varepsilon_p^* = \varepsilon - j\sigma/\omega$. Due to the DC offset of the applied voltage, particles in the medium will also experience electrophoresis ($F_{EP}$), which can be calculated from the following equation:

$$F_{EP} = \gamma \mu_{EP} E_{DC} \tag{11}$$

Here, $\mu_{EP}$ is electrophoretic mobility, and $\gamma = 6\pi\eta a$, is the friction coefficient for a spherical particle with radius $a$ in a medium with viscosity of $\eta$. Biomarker trapping occurs under electrokinetic force balance of $F_{nDEP}$ and $F_{EP}$, under a zero net force on the particle, $F_{net}$:

$$F_{net} = \alpha_2(x,y,\sigma_m) F_{EP} - \beta(x,y,\sigma_m) F_{DEP} \tag{12}$$

$\alpha_2(x,y)$ & $\beta(x,y)$ denote the respective levels of local ICP-induced enhancement as a function of position and medium conductivity. Since all steady-state electrical forces are conservative vector fields, we can calculate scalar potential energy fields, to describe the net electrokinetic force balance in terms of barriers and wells that affect particle trapping within the device. The potential field is given by the integral over the volume ($dV$) of the net electrokinetic force[18]:

$$U = \oint F_{net}.dV \tag{13}$$

This integral in the presence of AC and DC fields is:

$$U = \frac{\xi\beta E^2}{2} - \alpha\gamma\mu_{EP}.V_{DC} \tag{14}$$

$V_{DC}$ is the offset voltage to enable electrophoresis (EP).

## B. Experimental Methods

Details of experimental device geometry and operation are available in prior work[18-22] and in Supporting Information. Biomarkers were procured as follows: Alexa 488 labeled Streptavidin (~60 kDa) from Molecular Probes (Eugene, OR), Alexa 488 labeled Goat anti-mouse IgG antibody (~150kDa) from Invitrogen / Thermo Fisher Scientific, and Dylight 594 labeled Prostate Specific Antigen (~28 kDa) from Sigma Aldrich (St. Louis, MO).

# 1.2. Quantifying spatio-temporal dynamics of biomarker pre-concentration and depletion in microfluidic systems by intensity threshold analysis

**Introduction**

Sensitivity gains from the scale-down of various sensor paradigms can be realized only if the ensuing mass transport limitations are alleviated[38], since the slow analyte settling and binding kinetics at the sensor surface can substantially delay signal onset at micro/nanosensors[5]. Micro/nanofluidic device methodologies are routinely applied towards enabling analyte pre-concentration within physiologically relevant media[39-41] for reducing target diffusion time towards the sensor[16] and thereby enhancing detection sensitivity[42]. Selective pre-concentration also enables analyte enrichment over interfering proteins and small molecules[20], which is especially relevant given the wide concentration range of proteomic biomarkers within typical biofluids (mg/mL – pg/mL)[2]. In order to capitalize on the reduced diffusion lengths towards the sensor upon analyte pre-concentration within microfluidic and nanochannel geometries, there is a need to effectively overlap the pre-concentration and detection regions for ensuring enhanced target binding kinetics at the sensor. We describe herein a methodology for quantifying the spatio-temporal dynamics of biomarker preconcentration within microfluidic systems to ensure effective alignment of the sensor to pre-concentration within the microfluidic device, for guiding device fabrication.

A variety of force fields are utilized within microfluidics to realize varying degrees of pre-concentration, including electrokinetic, chemical, diffusional, inertial, and magnetic methods [43, 44]. The degree of pre-concentration is quantified by following the time evolution of fluorescence signals from labeled analyte biomarkers as a function of the force field driving the enrichment[45], [17]. Quantification of the spatio-temporal dynamics of pre-concentration has usually been accomplished by monitoring fluorescence alterations over time inside a particular region of interest (ROI), as specified by discrete points[46], [47], or along a representative line axis[48],[49], [50], or a statically defined box[51],[52] to identify the time and location of the peak intensity. All these prior methods for quantifying pre-concentration have been based on the application of a static ROI. One drawback of utilizing such passive methodologies to follow pre-concentration dynamics is the error that can be introduced due to exclusion of the full extent of pre-concentration or inclusion of the background intensity, thereby leading to underestimation or overestimation of the pre-concentration degree over time. Furthermore, static ROIs cannot capture the spatial dynamics of the pre-concentration zone, such as its extent or rate of growth. Finally, while such passive ROIs may be adequate in situations where pre-concentration is highly localized, they are not well suited for cases where pre-concentration has a more diffuse or delocalized nature, such as is common under a balance of opposing force fields.

Dielectrophoresis (DEP) enables the frequency-selective translation of polarized particles under a spatially non-uniform electric field[12, 14] and is routinely applied towards pre-concentration of biomolecules[13, 19, 36]. Positive DEP (pDEP) or particle translation towards localized regions of high field causes rapid pre-concentration. Quantification of pre-concentration dynamics under pDEP, as presented within prior work[53-58] is relatively straightforward due to the spatially localized nature of the pre-concentration region. Negative DEP (nDEP) or particle translation away from localized high field points can also cause rapid pre-concentration when it is applied in conjunction with electro-osmosis or electrophoresis[18]. However, since pre-concentration under the ensuing electrokinetic force balance is spatially delocalized, there is a need to develop methodologies that are based on the pixels over the entire extent of the pre-concentration zone, rather than utilizing methods based on pixels within a limited ROI, such as point, line, or boxed areas. Furthermore, given the wide variations in pre-concentration extent with force fields, there is a need for methodologies to dynamically define the ROI for characterizing the pre-concentration dynamics. Finally, since pre-concentration occurs chiefly due to the high spatial non-uniformities of the electrokinetic force fields, there is a need for criteria to weight each pixel within the pre-concentration zone based on its geometric location from the high field points that drive this force field, for accurately quantifying the relative particle polarizability and force dispersion dynamics.

Herein, we present a dynamic methodology for quantifying the pre-concentration dynamics of particles under an electrokinetic force balance, driven by nDEP away from insulating constriction tips, which causes a highly delocalized pre-concentration zone. This automated methodology for quantifying pre-concentration is based on defining a threshold intensity level for dynamic determination of the ROI, through a statistical description of particle distribution across the device geometry to enable the quantification of pre-concentration and depletion signals over the background level. We illustrate the need for such a method by utilizing it within two distinct applications: (a) for optimizing placement of the sensor electrode within a nanochannel device for overlapping the pre-concentration and sensor regions during the electrochemical detection of neuropeptide Y (NPY), a biomarker for stress; and (b) for quantifying the polarizability dispersion of silica nano-colloids by accounting for the field non-uniformities across the pre-concentration zone through weighting pixels in the ROI based on their geometric location from the force field driving the pre-concentration. In this manner, we are able to quantify the pre-concentration dynamics and the DEP force dispersion under both, positive and negative DEP. Given the ability of this image analysis methodology for automation, with no dependence on the pre-concentration position, profile or extent, we envision its universal utility in quantifying pre-concentration dynamics over 2D images, as well as from 3D Z-stack data acquired by confocal microscopy.

**Methods**

**A. Device geometry for pre-concentration**

Dielectrophoresis was conducted within an electrode-less device geometry, with external Pt electrodes (Alfa Aesar) that were driven using a function generator (Agilent 33220 A) and a voltage amplifier[59] for generating applied fields of ~300 $V_{pp}$/cm over a 10 kHz to 5 MHz frequency range, with sharp dielectric constrictions to create localized high field points for enabling DEP pre-concentration of polarized particles at the constriction tips by pDEP or away from constriction tips by nDEP. In this manner, fluorescently labeled NPY biomarkers (Phoenix) were pre-concentrated in PBS media (150 mM NaCl and 2 mM $NaN_3$) at pH 7.2 and $\sigma_m$ ~ 1.6 S/m, using AC fields in the 1-3 MHz range within a quartz nanochannel (200 nm depth) away from constrictions (30 μm to 30 nm over 3 μm extent). Silica nano-colloids (Corpuscular Inc., Cold Spring, NY; 80 nm sized) were pre-concentrated in 0.1x PBS media ($\sigma_m \sim 0.1$ S/m) in the 100 kHz – 2 MHz range using constrictions (1 mm to 1 μm over a 10 μm length extent) patterned on PDMS (poly-di-methyl-siloxane) channels (7 μm depth). The micro/nanofabricated device was bonded onto cover slip glass for imaging using an inverted microscope (Zeiss Observer A1) and an EMCCD camera (Hamamatsu). For the electrochemical detection of NPY, graphene-modified glassy carbon electrodes were patterned on the cover slip and bonded to the channel after alignment to the pre-concentration region[20], based on procedures described herein to quantify pre-concentration dynamics. To enhance particle accumulation under nDEP, a small DC offset (0.5-1 V/cm) was applied to the AC field (300 $V_{pp}$/cm). As per the schematic in Figure 5, electro-osmosis (EO) under the DC field dominates at all points within the device except in the vicinity of the constriction, where nDEP dominates. Upon the balancing of EO with nDEP, particle accumulation occurs on one side of the constriction, with no accumulation on the other side, since nDEP and EO are in the same direction. The degree of pre-concentration of biomarkers is quantified by ratio of intensity enhancement over the background.

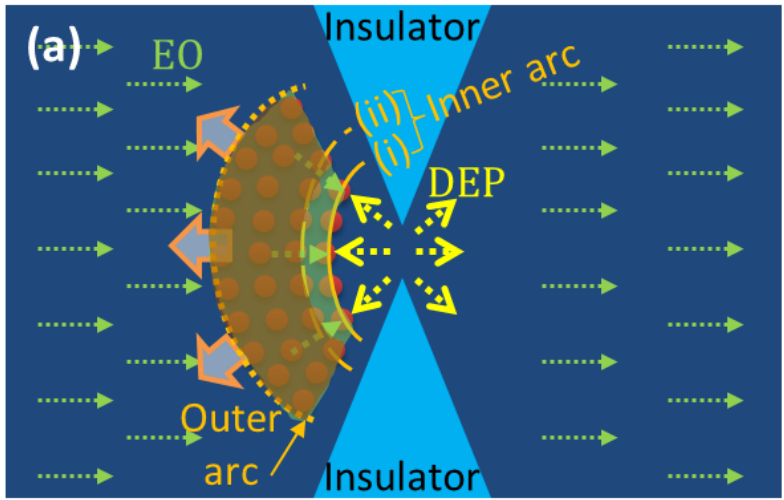

FIG. 5: Force balance under electro-osmosis (EO) versus negative dielectrophoresis at the insulator constriction (nDEP) causes pre-concentration of particles over a delocalized region of the microfluidic device geometry. The position of the inner arc, i.e. (i) vs. (ii) is determined by the force balance of nDEP and EO, with higher nDEP pushing the arc further from the constriction tip. The position of the outer arc is determined by the flux balance that includes diffusion. Fluorescence images of pre-concentration (greyscale) overlayed with equi-field lines (red lines) shows pre-concentration over a spread of field lines (one order of magnitude variation) under nDEP with EO.

## B. Dynamic determination of the region of interest (ROI)

Prior methods for quantifying pre-concentration dynamics were based on monitoring the time evolution of the cumulative or averaged florescence intensity of each pixel within a pre-defined ROI. This ROI, which is usually centered on the region with the highest pre-concentration, remains static irrespective of the time point of the image or alterations to the force field that cause the pre-concentration. The passive nature of this ROI determination method can cause errors in determination of the pre-concentration level, due to either signal underestimation caused by exclusion of the full extent of pre-concentration or overestimation caused by inclusion of background intensity levels from regions that do not experience pre-concentration. Figure 6 highlights this issue by presenting two cases: (i) an ROI chosen to cover the pre-concentration zone, including all of the collected time points (large ROI as in Figure 6d); and (ii) an ROI that is centered only on the region at a particular time point with the highest sample enrichment (small ROI as in Figure 6b). Since the pre-concentration zone tends to begin within a small region and gradually expand over time, the application of the large ROI of method (i) to measure enrichment level based on average pixel intensity within the ROI will lead to underestimation of the pre-concentration level and rate at the early time points, due to inclusion of pixels from regions not experiencing pre-concentration. On the other hand, if the pre-concentration level is quantified by summing the pixel intensities over the ROI, then the large ROI of method (i) will lead to overestimation of pre-concentration level at the early time points due to inclusion of pixel intensity from the background regions that do not experience pre-concentration. While this problem can be mitigated by using a background removal step to zero out the pixel intensity within the regions lacking pre-concentration, such an operation would make it difficult to detect particle depletion.

Similarly, the small ROI of method (ii) will lead to overestimation of the pre-concentration level if the signal averaging method is used, whereas it will lead to underestimation of pre-concentration level if the cumulative signal method is used, due to exclusion of some regions from the total signal.

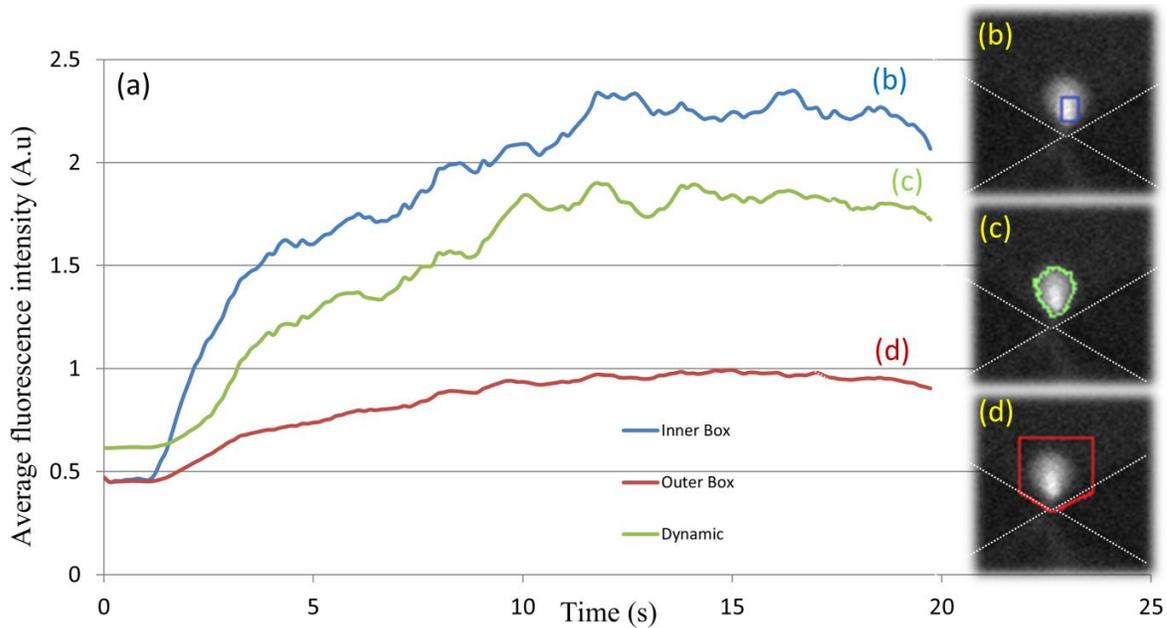

FIG. 6: (a) Comparing static methods for determination of region of interest (ROI) with dynamic methods for quantifying the pre-concentration from fluorescence signal levels. ROI based on the blue box inside the pre-concentration zone (b) versus the red box that surrounds the entire possible pre-concentration zone (d), is compared to the green curve that dynamically detects the boundaries of the pre-concentration region versus the background (c).

### C. Particle distribution statistics to determine signal and background thresholds

The criteria for setting the minimum and maximum threshold pixel intensity levels for determining sample pre-concentration and depletion, respectively, are described here. Since typical experiments to optimize pre-concentration levels are often performed under differing initial conditions, such as the starting particle concentration level, fluorescent labels of differing quantum efficiency, and variations in the intensity of the light source used for excitation, the threshold levels may need to be redefined for each experiment. Hence, we begin by delineating the background measured in a background region of interest (BROI), wherein the concentration of sample particles is uniform and not affected by pre-concentration or depletion, as shown in Figure 6a for our particular device geometry. The background intensity distribution will either be measured initially and remain constant through the duration of the pre-concentration experiment, or be continuously computed for each time point and/or varying force field. Accurate measurement of the background intensity can be challenging within some microfluidic device geometries, such as due to delocalized pre-concentration zones or overlapping pre-concentration and depletion regions, which can introduce substantial errors. In such situations, an ideal BROI is chosen to be in a region that is far enough from the pre-concentration zone, so that it is not affected by particle enrichment. In this background region, we can assume that particles are uniformly distributed in the media and consequently the light intensity distribution follows a normal distribution for which the mean intensity is defined as µ and the standard deviation is defined as σ. In such regions, only 2% of pixels should have intensity levels higher than two times the standard deviations above the mean intensity level: µ+2σ. If more than 2% of pixels in the BROI have intensity levels greater than µ+2σ or lower than µ-2σ, then it indicates that some portion of this BROI is experiencing either depletion or pre-concentration. Hence, the size of BROI is reduced row-by-row or column-by-column, starting from the side of the BROI that is closest to the likely pre-concentration region. After each reduction in size, the pixel intensity distribution is again measured and the process is repeated until the above criterion is satisfied. In this manner, the background region can be unequivocally assigned in an automated manner for each pre-concentration time point and force field variation. Once the background distribution has been delineated, it may be used to define threshold intensities for depletion and enrichment (Figure 6b).

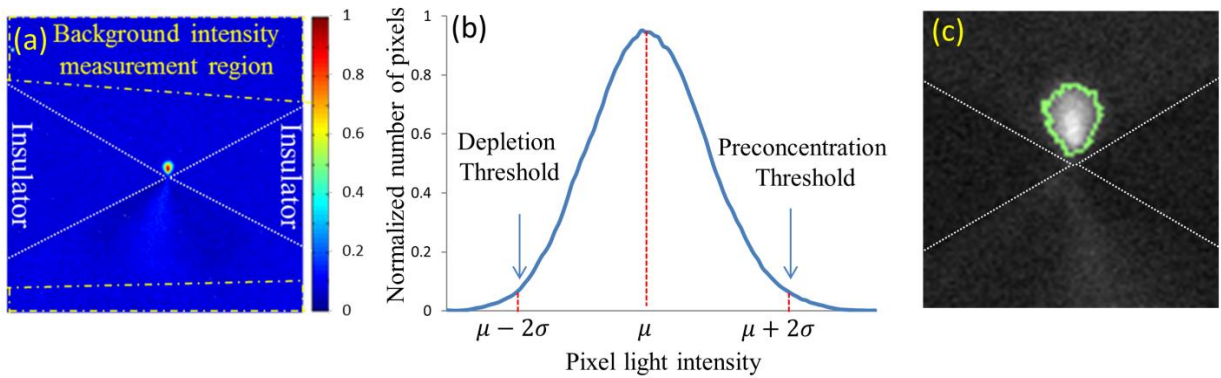

FIG. 6: (a) Defining the background region of interest (BROI) for measuring background intensity. (b) Normalized pixel intensity distribution in BROI and criterion for setting intensity threshold levels to detect pixels undergoing pre-concentration or depletion to determine the net pixel intensity distribution; (c) Dynamic determination of region of interest for quantifying pre-concentration. Light intensities are converted from a 16 bit scale to a scale of: 0-1.

Again, assuming the background light intensity distribution follows a normal distribution, there is a 98% chance that pixels with intensity levels greater than µ+2σ are experiencing particle enrichment, thereby delineating the threshold level for pre-concentration. In order to prevent false positives, our image analysis method labels a particular pixel as undergoing pre-concentration, only if one or more of its neighboring pixels are also above the threshold level. Similarly, since only 2% of pixels have intensity levels lower than twice the standard deviation below the mean intensity: µ-2σ, we delineate this as the threshold level for depletion. Using these thresholds, we can quantify the degree of pre-concentration or depletion of each pixel through intensity measurements.

**D. Computation of pre-concentration over 2D areas and 3D stacks**

Based on these threshold levels, an ROI of similar shape and extent as the pre-concentration zone can be dynamically defined (Figure 6c). Unlike the static ROI used in prior work, the dynamic ROI can adapt its shape, position and extent based upon alterations in the pre-concentration zone at different time points or under varying force fields, thereby enabling a more accurate computation of the cumulative and average pixel intensity at each time point of the pre-concentration. This dynamic method also allows for spatially characterizing the pre-concentration and depletion zones for computation of the rate of growth, direction of growth, and deviations from the previously defined ROI. Furthermore, it provides information on the statistics of particle distribution within the pre-concentration zone, which can be applied towards the design for localization of the sensor for particle detection. The threshold levels can also be defined based on sensitivity levels of a particular detection scheme to couple pre-concentration with detection. This dynamic method also enables accurate quantification of pre-concentration over a three dimensional volume. Images collected from a set of 3D z-stack images can be analyzed individually, each with its unique ROI, thereby allowing for accurate analysis of variations in the z direction. This would not be possible with a static ROI, since the varying shape, position and extent of the pre-concentration over the Z-stacks would lead to substantial error.

# E. Application towards aligning preconcentration and detection regions

We illustrate the relevance of this image analysis methodology for quantifying pre-concentration dynamics, by utilizing it to ensure optimal alignment of the electrochemical detection electrode for NPY biomarkers to its pre-concentration region under the nDEP force balance, which exhibits significant changes in shape and position over time. Figure 7a shows the highly delocalized pre-concentration profile of positively charged NPY under the electrokinetic force balance.

The image analysis methodology is utilized to quantify the position and shape of the regions that exhibit intensity enhancements up to 15x over the background. In fact, as per the time evolution of the intensity profiles in Figure 7c and 7d, the highest intensity region changes significantly in position and shape over the pre-concentration time, beginning close to the channel walls and evolving gradually towards the device centerline.

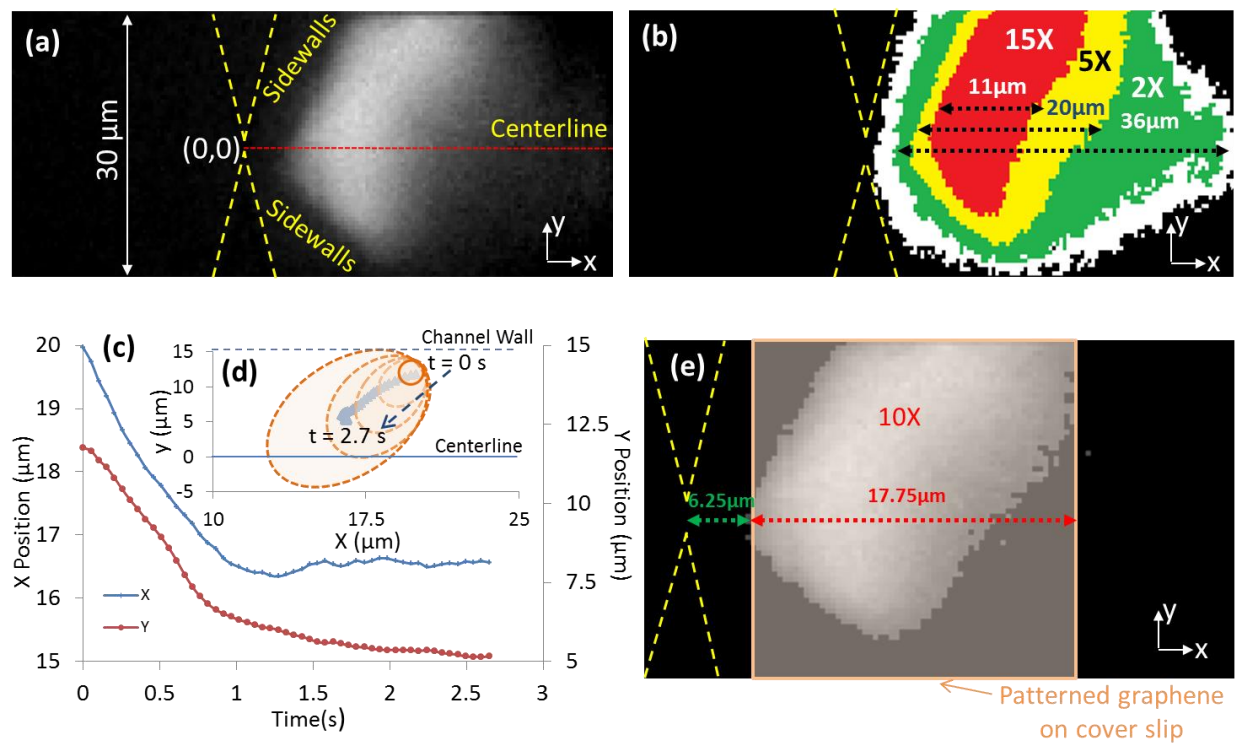

FIG. 7: Alignment of the graphene-modified electrode in the nanochannel based on spatio-temporal dynamics of NPY pre-concentration: (a) fluorescence image of NPY pre-concentration; (b) spatial spread of intensity enhancements over the active device region; (c) & (d) time evolution of pre-concentration region; (e) alignment of sensor electrode over the pre-concentration region for reducing diffusion lengths of pre-concentrated analytes.

In this manner, the edge position and width of graphene-modified glassy carbon electrodes patterned on the cover slip can be aligned to the constriction tip of the microfluidic channel prior to device bonding. This ensures effective coupling of NPY pre-concentration to electrochemical detection, as per Figure 7e, for covering the region with at least 10x intensity enhancements. Significant deviations in electrode alignment from this optimized position would reduce the binding kinetics of pre-concentrated NPY biomarkers on the sensor, thereby lowering signal from electroactive tyrosine groups, while the electrode area outside the pre-concentration region contributes to capacitive background without signal enhancement. In this manner, by utilizing this methodology to determine alignment of detection electrodes within the nanochannel of a constriction device, we have demonstrated the detection of neuropeptides at picomolar levels within a few seconds[20].

# 1.3. Frequency-selective enrichment of biomolecules in physiological media based on electrical double-layer induced polarization

**Introduction**

Biomolecular assays require the quantitative identification of trace levels of biomarkers within bio-fluids that also contain common interfering species, such as circulating antibodies, at million to billion-fold higher concentrations. Circulating human antibodies with receptor sites for animal proteins that arise upon exposure to particular antigens (so-called human anti-animal antibodies) have been recognized as a major source of interference to analyte detection using two-site or sandwich immunoassays[60],[4]. Specifically for the case of immunoassays for prostate specific antigen (PSA), human anti-mouse antibodies (HAMA) of immunoglobulins (IgG) can cause false positives, leading to misdiagnosis and overtreatment of patients[61],[62]. HAMA within serum can vary widely from μg/mL to g/mL levels, and can persist in blood for several months after antigen exposure. For the purpose of gauging post-operative cancer remission, PSA needs to be detected at sub-ng/mL levels[63]. Hence, the presence of PSA at nearly million to billion-fold lower levels than interfering HAMA species can lead to significant quantification errors, even at sub-1% interference levels. Current strategies to reduce interferences[64] include assay redesign, introduction of chemical modifications to HAMA and suppressing the patient's immune system. These approaches limit versatility of the immunoassay by requiring additional incubation steps and enhancing their costs due to the need for additional reagents, while their effectiveness is limited by the wide variations in HAMA levels within typical patients. An alternate strategy is to cause rapid and selective enrichment of the biomarker versus the interfering species[1], preferably within physiological media, to maintain its binding ability with receptors, without needing buffer changes that lead to dilution. Since antibody-based affinity methods that deplete the biomarker of interest cause only slow and mild levels of enrichment[2]; there is a need for complementary enrichment modalities.

Electrokinetic methods within nanochannels are commonly investigated for achieving highly enriched analyte plugs from dilute samples. Typically, a force balance under DC electrokinetics is coupled with the enhanced field arising from localized ion depletion in nanochannels to cause high degrees of biomarker enrichment[7]. However, due to the abrupt field profile, the trapped biomarkers are co-localized within a tightly confined region especially for molecules of like-charge, which limits the scope for selectivity based on spatially graded stacking. In this work, we instead utilize the frequency-selective features of AC electrokinetics, since the electrical double-layer around the biomolecule exhibits a characteristic time constant for surface conductance induced polarization. Specifically, the average distance required for surface conduction to induce biomolecular polarization differs based on its size and shape, while the sharp dependence of surface conductance on zeta potential of the biomolecule affects the magnitude of polarization.

Selective translation of particles occurs by dielectrophoresis (DEP) due to the characteristic frequency response of dielectric permittivity of the particle versus the medium[11],[12],[13],[14], which has been widely applied towards enriching µm-scale colloids. While DEP has been recently applied to the enrichment of nanoscale bio-colloids, such as ss-DNA[18],[16],[17] and proteins in physiological media[18],[19],[20],[21], there are no instances of its application in a frequency-selective manner towards biomarker enrichment and detection. The mechanistic basis of our work draws upon prior work[65] on sub-micron scale colloids within media of low conductivity, wherein colloidal size and surface charge characteristically alter the time constant for surface conduction in their electrical double-layer, thereby influencing dispersion of positive dielectrophoresis (pDEP). However, since we seek to selectively enrich and simultaneously bind proteomic biomolecules to receptors for enabling detection versus interfering species, we work within physiologically relevant media of high conductivity, wherein negative dielectrophoresis (nDEP) effects are highly significant. Hence, we focus instead surface conduction-induced alterations to the nDEP dispersion. Using DC-offset AC fields, these nDEP differences at the characteristic frequency of interest are balanced versus electrophoresis within nanochannels[66],[67], to cause rapid and selective biomarker enrichment versus interfering species. Based on prior work on PSA enrichment for speeding immunoassays[21], we enrich PSA with the necessary frequency-selectivity for reducing signal interferences on the immunoassay from circulating antibodies.

**Results and discussion**

**A. Physical basis for selective enrichment:** Polarized molecules are represented as spherical colloids of radius '*a*' within a media of permittivity: $\varepsilon_m$, experiencing dielectrophoretic behavior under an electric field ($E$) with a spatially varying gradient ($\nabla E$), caused by a trapping force ($F_{DEP}$) given by:

$$\vec{F}_{DEP} = 2\pi a^3 \varepsilon_m Re \underbrace{\left[\frac{\varepsilon_p'^{*} - \varepsilon_m^{*}}{\varepsilon_p'^{*} + 2\varepsilon_m^{*}}\right]}_{K_{i(\omega)}} |E|\nabla|E| \qquad (15)$$

Here, $K_{i(\omega)}$ is the frequency-dependent complex polarizability of the particle that depends on the complex permittivity of the medium ($\varepsilon_m^*$) and effective permittivity of the particle ($\varepsilon_p'^{*}$). The respective complex permittivities ($\varepsilon^*$) are defined in terms of dielectric permittivity ($\varepsilon$) and conductivity ($\sigma$):

$$\varepsilon^* = \varepsilon - j\sigma/\omega \qquad (16)$$

The effective complex permittivity of the colloid ($\varepsilon_p'^{*}$) is computed using a simplified dielectric model, wherein it is represented by an insulating core ($\varepsilon_{core}$ & $\sigma_{core}$) of radius: *a*, surrounded by a conductive ionic double-layer ($\varepsilon_{DL}$ & $\sigma_{DL}$) of thickness: $\kappa^{-1}$ and surface conductance: $K_s$, within media of relatively high conductivity ($\varepsilon_m$ & $\sigma_m$), as per Figure 8.

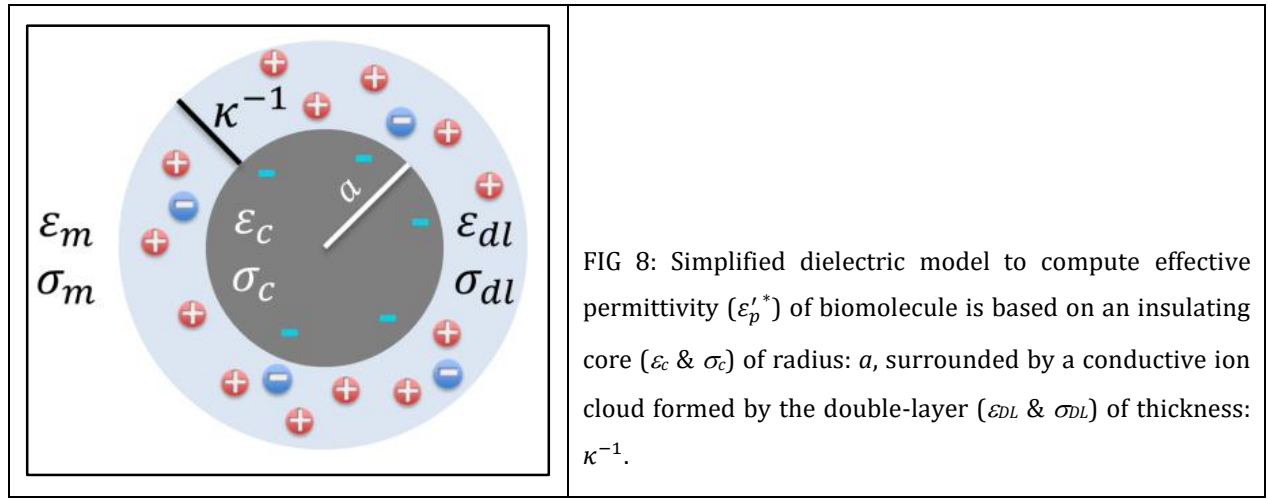

FIG 8: Simplified dielectric model to compute effective permittivity ($\varepsilon_p'^{*}$) of biomolecule is based on an insulating core ($\varepsilon_c$ & $\sigma_c$) of radius: *a*, surrounded by a conductive ion cloud formed by the double-layer ($\varepsilon_{DL}$ & $\sigma_{DL}$) of thickness: $\kappa^{-1}$.

Based on Eq. (16), polarization of the composite nanocolloid ($K_{i(\omega)}$) is determined at high frequencies (> 10 MHz) by the respective permittivity values: $(\varepsilon'_p - \varepsilon_m)/(\varepsilon'_p + 2\varepsilon_m)$, whereas at lower frequencies (<10 MHz) it is determined by the respective conductivity values: $(\sigma'_p - \sigma_m)/(\sigma'_p + 2\sigma_m)$. Due to the low net conductivity and permittivity levels of the colloid, we anticipate nDEP behavior or negative $K_{i(\omega)}$ values within conductive physiological media, since $\sigma'_p < \sigma_m$ and $\varepsilon'_p < \varepsilon_m$. Hence, within the schematic of Figure 9 that qualitatively describes the polarization dispersion versus colloid size, the nDEP behavior at high frequencies is represented (Fig 9a-9d) as medium-induced polarization anti-parallel to the applied field ($\vec{p} \uparrow\downarrow \vec{E}$), which is independent of particle size. At lower frequencies, the net particle conductivity ($\sigma'_p$) is chiefly determined by surface conductance ($K_S$) within the electrical double-layer of nanostructures, since $K_S/a$ can greatly exceed the conductivity of the particle core ($\sigma_c$), as per simplification:

$$\sigma'_p = \sigma_c + \sigma_{DL} = \sigma_c + \frac{K_S}{a} \tag{17}$$

As a result, the ion cloud around the particle can polarize parallel to the applied field ($\vec{p} \uparrow\uparrow \vec{E}$), thereby causing pDEP. This is especially apparent within smaller sized nanostructures (Fig. 9e vs. 9f), wherein the pDEP bandwidth is greater.

The polarization dispersion can be quantitatively computed using the simplified dielectric model in Figure 8, to compute the effective complex permittivity of the colloid ($\varepsilon'^*_p$) in terms of the dielectric properties of the insulating core and conducting double-layer, with the ratio of external (core plus double layer) to internal (core) radius represented by '$\gamma$'.

$$\varepsilon'^*_p = \varepsilon_{dl} \left( \frac{\gamma^3 + 2\left(\frac{\varepsilon_{core}^* - \varepsilon_{dl}^*}{\varepsilon_{core}^* + 2\varepsilon_{dl}^*}\right)}{\gamma^3 - \left(\frac{\varepsilon_{core}^* - \varepsilon_{dl}^*}{\varepsilon_{core}^* + 2\varepsilon_{dl}^*}\right)} \right) \tag{18}$$

As explained previously, the conductivity and permittivity of the core are insignificant versus the respective properties of the double layer. Hence, we focus on computing the represents the dielectric properties of the double layer ($\varepsilon_{dl}^*$):

$$\varepsilon_{dl}^* = \varepsilon_{dl} + \frac{\sigma_{dl}}{j\omega} \tag{19}$$

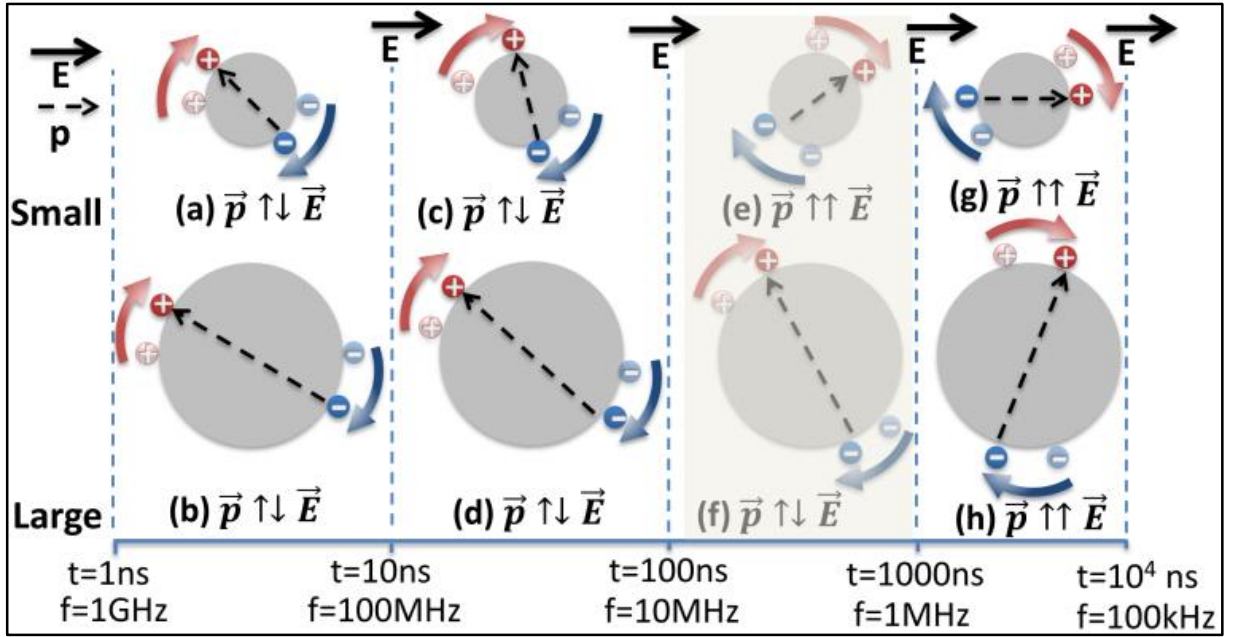

FIG 9: (a-d) At high frequencies or early time periods, nanocolloids are polarized anti-parallel to the field ($\vec{p} \uparrow\downarrow \vec{E}$), for small as well as large sized colloids. At lower frequencies (e-h), wherein sufficient time is available for surface conductance in the double-layer (red and blue arrows), the polarization on nanocolloids can be flipped parallel to the field ($\vec{p} \uparrow\uparrow \vec{E}$). This time constant ($\tau$) can differ for small versus large colloids (e vs. f) due to differences in average distance for surface conduction induced double-layer polarization.

For the DEP frequencies less than 10 MHz, it is reasonable to assume that $\varepsilon_{DL}$ does not vary too strongly with frequency, suggesting that the frequency variations arise chiefly due to $\sigma_{DL}$. As per Figure 9, the effectiveness of surface conduction in the double-layer to flip the polarization direction on the net dipole (from: $\vec{p} \uparrow\downarrow \vec{E}$ to $\vec{p} \uparrow\uparrow \vec{E}$) depends on the ability of ions within the double layer to traverse the full distance around the particle surface within the half cycle of the electric field frequency. Consequently, surface conduction-induced flipping is highly effective at low frequencies (sufficient time for conduction) and at smaller colloidal size. The frequency dispersion of double layer conductivity is given by[14]:

$$\sigma_{dl} = \left(\frac{2k_s}{a}\right)\left(1 + \frac{\omega\tau}{1+\omega^2\tau^2}\right) \qquad (20)$$

The time constant ($\tau$) required for surface conduction in the double-layer to flip the net polarization falls sharply ($a^2$) with colloidal size, as given by[14]:

$$\tau = \frac{2\pi(a+\kappa^{-1})^2}{D} \tag{21}$$

Here, $D$ is the diffusion coefficient and $\kappa^{-1}$ is the ionic strength-dependent length of the electrical double-layer. Hence, the polarization direction on smaller nanocolloids can be flipped to: $\vec{p} \uparrow\uparrow \vec{E}$ (Fig. 9e) at earlier time periods versus on larger nanocolloids (Fig. 9f: $\vec{p} \uparrow\downarrow \vec{E}$), due to the lower distance required for surface conductance-induced polarization of smaller particles. Along similar lines, the net polarization dispersion also depends on the shape of the colloid, since deviations from the spherical shape will cause its polarization to no longer be the same across all directions[12]. This directional influence of polarizability affects the so-called depolarization factor: $A_i$, where $i = x, y, z$; which is the ratio of the internal electric field induced by charges on the surface of a dielectric under external field to its net polarization (see Supporting Information). The computed polarization dispersion in Figure 10 shows pDEP at low frequencies, due to dominance of surface conduction; i.e. $\frac{K_S}{a} - \sigma_m > 0$. With increasing frequency, the effectiveness of surface conduction drops, thereby lowering pDEP polarization with frequency (Eq. (19)). Eventually, nDEP behavior is reached ($\frac{K_S}{a} - \sigma_m < 0$), with the transition determined by the time constant for polarization ($\tau$) of Eq. (20). At even higher frequencies, wherein permittivity differences dominate the polarization behavior, the small difference in permittivity of the double-layer around the colloid ($\varepsilon_{DL}$) versus permittivity of the media ($\varepsilon_m$) causes weaker levels of nDEP than observed due to the conductivity differences in the mid-frequency range.

The variation to the polarization dispersion induced by alteration of colloidal size is shown in Figure 11a, while that induced by surface conduction alterations for similar colloidal size, such as would occur for colloids of differing surface charge is computed in Figure 11b. Smaller colloidal size significantly lowers the polarization time constant ($\tau$) due to the quadratic dependence of $\tau$ on hydrodynamic radius ($a$), which can be observed as an upshifting in the crossover frequency.

Also, due to the highly effective surface conduction within colloids of smaller size, the overall bandwidth of the region wherein conductivity effects dominate over permittivity effects is enhanced. This means that the $\varepsilon_{DL}$-driven weak nDEP behavior is pushed to higher frequencies for smaller colloids. On the other hand, surface charge of the colloid influences the net polarization dispersion due to the sharp hyperbolic cosine dependence of surface conductance ($K_s$) on its zeta potential ($\xi$)[14]. As a result, the higher surface conduction raises the pDEP level and its bandwidth, while inhibiting the nDEP level at higher frequencies (Fig. 11b). In summary, these simulations of the polarization dispersion suggest that particle size and shape dictate the frequency for onset of nDEP (based on their sharp effect on $\tau$), whereas higher surface charge (i.e. higher $K_s/a$ for same $a$ values) inhibits the nDEP level.

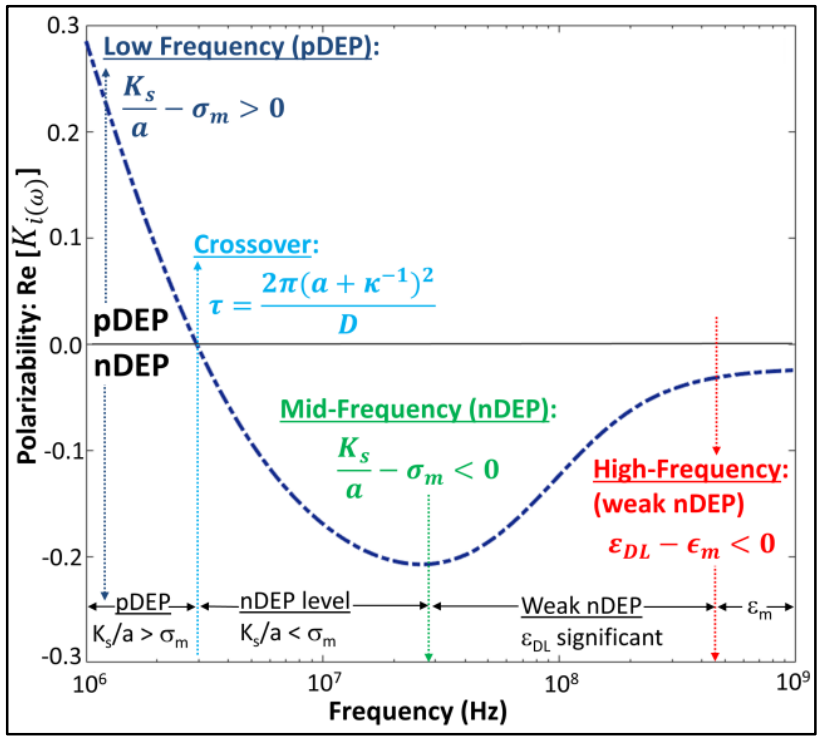

Fig. 10: Computed polarization dispersion (Real $[K_{i(\omega)}]$) shows pDEP at low frequencies due to high surface conduction in the double-layer, whereas less effective surface conduction at frequencies above the inverse time constant ($\tau$) leads to nDEP, especially within media of high $\sigma_m$. At even higher frequencies, the small differences between $\varepsilon_{DL}$ and $\varepsilon_m$ lead to weak nDEP behaviour. (Simulation conditions: $K_s$= 0.457 nS; $\sigma_{core}$=0.01 S/m; D=2×10$^{-9}$ m$^2$/s; $\kappa^{-1}$=2 nm)

**B. Influence of device geometry on force field:** Next we review the device geometry that is used to create the force balance towards causing enrichment of the nanocolloids. For the work presented here on biomolecules, we use the device structure as per Figure 12, wherein field non-uniformities are initiated within a set of nanoslits with lateral constructions based on an AC field offset by DC voltage applied to electrodes within the reservoir that lead to the microchannel and nanoslits.

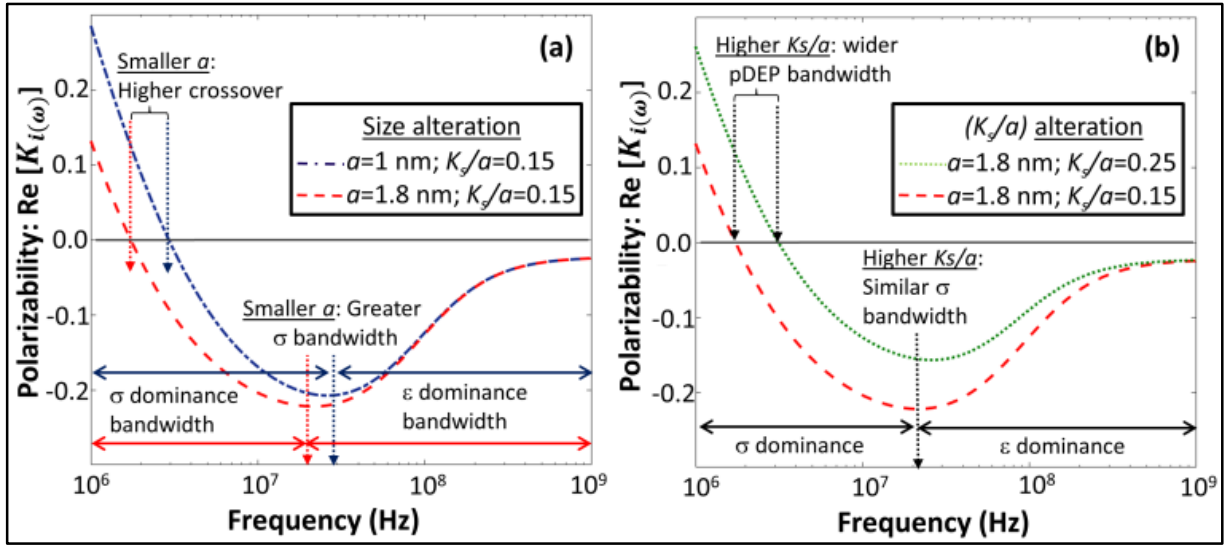

FIG. 11: Computed polarization dispersion for varying: (a) size; and (b) surface conduction represented as: $K_s/a$. (Simulation conditions similar to Fig. 10)

Based on this geometry, the net enrichment of negatively charged biomolecules in physiological media occurs in the nanoslit along the cathodic region away from the lateral constriction. We choose to utilize the enrichment under nDEP rather than that under pDEP, since the DC-offset nDEP profile is sufficiently delocalized to enable facile alignment of a sensor surface immobilized with receptors for capture of the target.

Also, pDEP is less likely within the physiological media used to capture target molecules. Enrichment occurs due to the force balance of nDEP versus electrophoresis (EP):

$$F_{net} = F_{EP} - F_{nDEP} \qquad (22)$$

Here, $F_{net}$ denotes the net force causing the enrichment, under $F_{nDEP}$ (given by Eq. (15)) and $F_{EP}$ that is given by:

$$F_{EP} = 6\pi\eta a\mu E; \ wherein \ \mu = \frac{2\varepsilon\zeta}{3\eta} \qquad (23)$$

Here, $\mu$ is the electrophoretic mobility that is dependent on zeta potential ($\zeta$) and viscosity ($\eta$). Hence, while colloidal size affects the respective magnitudes of $F_{nDEP}$ and $F_{EP}$, the difference in enrichment levels between two molecules of differing sizes ($F_{net}^1$ vs. $F_{net}^2$) is less strongly dependent on colloidal size and more strongly dependent on the differences in spatial profiles of $F_{nDEP}$ vs. $F_{EP}$. For instance, $F_{EP}$ dominates over much of the nano-slit region, except in the immediate vicinity of lateral constriction, wherein $F_{nDEP}$ dominates. Due to the highly localized spatial profile of nDEP, the enrichment initiation boundary within the nanoslit (Fig. 12) is determined by $F_{EP}$. Hence, biomolecules with lower $F_{EP}$ exhibit an enrichment initiation boundary that is further away from the lateral constriction and the enrichment is more spread-outwards, whereas for molecules with higher electrophoretic mobility, the enrichment initiates closer to the lateral constriction and forms over a relatively narrow band, due to the exponential drop in $F_{nDEP}$ away from the constriction.

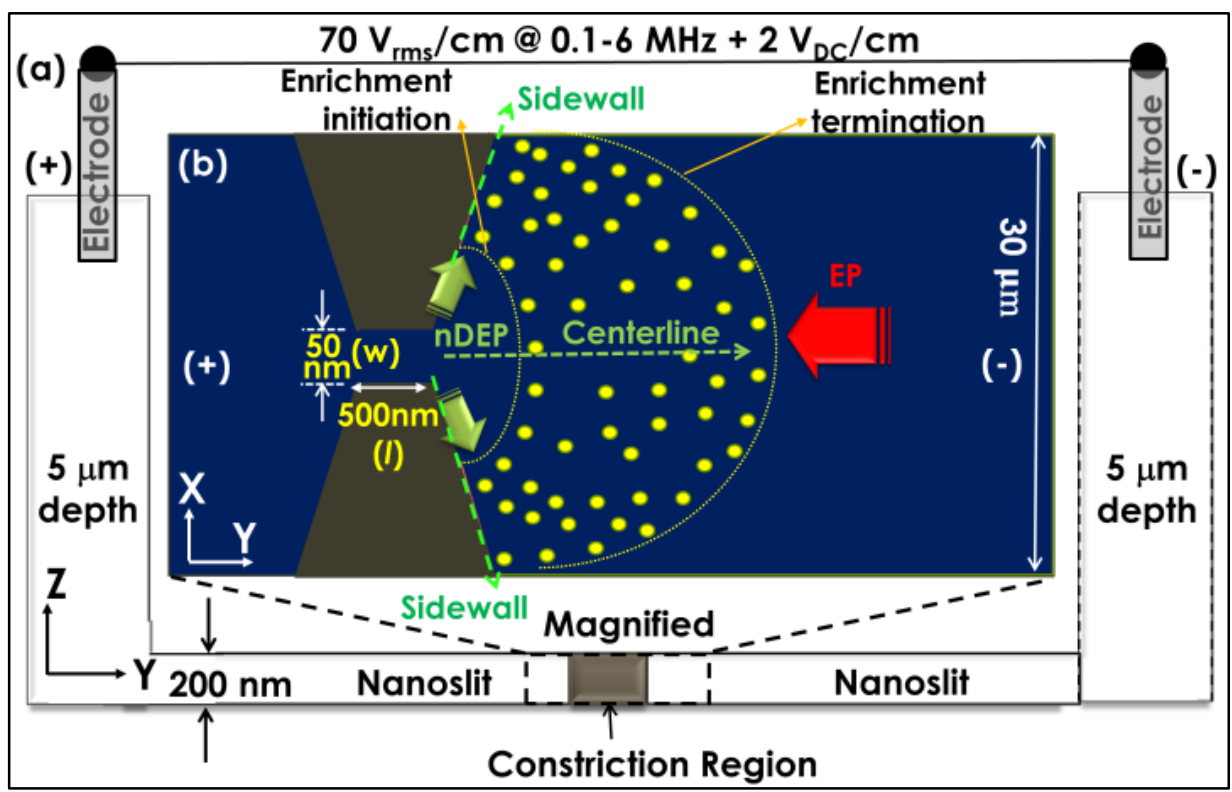

FIG. 12: Biomolecular enrichment (negative charged) occurs inside the nanoslit within the cathodic interface of the constriction due to a force balance of nDEP (green arrows) versus EP or electrophoresis (red arrow).

## C. Surface conduction induced polarization time constant

We validate the influence of colloidal size and surface charge on surface conduction in the double-layer by studying the polarization time constant of silica nanocolloids. Specifically, differing colloidal size and surface charge alters the surface conduction in the double-layer, thereby altering the crossover frequency from pDEP to nDEP, due to flipping of the net dipole. In this case, the device geometry shows lateral insulator constrictions of 1 μm gap (Fig. 13a), with the indicated positions for pDEP (Fig. 13b) and nDEP enrichment (Fig. 13c).

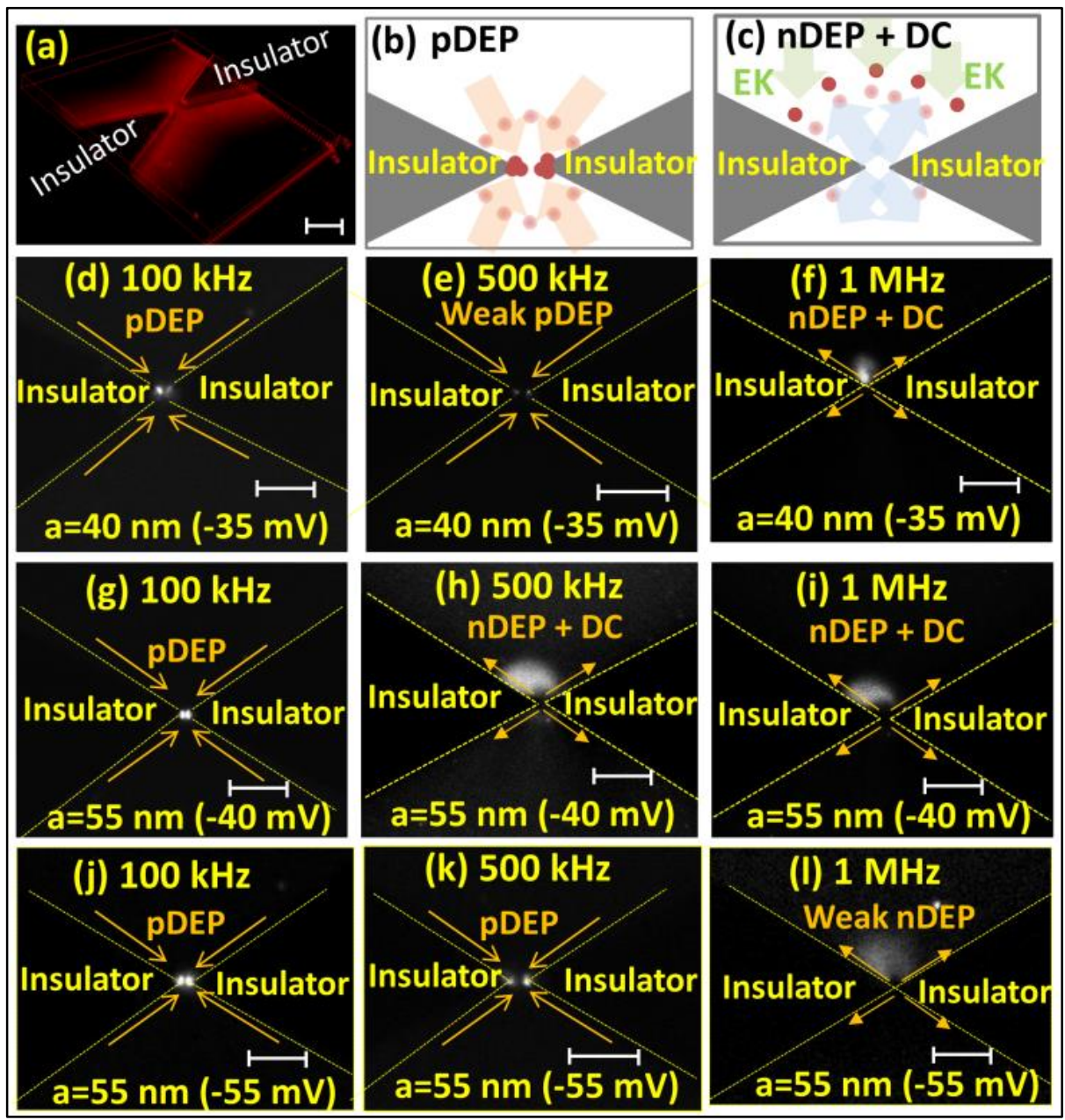

FIG. 13: (a) Lateral insulator constriction and schematic enrichment profile under: (b) pDEP; (c) nDEP (EK denotes DC-electrokinetics);. Frequency-dependent DEP after 20s for: (d, e & f) small (a=40 nm) versus (d, f & h) large (a=55 nm) colloids at ξ=-40 mV versus (j, k & l) ξ=-55 mV. Scale bar=4 μm and ξ potential in brackets.

At low frequencies (100 kHz), surface conduction in the double-layer has sufficient time to polarize the nanocolloid in the field direction, thereby exhibiting pDEP behavior for all types of nanocolloids (Fig. 13d, 13g & 13j). At higher frequencies (500 kHz), while smaller nanocolloids (a=40 nm) continue to exhibit polarization parallel to the field, as evident from the weak pDEP behavior, the surface conduction on larger nanocolloids (a=55 nm) is insufficient to induce polarization, thereby causing nDEP. However, for the case of larger nanocolloids (a=55 nm) with a significantly higher zeta potential ($\xi$=-55 mV) that is created by carboxyl functional groups, the pDEP behavior continues to be apparent at 500 kHz. At 1 MHz, the insufficient time for surface conduction drives all of three nanocolloid-types to nDEP behavior. However, the larger nanocolloid (a=55 nm) with the higher zeta potential ($\xi$=-55 mV) shows the weakest nDEP behavior of all three nanocolloid-types, consistent with the computed inhibition of the nDEP behavior with increasing surface conduction (Fig. 11b). Hence, enhanced surface conduction on smaller colloids up-shifts the frequency for nDEP onset, while higher surface charge on colloids inhibits the nDEP level.

## D. Frequency-selective enrichment molecular biomarkers

Next, we utilize the frequency selectivity offered by this polarization method to cause selective enrichment of the biomarker of interest, namely PSA (Fig. 14(i)) versus circulating antibodies of anti-mouse immunoglobulin (IgG) (Fig. 14(ii)), in the 0.8-6 MHz range. The fluorescence intensities for varying frequency conditions of nDEP induced electrokinetic trapping are determined by averaging the maximum intensities from 20 pixels across the image and normalizing these levels to those obtained under purely DC fields, where no perceptible rise in fluorescence is apparent.

These IgG type antibodies interfere with analyte detection using sandwich immunoassays, by causing false positives. While the IgG antibody (~150 kDa) is significantly larger than free PSA (33 kDa), it is composed of various heavy (~50 kDa) and light (~25 kDa) peptide fragments that are comparable in size to serum PSA present within its free and bound forms (~90 kDa). Given these different forms of the biomolecules in the body, there is a need for enrichment methods other than size-based filtration. Additionally, we seek to enrich and detect the smaller molecule i.e. PSA, versus the larger interfering molecule (anti-mouse IgG), which is not easily accomplished by standard separation techniques.

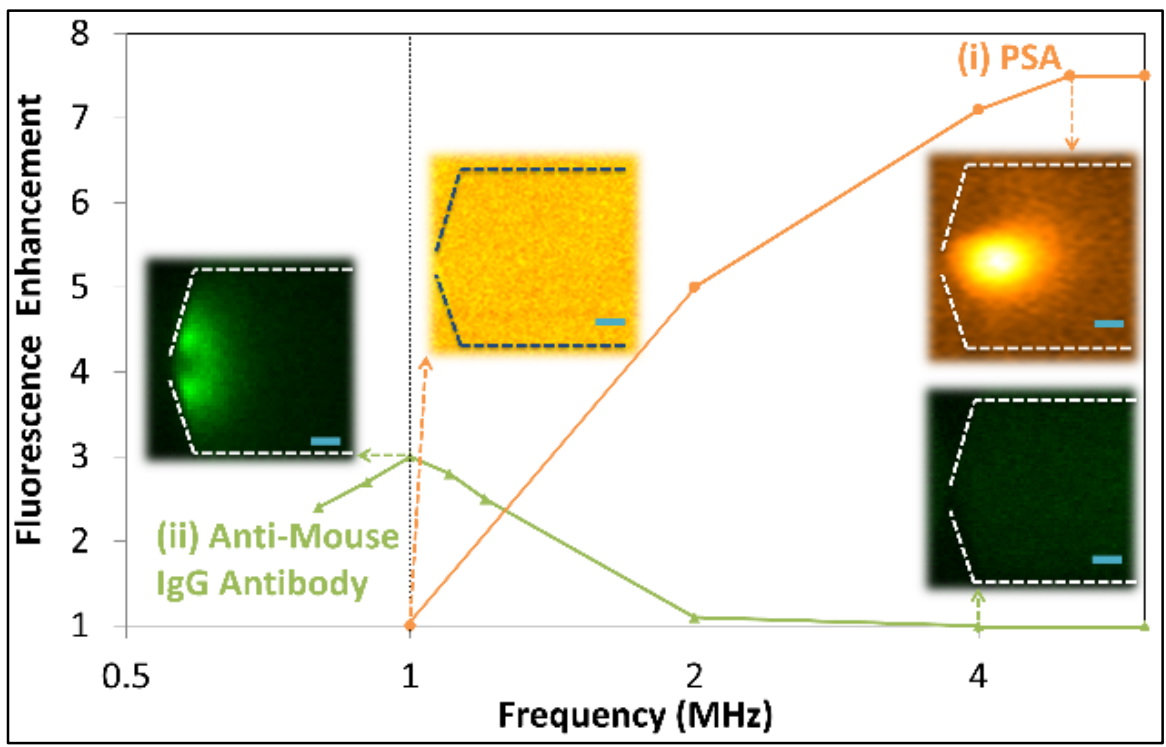

FIG. 14: Frequency-dependence of nDEP enrichment of PSA versus anti-mouse IgG in the 0.8-6 MHz range in a nanoslit device with lateral constrictions (dashed line). Scale bar is 10 μm.

Figure 14 shows that PSA enrichment under nDEP requires AC-fields of 4-6 MHz, whereas anti-mouse IgG antibodies show only weak levels of enrichment in 0.8-1.2 MHz, with no enrichment beyond 2 MHz. Based on the simulations of polarization dispersion of nanocolloids in Figure 11, we infer that the time constant for surface conduction induced double-layer polarization is significantly higher for the larger anti-mouse IgG species versus smaller PSA molecules, thereby leading to nDEP onset at lower frequencies versus that for the case of smaller PSA molecules. This is also consistent with the observations in Figure 13 showing that rapid double-layer polarization of smaller nanocolloids leads to nDEP behavior only at higher frequencies.

Furthermore, the open Y-shaped structure of IgG species, with their highly charged groups likely enables significantly higher levels of surface charge versus that for the coiled structure of PSA. Hence, the enhanced surface conductance likely inhibits nDEP levels, consistent with the simulations in Figure 11b and the observations for the high zeta potential nanocolloids (Fig. 13l versus 13i). With PSA molecules, on the other hand, while nDEP onset requires higher frequencies (few MHz) due to the lower double-layer polarization time-constant for smaller molecules (Eq. 20), the nDEP level is not inhibited). Also, the higher surface charge on the anti-mouse IgG species leads to higher $\mu_{EP}$ and $F_{EP}$, based on Eq. 22. The higher $F_{EP}$ versus $F_{nDEP}$ over much of the nanoslit region limits IgG enrichment to the immediate vicinity of the constriction (Fig. 14(ii)), wherein $F_{nDEP}$ becomes significant. On the other hand, the lower surface charge and $F_{EP}$ on PSA molecules enable enrichment over a wider region, as per the image of Figure 14(i), since $F_{nDEP}$ is comparable to $F_{EP}$ over a greater region of the nanoslit. All of this explains the selective-enrichment of PSA in the 4-6 MHz range, wherein anti-mouse IgG is not enriched.

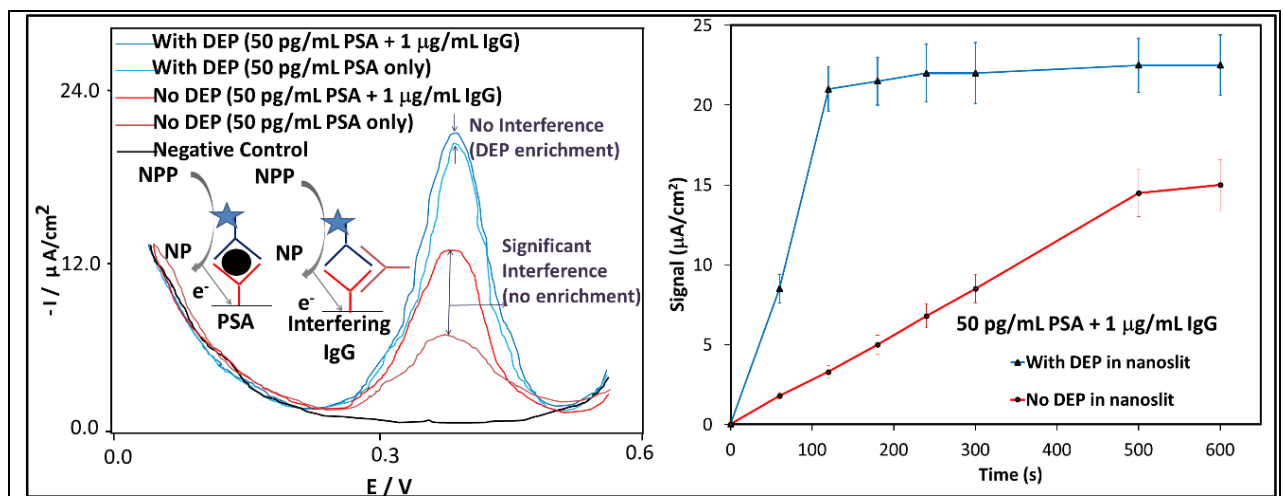

FIG 15. (a) Anti-mouse immunoglobulin (IgG) antibodies cause false positives to PSA immunoassay, as per the electron transfer schematics shown in the inset (assay details in reference[21]). In absence of nDEP enrichment in the nanoslit, signals from 50 pg/mL of PSA are almost two-fold higher in presence of 1 μg/mL IgG, whereas with nDEP enrichment, these false positives are insignificant. (b) Signal versus time shows rapid enhancement in binding kinetics of PSA, causing near steady-state signals within 2 minutes under surface conduction enhanced nDEP enrichment.

**E. Eliminating false-positives on PSA immunoassay**

Based on the selective enrichment characteristics of PSA versus interfering anti-mouse IgG antibodies under nDEP enrichment within the nanoslit at 6 MHz, we demonstrate its coupling to immobilized receptors on graphene-modified surfaces to study the false positives within immunoassays. Electron transfer schematics for the electrochemical detection platform for PSA and pathway by which the anti-mouse IgG antibody interferes with the sandwich immunoassay are shown in the inset of Figure 15a. As per voltammograms in Figure 15a, in the absence of nDEP enrichment inside the nanoslit device, a significant level of false positive signal due to immunoassay interference is observed within a few minutes of binding. With nDEP enrichment inside the nanoslit device, no significant difference is observed between the immunoassay with just 50 pg/mL PSA versus that with an additional level ($2\times10^4$-fold higher levels) of interfering anti-mouse IgG antibodies. The kinetic plot in Figure 11b shows that steady-state signals can be reached within 2 minutes of binding under nDEP enrichment of PSA, whereas in the absence of nDEP enrichment the signal does not reach a steady-state level, presumably due to anti-mouse IgG induced non-specific binding that causes false positives[68].

## F. Experimental Methods

Details of the device geometry and operation are available in prior work[21],[66] and in Supporting Information. Briefly, four reservoirs lead to a microchannel of 5 μm depth and several nanoslits of 200 nm depth, 30 μm width and 300 μm length. Within each nanoslit, a sharp lateral constriction is used to create spatial non-uniformities in the field. All experiments were conducted in a supporting electrolyte of phosphate buffer saline (PBS; 0.1 M, pH 7) of ~1.6 S/m conductivity. Glassy carbon modified Pt electrodes were used to apply an AC field of 70 $V_{rms}$/cm in the 1-6 MHz range, with a 1.5 $V_{DC}$/cm offset field to localize the nDEP enrichment. The data described herein is shown in the vicinity of the constriction region. Biomarkers were procured as follows: Alexa 488 labeled Goat anti-mouse IgG antibody (~150kDa) from Invitrogen / Thermo Fisher Scientific, Dylight 594 labeled Prostate Specific Antigen (~30 kDa) from Sigma Aldrich (St. Louis, MO) and silica nano-colloids (Corpuscular Inc., Cold Spring, NY). Method for altering surface charge on nanocolloids. The nDEP enrichment is coupled to the PSA immunoassay immobilized on graphene and gold nanoparticle-modified working electrodes, alongside reference and counter electrodes for voltammetric detection. The PSA antibody immobilized on the detection electrode was orthogonally aligned to the nDEP enrichment region. The sandwich assay uses Alkaline phosphatase (ALP) tagged to secondary PSA antibody for selective catalysis of α-naphthyl phosphate (NPP) to the electroactive α-naphthol (NP) product (see detection scheme in inset of Fig. 4a). Square-wave voltammetric (SWV) scans of α-NP were performed in PBS media by scanning from 0.1 to 0.4 $V_{DC}$, using a 100 Hz frequency; 4 mV step potential; and 25 mV AC amplitude. Standard deviations (±σ) were based on three different electrochemical measurements at each time point.

**Conclusion**

In the first part of this chapter we introduced a methodology based on AC electrokinetics for initiating frequency-selective and ultrafast enrichment of biomarkers within physiological media inside nanoslit devices. Using AC fields to generate negative dielectrophoresis and an offset DC field to initiate ion concentration polarization at surface charge non-uniformities in the nanoslit, the localized field intensity is enhanced due to ion depletion and the field gradient is enhanced due to ion accumulation, thereby enabling biomarker enrichment in physiological media. The optimal device geometry for creating the charge non-uniformity inside the nanoslit is a sharp lateral constriction with a gap of ~40 nm and a length of ~300 nm, to enable an extension of the ion depletion region from the anodic interface up to the cathodic interface. As a result, ultrafast electrokinetic enrichment of biomarkers occurs within physiological media due to the highly focused trapping profile, whereas the trapping profiles are more dispersed within media of lower bulk conductivity, due to the lowered biomarker nDEP alongside the enhanced level of disruptions from electro-osmotic flow. Using a conservative estimate for the surface charge non-uniformity ($\rho_c = -0.08\ Cm^{-2}$), the ICP-induced enhancement in the potential barrier for electrokinetic trapping within physiological media ($\sigma_m$=1.6 S/m) is ~5-fold higher than in the absence of ICP. Using a Boltzmann distribution for biomarkers, this suggests >100x higher levels of enrichment than obtained solely under nDEP, as validated by experimental observations. Finally, since this AC electrokinetic methodology causes ultrafast biomarker enrichment in conductive physiological media, wherein DC electrokinetic enrichment is limited, we envision its application for biomarker discovery, protein crystallization and in biosensing for speeding assay kinetics and reducing interferences.

In the second part of this chapter we presented a dynamic methodology for defining the region of interest towards quantifying the pre-concentration dynamics of fluorescently labeled biomarkers over delocalized zones, as obtained under a balance of varying force fields. Unlike prior methods based on a statically defined region of interest, our method uses a statistical description of particle distribution across the device geometry to determine threshold intensity levels for pre-concentration and depletion, thereby dynamically defining the region of interest. This automated methodology is applied to study the pre-concentration dynamics of Neuropeptide Y biomarkers and silica nano-colloids under an electrokinetic force balance, driven by negative dielectrophoresis from insulating constriction tips, which causes a highly delocalized pre-concentration zone. The spatio-temporal evolution of the preconcentration dynamics is used to optimize alignment of the electrochemical detection electrode to the region with a threshold preconcentration level, to enhance analyte-binding kinetics to the sensor.

In the third part of this chapter we used our developed platform to enable frequency-selective enrichment of molecular biomarkers within physiological media. Size differences between biomolecules alter their time constant for surface conduction-induced double-layer polarization, thereby tuning the onset frequency for nDEP, whereas surface charge differences between biomolecules alter the opposing electrophoretic force field, thereby modulating enrichment levels. In this manner, we show the frequency-selective enrichment of PSA versus anti-mouse IgG antibodies in the 4-6 MHz range. By coupling enrichment to voltammetric detection using immobilized receptors on graphene-modified surfaces, we show that the false positives usually obtained with >$10^4$-fold higher levels of interfering anti-mouse IgG antibodies can be eliminated. Based on the rapid and selective biomarker enrichment characteristics of this AC electrokinetic methodology in conductive physiological media, we envision its application for biomarker discovery, protein crystallization and for speeding bio-assay kinetics and reducing interferences.

# 2

**Predictive analytics platform for studying cancer and neurodegenerative diseases by probing mitochondrial morphological patterns as a new biomarker**

Recent years have seen an explosion in the amount of genetic, proteomic and metabolomic information available for a wide variety of human tumors and cancer cell lines. This information can be used to associate various patient outcomes with specific genetic aberrations and to potentially guide therapeutic decisions. However, specific genetic, proteomic and metabolic states can often interact in complex and unpredictable ways to alter cellular phenotypes and these phenotypes are what ultimately drive the tumorigenic process. To that end, it is important that we begin to amass phenotypic data on a large number of human tumor samples so that we can (1) determine the relationship between the genetic, proteomic and metabolic state of the tumor cells with the key phenotypes that drive tumor growth, and (2) begin to understand the relationship between particular cellular phenotypes and patient outcomes. This approach is especially significant for optimizing therapeutic strategies, since current therapeutic approaches benefit only a subgroup of cancer patients due to both genetic and phenotypic heterogeneity of the tumors. Herein, we develop a machine learning based technique to predict certain proteomic outcomes based on phenotypic and morphological heterogeneity of mitochondrial networks within cells, which we envision as a new biomarker to inform therapeutic strategies. To achieve this goal, we developed a robust image processing methodology to quantify mitochondrial networks and extract different measured properties from them.

In the second part of this chapter, in an attempt to better understand the role of alterations in mitochondrial dynamics within several pathological conditions, identifying therapies that target aberrant regulation of the mitochondrial dynamics machinery, and characterizing the regulating signaling pathways, we aimed at developing a label-free means to quantify the dynamic alterations in mitochondrial morphology. In particular we examine the ability of conductivity alterations to the cell interior as a label-free method to track intracellular mitochondrial modifications, based on benchmarking mitochondrial features using label-based image analysis. This method is validated using independent mitochondrial modification methods that are carried out on two different cell lines, namely human embryonic kidney cells and mouse embryonic fibroblasts. In both cell lines, the inhibition of mitochondrial fission that leads to a mitochondrial structure of higher connectivity is shown to substantially enhance conductivity of the cell interior, as apparent from the significantly higher positive dielectrophoresis levels in the 0.5-15 MHz range. Using single-cell velocity tracking, we show ~10-fold higher positive dielectrophoresis levels at 0.5 MHz for cells with a highly connected versus those with a highly fragmented mitochondrial structure, suggesting the feasibility for frequency-selective dielectrophoretic isolation of cells to aid the discovery process for development of therapeutics targeting the mitochondrial machinery.

## 2.1. Predictive analytics platform for studying cancer and neurodegenerative diseases by probing mitochondrial morphological patterns as a new biomarker

**Introduction**

Mitochondria are double membrane organelles in eukaryotic cells that are highly dynamic and primarily responsible for meeting the energy requirements of the cell[69-71]. An important requirement for proper mitochondrial function is a tightly regulated mitochondrial morphology, which is characterized by continuous cycles of fusion and fission. Aberrant regulation of mitochondrial machinery is associated with a number of human diseases, including Parkinson's disease, Alzheimer's disease and diabetes[71], as well as serving as a classic "hallmarks" of human cancer[72], including apoptosis, proliferation, autophagy and metabolism. All of this underscores the need for means to characterize mitochondrial morphology and utilize its quantitative characteristics to selectively collect or bin cells for understanding the underlying molecular mechanisms and developing novel therapeutics that utilize this tumor-specific phenomenon.

Pancreatic cancer is the fourth causes of cancer-related deaths for men and women in the United States, which also has the lowest 5-year survival of any cancer. One third of human tumors and up to 90% of pancreatic tumors harbor mutations in the Ras genes. The activation of Ras or MAPK is shown to promote mitochondrial fission through phosphorylation of the fission GTPase Drp1[73]. Activation of the Ras-MAPK (mitogen-activated protein kinase) pathway promotes phosphorylation of the mitochondrial fission GTPase Drp1 (dynamin-related protein 1), which subsequently induces mitochondrial fission. On the other hand, mitofusin proteins 1 and 2 (MFN1/2) on the outer mitochondrial membrane are responsible for reversing the mitochondrial fission pathway[74]. In recent work[75], our collaborators (Kashatus et al) demonstrated that mitochondrial fission was required for tumor growth in a xenograft model of pancreatic cancer, since its inhibition through shRNA-mediated knockdown of Drp1 was shown to block tumor growth. It has also been shown that mitochondrial fission promotes maintenance of stem cells in glioma and that inhibition of Drp1-dependent mitochondrial fission can effectively block tumor growth in a mouse model of gliomagenesis[76]. Despite the wealth of data linking mitochondrial fission to tumor growth, the discovery of therapeutics targeting the mitochondrial machinery has been limited. One reason is the lack of methodologies capable of quantitatively distinguishing phenotypic differences in mitochondrial morphology and relating them to their genetic heterogeneity, which impedes systematic investigation of the role of mitochondrial dynamics in tumor development. This requires data analysis techniques that step beyond single cell phenotypes to techniques that can decipher patterns between clusters of data sets. Hence, there is a need for means to characterize mitochondrial morphology at different levels and utilize its quantitative characteristics to selectively collect or bin cells for understanding the underlying molecular mechanisms that regulate the morphology and developing novel therapeutics that utilize this disease-specific phenomenon.

In recent years there is a huge interest in developing image analysis methodologies for morphological evaluation of mitochondrial network. Most of these methods utilize high-content fluorescence microscopy to evaluate mitochondrial morphology evaluation[77-82]. However, since most of prior methods treat all types of mitochondria as similar objects, they lack the ability to detect subtle changes in interconnectedness of mitochondrial network, and are not well suited towards quantifying the different mitochondrial morphological shapes that occur much less frequently than others within the mitochondrial networks.

To overcome the limitations of current methodologies, we developed an image analysis routine (MyMiA Project) to quantify key features of mitochondrial morphology, such as branch-level and cell-level measurements that can be utilized to compute the connectedness due to mitochondrial morphologies. Next, we developed and utilized a robust machine learning approach (Sibyl Project) to leverage the rich mitochondrial data produced by MyMiA and the extensive physiological, metabolomic, transcriptomic and other data generated by two relevant mouse models of pancreatic cancer: A well-validated, tractable and inducible genetically engineered model of pancreatic ductal adenocarcinoma that recapitulates the full progression of the human disease and a robust and biologically relevant orthotopic mouse model using patient-derived pancreatic tumors that faithfully models the full heterogeneity of the human disease. Sibyl project aims at correlating mitochondrial patterns across various tumor subtypes to metabolomic profiles through supervised learning routines and identifying morphological patterns between cells with similar physiological traits by using statistical analysis and unsupervised learning methods.

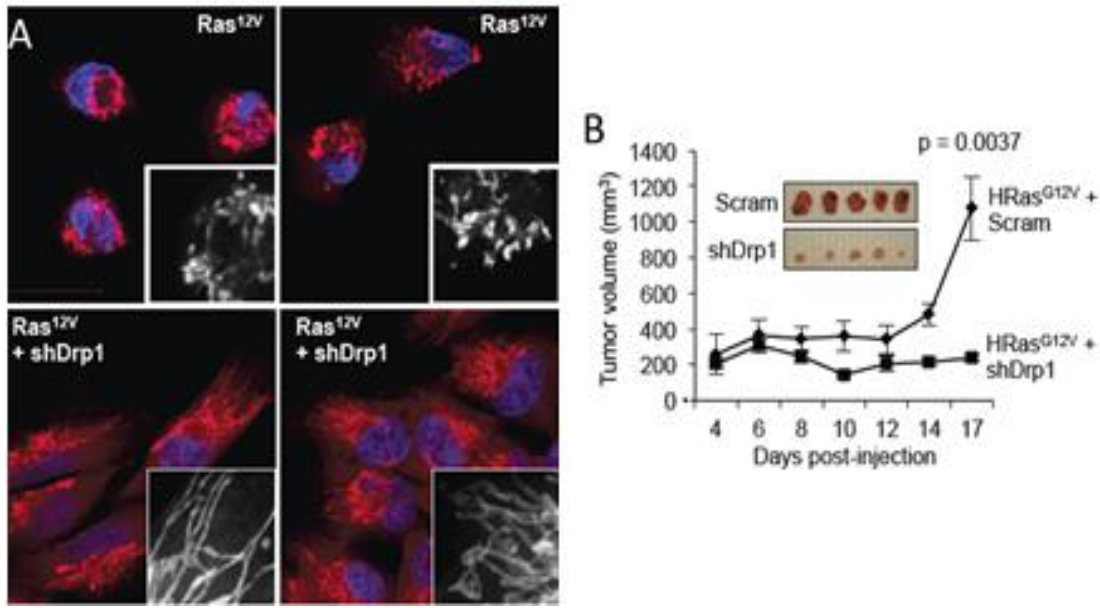

FIG 1. Inhibition of Drp1 reverses Ras-induced mitochondrial fragmentation and blocks tumor growth. A. Mitochondrial morphology of HEK-TtH cells transiently expressing HRasG12V in the presence or absence of Drp1 shRNA. B. HEK-TtH cells stably expressing HRasG12V and either scramble (u) or Drp1 shRNA (n) were injected into immuno-compromised mice and tumor volume was determined on indicated days post-injection (p-value for day 17).[73]

All the mouse models and related experiments were conducted in "Kashatus Lab" in department of immunology and cancer biology at University of Virginia's School of Medicine by their lab students and staff. The image analysis and microfluidics work was conducted in the "Swami Lab".

## Cell lines and imaging

### A. Cell Lines

Generation of HEK cells expressing HRasG12V plus Drp1 shRNA or shScramble control are described in detail in the previous work by Kashatus. et.al [75]. Mouse embryonic fibroblasts (MEFs) with Mfn1/2 knockout (MfnKO MEFs) were purchased from American Type Culture Collection (ATCC). To generate Drp1 knockout MEFs, Drp1$^{flox/flox}$ mice[83] were bred to TP53$^{flox/flox}$ mice[84]. MEFs were generated from Drp1$^{flox/flox}$; TP53$^{flox/flox}$ embryos and subsequently infected with adeno-associated-CMV-Cre-GFP (AAV-CMV-Cre-GFP, University of North Carolina at Chapel Hill Vector Core). Single cell clones were isolated and recombination of both alleles of Drp1 and p53 was confirmed by PCR.

In order to determine the requirement of Drp1-mediated mitochondrial fission for pancreatic cancer growth, we proposed to cross *Drp1$^{flox/flox}$* mice[36] with *LSL-Kras$^{G12D/+}$; Trp53$^{flox/flox}$; Pdx-1-CreER$^{Tg/+}$* mice, resulting in mice in which intraperitoneal (i.p.) injection of tamoxifen (TM) would lead to expression active KRas$^{G12D}$ and deletion of p53 and Drp1 in adult pancreatic cells. In this model, inducible expression of oncogenic KRas and loss of p53 causes a high percentage of mice to develop an invasive and poorly differentiated pancreatic ductal adenocarcinoma that closely recapitulates the clinical and pathological features of the human disease.

**B. Cell imaging for analysis of mitochondrial morphology**

The described HEK and MEF cell lines were plated on glass microslides a day prior to visualization of their mitochondria by one of the following methods: (1) Cells were fixed, permeabilized and incubated with α-Tom20 primary antibody (Santa Cruz Biotechnology) in conjunction with an α-rabbit Alexa-488 secondary antibody (Life Technologies); (2) Cells were engineered to stably express mitochondria-targeted YFP[85] (BD Biosciences). A Zeiss LSM 700 confocal microscope with 63X oil objective was used for imaging.

**Methods**

**A. Image analysis methodology to quantify mitochondrial connectivity (MyMiA)**

The panel of cell lines isolated from our tumor model exhibits a range of Drp1 expression levels that ultimately result in a range of mitochondrial morphologies. To quantify different mitochondrial topological structures in these cells, we developed an image analysis platform, MyMiA, which uses color separation and shape detection methodologies on the stained cell images to segment the cell nucleus, membrane and its mitochondrial network. Then the skeleton of mitochondrial network in each cell is extracted and mapped to the original network to enable characterization and quantification of that mitochondrial network of each cell. Image analysis pipeline is composed of several steps, as discussed below. Figure 2 shows the image-processing pipeline in MyMiA.

**Image preparation**

### i. Color separation and object detection

Different stains are used for fluorescence labeling of different structures in the cells (nucleus, mitochondria, membrane). In order to analyze the structure of mitochondrial network both in intercellular and intra cellular scopes, first it required to distinguish and separate each structure from one another. To achieve an automated separation and identification of cell structures there should be no pre-assumption on colors used for each structure. To identify the colors used for staining the different structures in the cells, an unsupervised machine learning algorithm, Mean-shift clustering algorithms are used to identify the colors used for staining, as shown in Figure 3(1).

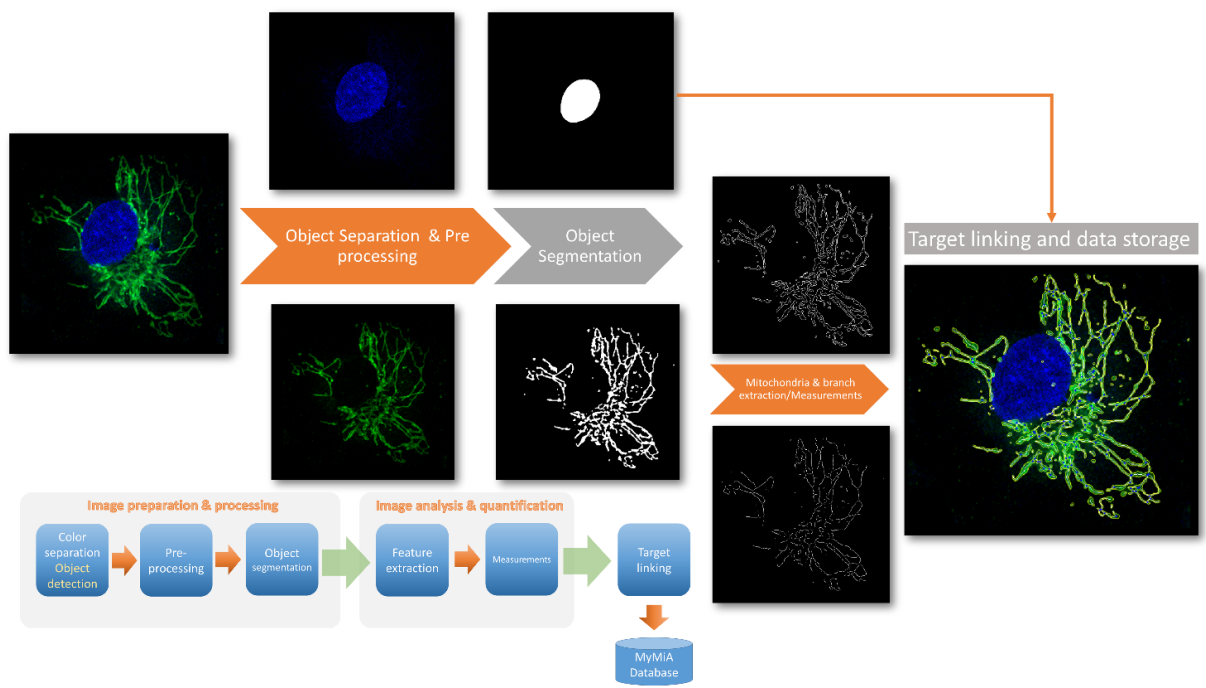

Figure 2. Image processing pipeline in MyMiA

After color separation, to identify each structural element of the cell, as shown in figure 3(2), a simple decision tree is used to classify each element into its object group, (nucleus, mitochondria, membrane) based on its structural features. (Nucleus is big and round, membrane is big, and has low surface to perimeter ratio, mitochondria are small, and/or have tree like structure).

**ii. Pre-processing**

For high accuracy characterization of mitochondrial networks, high quality and noise-free images are essential. However most of microscopy images usually suffer from different sources of noise, such as uneven illumination and presence of dim, out-of-plane objects. Consequently, preprocessing is vital for accurate analysis of the images, and is one of the essential parts of images-processing pipeline in MyMiA.

First, a Median filter is used as a de-noising tool to deal with some of aforementioned sources of noise. The size of filter should be chosen meticulously. It should be large enough to remove the noise inside mitochondrial network, while preventing blurring at the edges[80]. Width of the mitochondria, as its smallest feature, can be a good measure for the size of this filter.

In the second step a "top hat" filter[86] is applied to the image, to remove non-uniform background from the image. This step is crucial for creating a high quality binary mask of the mitochondrial network, and its segmentation. To further remove the noise, an adaptive noise-removal Wiener filter is used. Since this filter can tailor itself to the local image variance (little smoothing in large variance, and vice versa), it well preserves the edges.

In the third step of pre-processing, a median filter is applied, followed by intensity normalization. This step is necessary to avoid any possible noise after applying the top hat filter, and also eliminating unwanted smoothing effects of previously applied filters.

After this three-step process, a binary mask of mitochondrial network is created using the Bradley adaptive thresholding. The idea behind this is that one global threshold does not work well for detecting and detaching the objects of interest from their underlying background, since the background intensity is not same at different areas of an image. Local adaptive thresholding uses different thresholds to separate similar objects (mitochondrial network, nucleus) from background at different regions of the image.

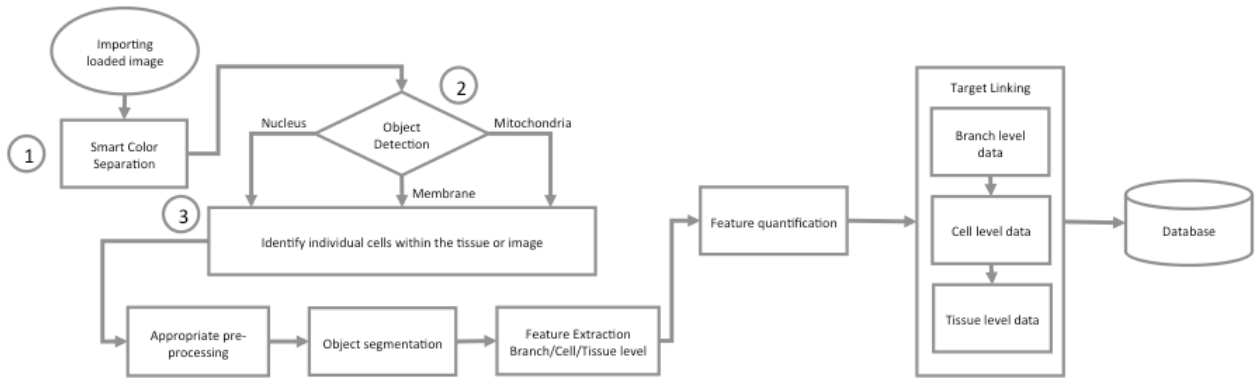

FIG 3. Mitochondrial network quantification process flowchart.

**Object segmentation**

The second stage of the image-processing pipeline in MyMiA is segmenting different objects, such as nucleus and mitochondria in the cell.

**i. Nucleus**

In order to segment the nucleus another step of pre-processing is required. First the relevant color channel is selected. Next, a large smoothing Gaussian filter is applied to reduce the intensity variation inside nucleus. Then size filtering is used to correctly choose the nucleus as the largest element in that specific color channel. MyMiA keeps track of each nucleus and its features, such as size and shape and saves them for the next step of the analysis.

**ii. Mitochondria**

Identifying individual branches in mitochondrial network is a multi-step process with the following steps: (i) selecting the relevant color channel. (ii) using a thinning process to create the skeleton of the mitochondrial network from it binary mask. (iii) finding all mitochondria and their respective branches, branch points and end points in the mitochondrial network and keeping track of each of them. (iv) creating mitochondrial boundary and individual branch (within each mitochondrion) masks and mapping to the original image to find the properties of each individual branch. (v) finding different properties of each mitochondrion, such as area, perimeter, major axis, and minor axis, and also properties of its respective branches, such as length, width, area, angle, etc. and saving them for the next step of the analysis.

One of the unique features of mitochondrial network is its degree of connectedness, which is not well defined within prior work. In MyMiA we proposed and use a unique methodology to measure the degree of connectedness in mitochondrial network of the cell. In each cell we count the total number of branch points and number of branches.

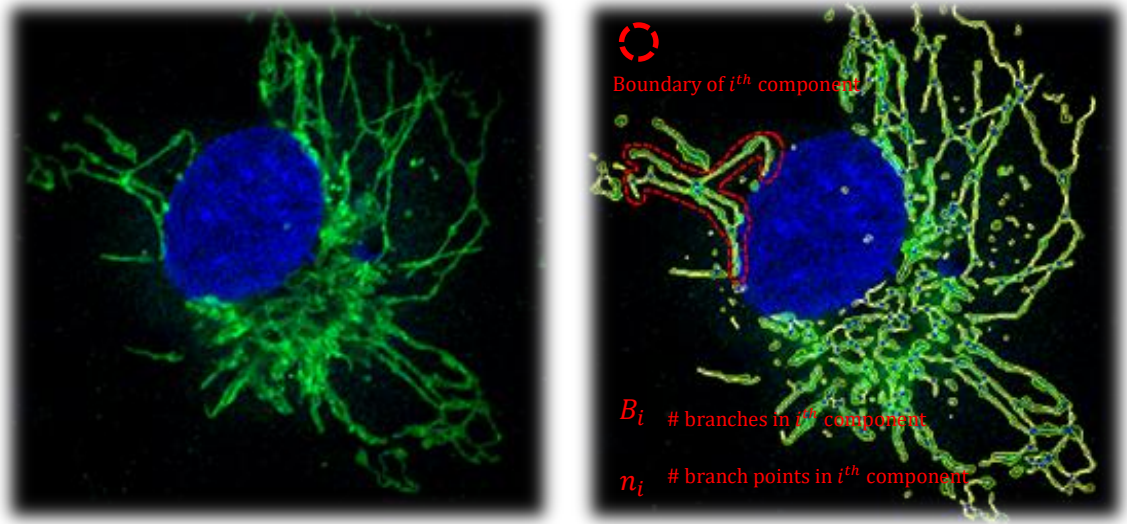

FIG 4. Concept of network components and strength of connection in a component

Based on this data and considering possible connections in a mitochondrial network, we create an ideal network with maximum connectivity, as if all the mitochondria in the cell where connected together. Then to compute connectedness, within each individual component of our cell (the component which is isolated from other components as shown in Figure 8) we define the strength of the connection as:

$$\text{Connection strength} = B_i n_i \qquad (1)$$

$B_i$ is the number of branches in i$^{th}$ component in the mitochondrial network of the cell, and $n_i$ is the number of branch points in the same component. Then the connectedness of the network is defined as the following:

$$C_f = \frac{2 \sum_i B_i n_i}{((B_{ideal}-1) \times (B_{ideal}-2)) + 2B_{ideal} - 2} \qquad (2)$$

where $B_{ideal} = (\sum_i n_i - (Max(i) - 1))$ is the possible number of branches in a network if all components in our cell where connected together to form a fully connected network.

**Target Linking and storing data**

The next step of the analysis is target linking, which links each nucleus, and mitochondrial network to its surrounding membrane (or a Region of Interest (ROI) defined by user). As a result of this step, properties of each element of that cell, measured in previous steps, are linked to that particular cell, thereby assigning every nucleus or mitochondrial network to a specific cytoplasmic object. This can be used to compare mitochondrial morphological and intensity-based measures across mitochondrial subtypes within individual cells. Following this step, all the relevant data are stored in a database and can be exported to be used in other data analysis software such as Microsoft excel, or more rigorously with R or Python. Figure 5 shows screen shots from different parts of MyMiA software.

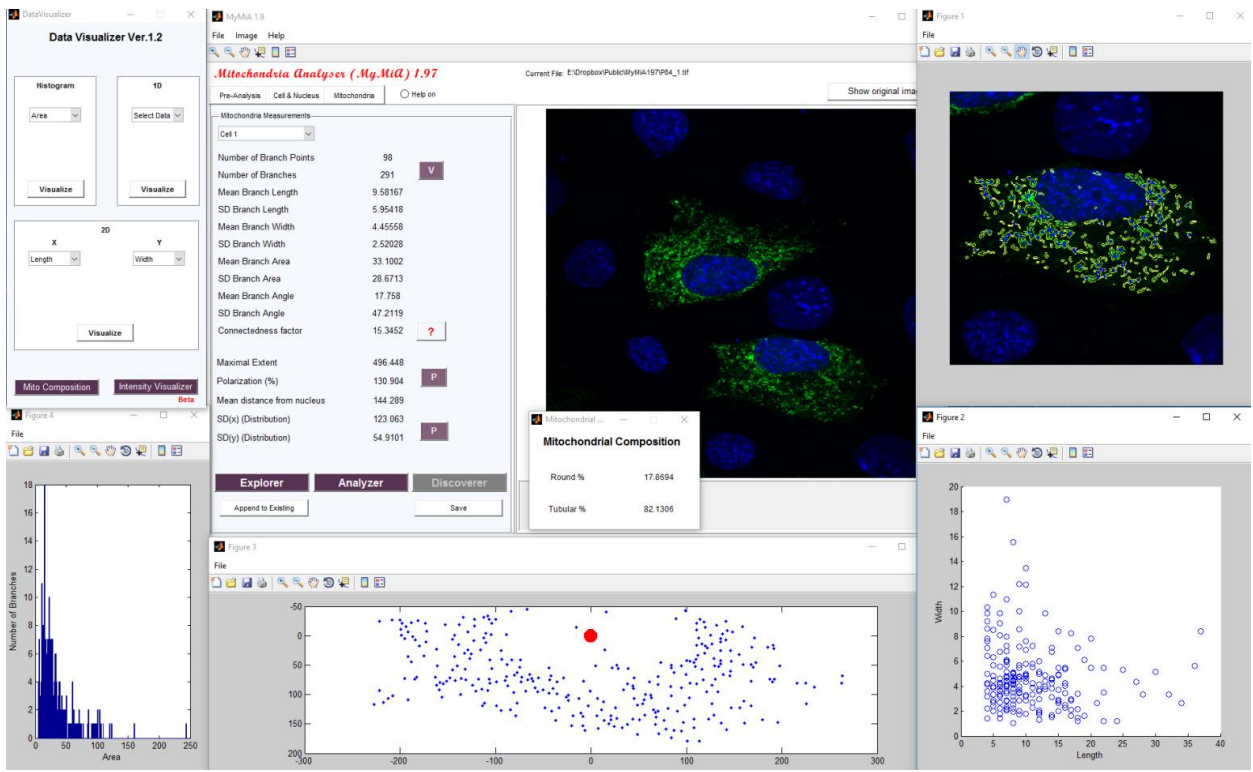

FIG 5: Screenshots from MyMiA Image analysis platform.

**Measured quantities by MyMiA**

MyMiA measures 25 different morphological features in each cell including both cell level measurements and branch level measurements (See table 1).

| Cell level measurement | Branch level measurement |
|---|---|
| Number of mitochondria | Connectedness |
| Average mitochondria Area | Number of Branch Points |
| Mitochondria Area STD | Number of Branches |
| Mitochondria major axis | Branch Distribution STD |
| Mitochondria minor axis | Mean Branch Length |
| Polarization degree | Branch Length STD |
| Extent of mitochondrial network | Mean Branch Width |
| Mean Distance from nucleus | Branch Width STD |
| Distribution Skewness | Mean Branch Angle |
| Distribution Kurtosis | Mean Aspect Ratio |
| | Aspect Ratio STD |
| | Mean Branch Area |
| | Branch Area STD |
| | Branch Angle STD |

Table 1: List of morphological features measured by MyMiA, based on the scope of measurement.

## B. Predictive analytics platform for cancer and neurodegenerative diseases studies using mitochondrial morphology (Sibyl Project)

By observing differences in mitochondrial networks we hypothesized that differences or combinations of differences in morphological features between different tumor cells are correlated to differences in physiological features that drive the tumorigenic phenotype within these cells. To test our hypothesis, we used six cell lines that exhibit a range of mitochondrial morphologies and a wealth of phenotypic expression patterns. With that in mind, we used our data to develop a machine-learning classifier that, given the quantified mitochondrial features of a cell (using MyMiA), can accurately predict their cell line. Once the system is set up, it can be used to establish correlations between mitochondrial morphology and key physiological features (for example: survival…) we have measured for these or other cell lines. Ultimately, this would provide us with testable predictions about how specific manipulations of mitochondrial morphology in these cells would alter cell physiology and provide a framework for the analysis of tumor cell populations with inherent heterogeneity in mitochondrial morphology, rather than that obtained through genetic manipulation; with a possibility for utilizing patient-derived tumors and xenografts within future iterations. For this analysis, we generated a database of morphological features from 1320 different cells, obtained from 6 different cell lines that represent 3 distinct Drp1 genotypes.

**Data Preparation**

Good data is a pre-requisite of a good model. To improve the performance of our machine-learning algorithm we employed multiple pre-processing and data preparation steps.

**A. Data exploration**

The first and most important step of the analysis is aimed at understanding the data. This step is key to ensure that our data is relevant to the questions we are trying to answer. In our case, MyMiA's data was generated for the purpose of correlating the phenotypic features to genotypic and proteomic targets, and since we had all these information available for different cells, our data was sufficient to support our analysis. We performed some standard data exploration routines like plotting the distribution of values for different features, to get a sense of the distribution of the data, and visually inspect for outliers in the data. While there was no visual outlier in the data, to be more quantitative, for each feature inter-quintile test of outlier detection was used to detect possible outliers, but none found in our data set.

**B. Feature engineering**

As mentioned in the previous section, MyMiA extracts 25 different aggregate features of each cell. However along with these set of features, MyMiA generates the vectors of the raw data that are used to create these features. This provides a large flexibility to create new features based on the need, along the way where they seem necessary. For instance, with more experience and visual inspection of images, it seemed that Wriggleness or twist of the mitochondria is a feature that can be a good predictor of the cell genotypes, since it widely varies among different genetic groups. While this feature was not initially quantified by MyMiA, the raw data could be used to create this feature as the following:

$W = 4\pi A/P^2$, Where, A is the area of the mitochondria and P is the perimeter.

**C. Feature selection**

While all extracted mitochondrial features are important to best describe the unique mitochondrial network in each individual cell and differentiate it from other cells, not all of them are required to predict certain physiologies in the cells. Some mitochondrial features may become more highly influenced by a particular cellular process or activity, while not changing dramatically with other cellular processes.

To choose the most relevant set of features to predict the genetic group of the cell, we used a recursive back step feature selection. In this approach we started by building a model with all the possible features, and then removed the features one by one, to the point where removing further features resulted in a big drop in the accuracy of predictions. This method proved to be very effective and did not neglect the collinearity of the features as shown in Figure 6.

**B. Model Selection and validation**

Our main goal is to develop a set of models that can use mitochondrial morphology data to classify cells, with high performance and accuracy, based on an additional defining variable such as cell type, genotype, or some relevant physiological readout. Successful generation of these models will allow us both to prospectively predict key physiological features of a tumor (eg - likelihood that a tumor will metastasize or respond to treatment) and to generate testable hypotheses about the relationship between mitochondrial morphology and physiology (eg - connectedness supports lipid metabolism, distance from nucleus supports migration and invasion, etc.).

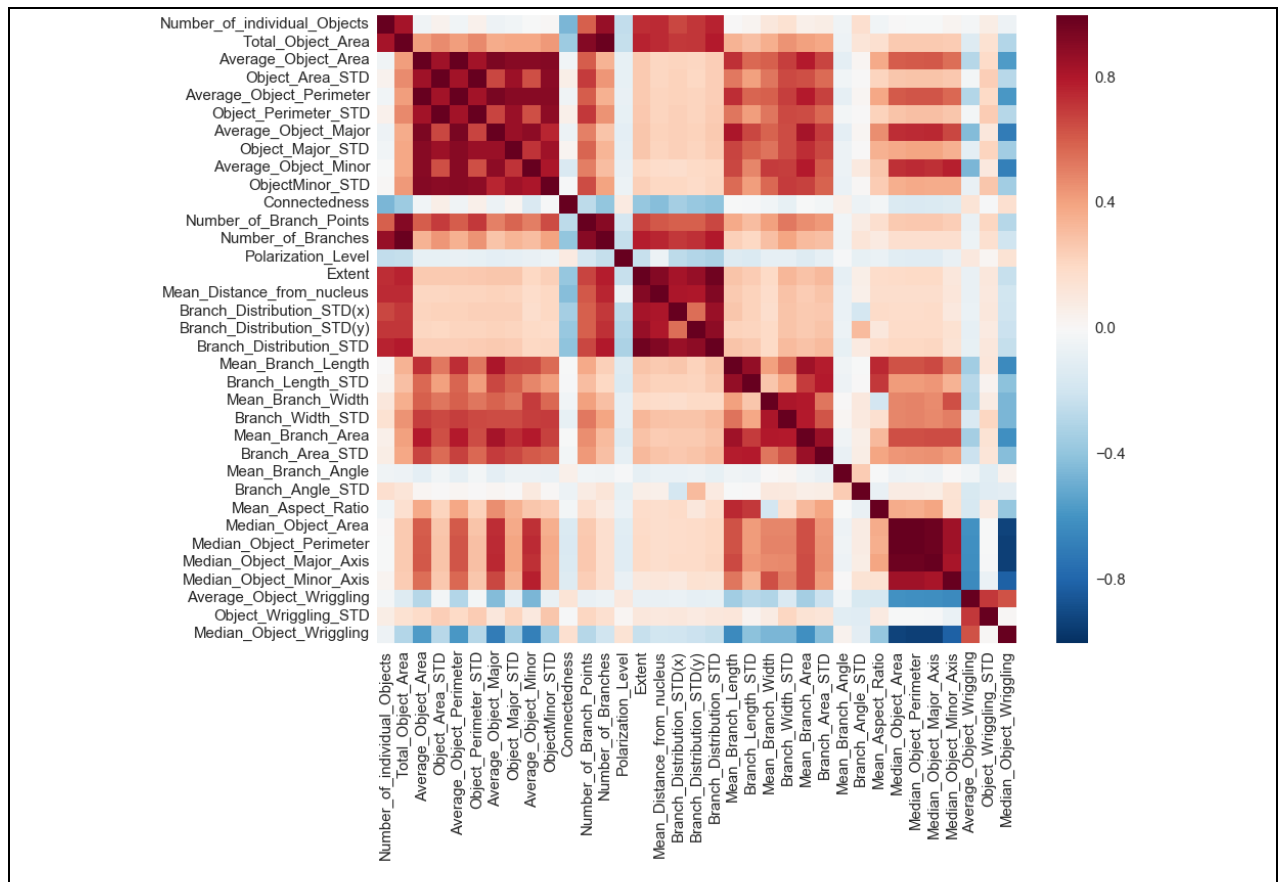
FIG 6. Correlation matrix of features. It is clear that there is a strong correlation between some of the features.

In order to develop this model, our initial approach was to attempt several different machine learning models using the data from our six tumor cell lines and train them to classify the cells into one of several predefined classes (i.e - classify each cell into one of six cell lines or one of three genotypes). Each potential model was tested using 10-fold cross validation. Briefly, the data were randomly divided into 10 equal portions. One of the 10 portions was held out as the validation data to later test the performance of the model, while the remaining 9 portions were mixed and used to train the system. At the end of this stage, performance of the machine-learning model was evaluated by its ability to predict the class of the held out sample, which was not used during the training phase. This process was then repeated 9 times (one time for each remaining portion) for each of the validations sets, so that each of the 10 portions was only used once for evaluating the model's performance. The average of the 10 performance measures, one from each validation set, were then reported for each model. In this method, data from each cell was used 9 times for training and only one time for testing, and thus the entire dataset was used for both training and testing the model. Performance measures were defined as follows, where TP=True Positive, TN=True Negative, FP=False Positive and FN=False Negative:

$$Accuracy = \frac{TP + TN}{TP + TN + FP + FN} \qquad Recall = \frac{TP}{TP + FN} \qquad Precision = \frac{TP}{TP + FP}$$

**Results and conclusion**

We used different classification models for our purpose using the above mentioned methodology to classify different cells for different genetic groups and cells lines. Among the different classification models tested, Adaptive Boosting (AdaBoost) with decision tree solver, was the model with highest accuracy, recall and precision in predicting the cell line and the genotype. In brief, AdaBoost combines multiple simple classifiers, with performance slightly better than random selection, for improved performance. The output is a weighted sum of outputs from different classifiers. Each classifier is tweaked in favor of misclassified instances by previous classifiers to further enhance the performance by making the mistakes bolder. The base weak learner in our AdaBoost algorithm is decision tree, which is a classification method that breaks down the dataset into smaller and smaller subsets while simultaneously and incrementally growing an associated decision tree. The basis for breaking down data is information gain, i.e. how much does the uncertainty in the data decrease if it is split based on one of the features. The goal after the split is to achieve the most homogenous set of subsets, i.e. subsets with the most similar values or least impurity. The data are initially split into smaller subsets based on the feature that reduces the uncertainty the most. Each of the created subsets is then divided into smaller subsets with even more homogeneity. This process is continued on each subset, up to a point where further splitting is not possible.

Application of the optimized AdaBoost algorithm on our set of 1320 analyzed cells, on the test set (25% of data) resulted in *80% average accuracy*, *79% average precision* & *79% average recall* in predicting the genotype of the cell, where random selection would result in 33% (since we had a fairly balanced dataset with equal contribution from each of the 3 different genotype samples). Notably, it also predicted the correct cell line with *64%* *average accuracy, 62% average precision & 62% average recall*, where random selection would result in 16%.

In this section, we developed an image-processing platform MyMiA to quantify the mitochondrial images, and extract a vector of morphometric features from each cell. This valuable set of data could be potentially used to find correlations between phenotype and genotype variables in cancers cells to enable us to make predictions about the cell behavior based on cell images. We envision this platform and methodology being used in the future for even more biologically relevant questions in both the hospital clinic and in basic research. For instance, if we analyzed libraries of healthy and tumor cells stratified by tumorgenicity, we may be able to predict clinical measures such as metastatic potential and aggressiveness based on mitochondrial features. This could be used to decrease the need for more invasive procedures such as open surgery biopsy needed for many traditional histopathology methods and using fine needle aspirate instead to obtain tissue which may help diagnose and predict metastatic potential. In basic research, if we have knowledge about a cancer cell's genetic and metabolic features we can use MyMia to characterize and Sybil to predict genetic and metabolic features of newly derived cell lines without expensive testing, as well as understand what mitochondrial morphology phenotypes result in metabolic changes within the cell. For instance, due to changes in vascularization and oxygen depletion, tumor cells shift their metabolic profile towards glycolysis and away from respiration. Associated mitochondrial morphology changes may be evident as this shift occurs, and associated pathways could be targeted in looking at new therapies.

## 2.2. Label-free quantification of alterations in intracellular mitochondrial dynamics by monitoring cytoplasmic conductivity

**Introduction**

Mitochondria, which are key regulators of metabolism and cell death within eukaryotic cells, undergo constant cycles of fusion and fission, thereby allowing the cell to quickly adapt to environmental conditions for promoting cellular health [87],[88],[89]. Mutation and aberrant regulation of the mitochondrial fusion and fission machinery is associated with a number of human diseases[90],[91],[92], including Parkinson's disease, Alzheimer's disease and diabetes[93], as well as in physiological processes whose dysregulation are classical hallmarks of human cancer[90]. Since increased mitochondrial fission can promote glycolysis[94], it has been postulated that tumors may increase mitochondrial fission activity to promote the metabolic shifts[95] that create the molecular building blocks required for rapid proliferation[96]. Hence, analysis of phenotypes caused by mitochondria shaping proteins can help uncover new diagnostic and therapeutic strategies for disease states, especially in conjunction with tools that monitor specificity of the subcellular alterations.

Despite the wealth of data linking mitochondrial fission to tumor growth, the discovery of therapeutics targeting the mitochondrial machinery has been limited. One reason is a lack of robust methods to analyze dynamic changes in mitochondrial morphology. Genetic and pharmacological screens are powerful tools to identify novel signaling pathways and new inhibitors, but they require unbiased and quantitative readouts. Previous quantification attempts based on high content image analysis require stains or markers to identify mitochondria and also require the fixing of cells for optimal images[97]. Apart from being time-consuming, this methodology cannot easily be applied to non-adherent cell types such as immune cells and is not capable of isolating live cell populations based on their mitochondrial structure, especially from heterogeneous samples, for downstream quantitative analysis.

In this work, we explore a label-free approach based on cell electrophysiology to quantify alterations to mitochondrial structure induced by Drp1 and MFN1/2, which are independently quantified by mitochondrial image analysis of the respective labeled cells. Electrophysiology-based methods are of interest since they can characterize the mitochondrial alterations, as well as enable frequency-selective dielectrophoretic isolation of cells with particular mitochondrial morphology for downstream analysis. Dielectrophoresis (DEP) causes the frequency-selective translation of polarized bio-particles under a spatially non-uniform electric field, either towards the high field region by positive DEP (pDEP) for highly polarizable particles versus the media, or away from high field region by negative DEP (nDEP) for particles within highly polarizable media[98],[14]. The DEP frequency spectra can be fit using a standard shell dielectric model to compute conductivity of cell interior[12], which strongly depends on the maximum level of pDEP of a cell in the MHz range, at a given media conductivity[99],[100],[101]. Herein, we show a significant enhancement in cellular pDEP levels in the 0.5-15 MHz range after genetic manipulations that inhibit mitochondrial fission, as validated using independent mitochondrial modification methods that are carried out on two different cell lines: Drp1 knockdown on human embryonic kidney (HEK) cells and Drp1 knockout on mouse embryonic fibroblasts (MEFs). Based on this, we infer that significant alterations in intracellular mitochondrial structure can be identified and quantified, suggesting feasibility for utilizing label-free dielectrophoretic methods to selectively isolate cells based on their mitochondrial morphology.

**Methods**

**A. Dielectrophoretic spectral measurements**

DEP spectral analysis was performed with unlabeled cells (~$10^6$/mL). Prior to the DEP experiments, the cell media was replaced with 8.8% sucrose water, with media conductivity of 150 ± 5% mS/m, as adjusted by its own culture media. Viability of cells within this altered media was verified over a period of one hour, as assessed by trypan blue exclusion and by stability of their mitochondrial morphologies by immunofluorescence (see Supporting Information). DEP spectral measurements were conducted on a 3DEP dielectrophoretic analyzer (DepTech, Uckfield, UK) using a recording interval set to 30 seconds at 10 $V_{pp}$, with data collected over 20 points between 100 kHz to 45 MHz. In this 3DEP reader, the electric field is applied to gold-plated conducting electrode stripes inside the wall of each well, with the DEP response measured at 20 different frequencies that are applied individually within each well. The relative DEP force at each frequency is obtained by analyzing spatio-temporal variations in light intensity from particle scattering using particular bands in each of the 20 wells, after normalization to the background at zero field (time = 0), by accounting for the electric field distribution in the wells[102],[31]. For each batch of cells, the relative DEP force at each frequency was obtained using 2 independent measurements on the respective cell samples and this process was repeated for 5 separate batches of the same cell type to ensure reproducibility, as per the error bars (Figure 10a & 10c). The maximum pDEP force level in the MHz range for each cell line (HEK and MEFs) was used as the basis to normalize all other measured DEP force levels for the respective cell line. This approach assumes that the maximum pDEP level for each cell line occurs when the cell achieves its maximum polarization level at a particular modification.

**B. Fitting DEP spectra using a multi-shell dielectric model**

In order to discern the particular subcellular regions influenced by the mitochondrial alterations carried out herein, the acquired DEP spectra over the 0.1-40 MHz range was fit to a standard three-shell or four layer dielectric model of the cell. The layers from cell interior to its envelope include: nucleus, nucleus envelope, cytoplasm and cell membrane [12] (see Figure 7).

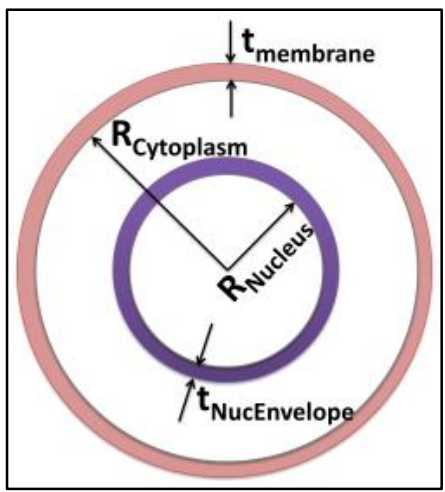

FIG 7: Geometry of the four-layer dielectric model (each layer with permittivity or $\varepsilon$ and conductivity or $\sigma$) that is used to fit the DEP spectra, including membrane ($\sigma_{mem}$ and $\varepsilon_{mem}$), cytoplasm ($\sigma_{cyto}$ and $\varepsilon_{cyto}$), nucleus envelope ($\sigma_{NucEnvelope}$ and $\varepsilon_{nucEnvelope}$), and nucleus ($\sigma_{nucleus}$ and $\varepsilon_{nucleus}$)

The DEP force ($F_{DEP}$) on spherical particles is given by:

$$F_{DEP} = 2\pi\varepsilon_m R^3 Re[K_{i(\omega)}] \cdot \nabla(|E|^2) \tag{1}$$

Here, $\varepsilon_m$ is the permittivity of the medium surrounding the cell, R is the radius of the cell, ω is the radian frequency of the applied field, and $E$ is the electric field in the region where the cell is located. The Clausius-Mossotti factor ($K_{i(\omega)}$) is a frequency dependent measure of cell polarizability determining force direction as:

$$K_{i(\omega)} = \left[\frac{{\varepsilon'_c}^* - {\varepsilon_m}^*}{{\varepsilon'_c}^* + 2{\varepsilon_m}^*}\right] \tag{2}$$

Here, ${\varepsilon_m}^*$ and ${\varepsilon'_c}^*$ are the complex permittivity of the medium and effective permittivity of cell respectively. In each case, the complex permittivity depends on the respective permittivity (ε) and conductivity (σ) values as per the following frequency dispersion:

$$\varepsilon^* = \varepsilon + \sigma/j\omega \tag{3}$$

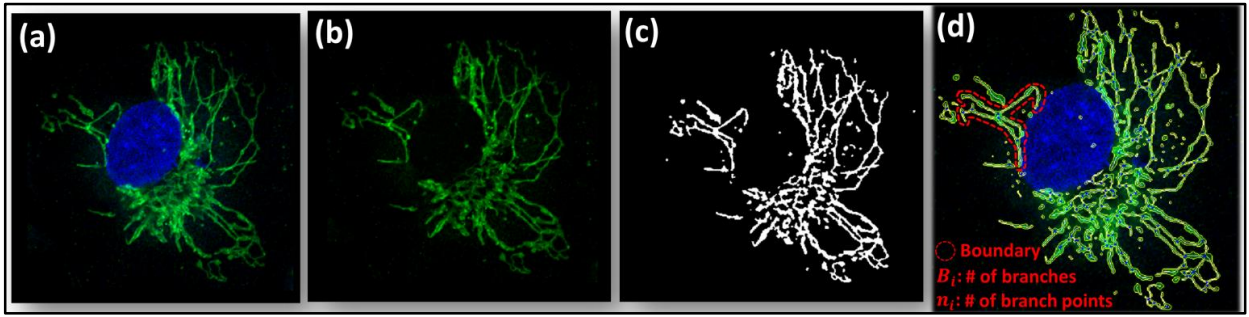

FIG 8. Methodology used to measure mitochondrial structure through computing a connectedness factor: (a) Color channeling on the image of the stained cell followed by a decision tree is used to separate out the mitochondrial features the features (b). Pre-processing steps followed by adaptive thresholding is used for segmentation of the mitochondrial network to identify individual mitochondria (c), for identifying branches (individual mitochondrion) and branch points (mitochondrial joints). Each branch is used as a mask to perform measurements such as length, width, area on each individual mitochondrion, so that the connectedness factor can be computer (Eq. (6)), as indicated by the red dashed line in (d).

The effective permittivity of cell (${\varepsilon'_c}^*$) is calculated in the following iterative manner. First, the innermost layer of the cell, i.e. the nucleus, and its surrounding envelope are approximated and replaced by a sphere of the same size (nucleus radius plus thickness of the nucleus membrane). The effective permittivity is expressed in terms of the respective complex permittivities: $\varepsilon_{nucleus}^*$ and $\varepsilon_{nucEnvelope}^*$, while '$a$' is the ratio of external (nucleus plus envelope) to internal (nucleus) radius of the combined layers:

$$\varepsilon'^*_{NucEnvelope} = \varepsilon_{nucEnvelope} \left( \frac{a^3 + 2\left(\frac{\varepsilon_{nucleus}^* - \varepsilon_{nucEnvelope}^*}{\varepsilon_{nucleus}^* + 2\varepsilon_{nucEnvelope}^*}\right)}{a^3 - \left(\frac{\varepsilon_{nucleus}^* - \varepsilon_{nucEnvelope}^*}{\varepsilon_{nucleus}^* + 2\varepsilon_{nucEnvelope}^*}\right)} \right) \quad (4)$$

Next, the same process is repeated until the whole cell is replaced by a single sphere to consider the 4 layer spherical model (including nucleus, nucleus envelope, cytoplasm, and cytoplasmic membrane). In this manner, the effective permittivity of the nucleus layer in Eq. (4), is combined with its surrounding cytoplasm to give effective cytoplasmic permittivity and with cell membrane to give effective cell permittivity. Based on this, Eq. (2) is used to fit the experimental DEP response of each cell modification. Parameters of the closest fit to the data are used to represent dielectric properties of cell elements.

## Results

### A. Images analysis to quantify mitochondrial structure

In order to separate mitochondrial features from those of the nucleus, we use color channeling, since they are differentially stained. This is achieved by maintaining one channel with each of RGB colors at a particular time and detecting objects with that specific color, followed by a decision tree to separate the features (Fig. 8a versus 8b). Following this, multiple pre-processing steps, including median, top-hat, and adaptive wiener filtering are applied to remove noise and other imperfections in the image. After pre-processing, an adaptive thresholding method is used for segmentation of the mitochondrial network, to identify individual mitochondria, as shown in Fig. 2c. This structure is then used to identify branches (individual mitochondrion) and branch points (mitochondrial joints). Each branch is used as a mask to perform measurements such as length, width, area on each individual mitochondrion (Fig. 8c versus Fig. 8d). The connectedness factor ($C_f$) is a measure of connectivity of the mitochondrial network based on its comparison to an ideal network, as given by:

$$C_f = \frac{2\sum_i B_i n_i}{((B_{ideal}-1)\times(B_{ideal}-2)) + 2B_{ideal} - 2} \quad (6)$$

Here, $B_i$ is the number of branches in each connected group in the mitochondrial network of the cell, with the connected group representing a group of two or more mitochondria that are connected together, and $n_i$ is the number of branch points in the same connected group. $B_{ideal}$: is the number of the branches within the fully connected ideal network, as given by:

$$B_{ideal} = (\sum_i n_i - Max(i) - 1) \qquad (7)$$

As per the red dashed line in Fig. 8d, the strength of a particular connection for each mitochondrial component of the cell is defined as: $B_i n_i$; where $B_i$ is the number of branches in i$^{th}$ component in the mitochondrial network of the cell, and $n_i$ is the number of branch points in the same component.

**Genetic manipulation of mitochondrial morphology**

HRas is a proto-oncogene whose activation results in many downstream physiological changes, including altered metabolism, increased proliferation, blunted apoptosis, and changes in gene expression[103]. In order to generate cell lines that exhibit significant differences in mitochondrial morphology, but with minimal changes to other aspects of cellular physiology, we carried out the following.

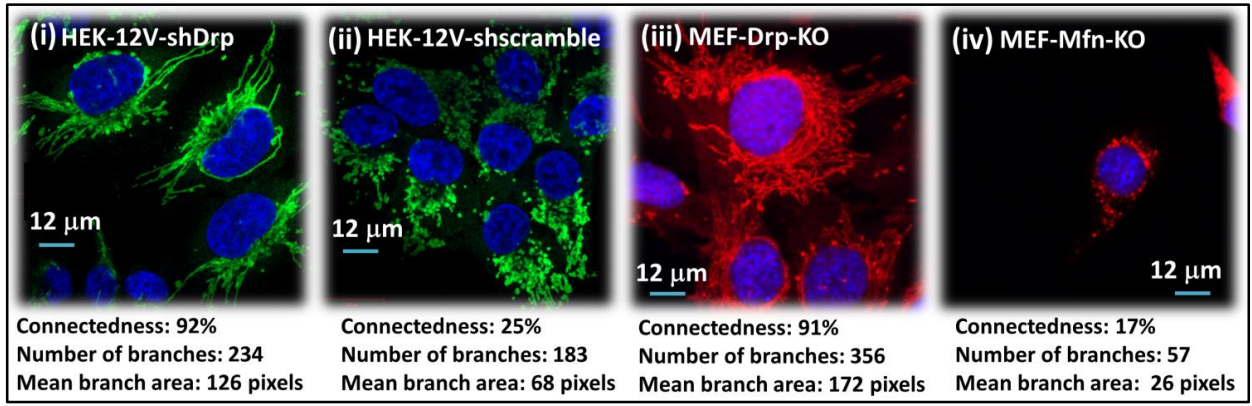

FIG 9: Fluoresence images of fixed cells using anti-Tom20-labelling to reveal mitochondrial (green and red for HEK and MEFs, respectively) features (nucleus labeled blue). For mitochondrial features, the described image analysis method is used to compute the number of branches, branch area (in pixels) and connectedness factor as per Eq. (5), as indicated for each cell type. Note that while the respective fixed cells appear to differ in size due to well-known interfacial interactions with the substrate, the size variations of suspended cells are minimal based on flow cytometry data (Supporting Information section)

First, we stably expressed a constitutively active version of HRas called: HRasG12V in immortalized HEK cells, which we previously demonstrated to result in a highly fragmented mitochondrial network[75]. Next, we stably expressed shRNA targeting either Drp1 (HEK-RAS-12V-shDrp) or a scramble control sequence (HEK-RAS-12V-shScramble). As expected, shRNA-mediated knockdown of Drp1 results in a complete reversal of the mitochondrial fragmentation, indicating an extremely interconnected phenotype (92% connectedness, Figure 9(i)). Conversely, expression of the scramble control sequence had no effect on the HRas-induced mitochondrial fission, thereby causing these cells to maintain a highly fragmented phenotype (25% connectedness, Figure 9(ii)). Importantly, because we are directly targeting the mitochondrial fission machinery (Drp1) in cells that start with a fragmented phenotype, this approach allows us to generate two extreme mitochondrial phenotypes (i.e. highly fragmented and highly connected) with minimal differences in all other aspects of cellular physiology. Furthermore, we compare these "extreme" mitochondrial morphologies using a different cell type: mouse embryonic fibroblasts (MEFs) generated by a different method, i.e. genetic knockout of the mitochondrial fission (Drp1)[83] and fusion (Mfn1/2)[104] machinery, respectively. Since Drp1 and MFN1/2 primarily influence mitochondrial dynamics, their deletion should induce minimal changes in overall cellular physiology, aside from mitochondrial fission and fusion[105]. We generated immortalized Drp1-/- MEFs by infecting cells isolated from Drp1flox/flox; TP53flox/flox embryos with adenoviral Cre recombinase. As expected, due to their inability to perform mitochondrial fission, these cells exhibit extreme mitochondrial connectedness (91% connectedness, Figure 9(iii)). To generate the opposite phenotype, we obtained immortalized Mfn1/Mfn2 double knockout cells that are deficient in mitochondrial outer membrane fusion[104]. As expected, these MEFs exhibit highly fragmented mitochondrial network, due to unopposed mitochondrial fission (17% connectedness, Figure 9(iv)). Notably, both sets of MEFs are from the same strain of inbred mice (129/Sv) and both are p53 deficient (Mfn1/2-/- MEFs express SV40 Large T-antigen, which inhibits p53[106]). Hence, these cells should exhibit minimal changes in overall physiology, independent of mitochondrial morphology. In this manner, using two different cell types and two separate methods to alter mitochondrial morphology, we envision that the measured alterations to cellular physiology arise solely due to mitochondrial phenotype, rather than due to off-target changes associated with our method.

**Mitochondrial phenotype induced DEP alterations**

Modifications to the mitochondrial structure influence cellular physiology, which is measured in this work, based on alterations to cytoplasmic electrophysiology. Each subcellular region dominates the DEP-induced cell polarizability response over a particular frequency range, depending on the contributions of its electrophysiology to the net conductivity and permittivity dispersion[107]. Hence, modifications to the mitochondrial structure can be selectively probed based on alterations to the DEP frequency response in the frequency region wherein the electrophysiology due to mitochondrial structure dominates the net polarizability dispersion. It is also noteworthy that a critical level of media conductivity ($\sigma_m$) is required for optimal distinction of alterations in DEP spectra for particular cell types with differing mitochondrial connectedness. For instance, low $\sigma_m$ levels enhance the contrast for discerning alterations in crossover to pDEP behavior due to morphological modifications to the cell envelope [108],[109],[110], which influence the respective permittivity values. Similarly, as per the spectral simulations and data in Supporting Information section, relatively higher $\sigma_m$ levels that are closer to the conductivity of the cytoplasm ($\sigma_{cyto}$) wherein the mitochondria are situated, should enable distinctions in mitochondrial phenotype. In this manner, alterations in the bulk cell DEP response can be indicative of modifications to mitochondrial features, as long as the appropriate $\sigma_m$ level and DEP frequency range are chosen for these comparisons. Based on this, the spectra are measured at $\sigma_m$ of 0.15 S/m, since it is close to the anticipated range for $\sigma_{cyto}$ (0.5-1 S/m), while being low enough for pDEP behavior (i.e. $\sigma_{cyto} > \sigma_m$ ). Comparing the DEP response of HEK-RAS-12V-shDrp (Fig. 10a(i)) that exhibits a highly connected mitochondrial structure to that of HEK-RAS-12V-shScramble (Fig. 10a(ii)) that exhibits highly fragmented mitochondria, the chief differences are in the 0.5-15 MHz region, as verified by a t-test at each frequency to confirm well-separated mean values at 95% significance level.

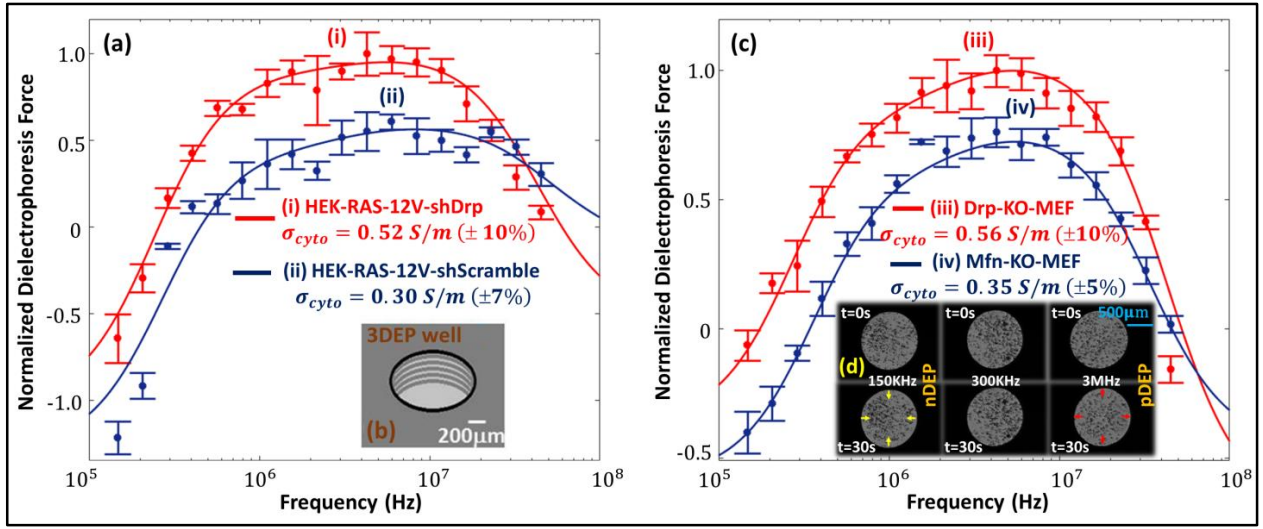

FIG 10. Dielectrophoretic frequency spectra of: (a) oncogenic HRas modified HEK cells expressing Drp1 shRNA (i) versus scramble control shRNA (ii); (c) MEFs lacking Drp1 (iii) versus those lacking mitochondrial fusion GTPases Mfn1 and Mfn2 (iv). Respective fixed cell images are in Fig. 3. DEP spectra were measured at $\sigma_m$ of 0.15 S/m in a 3DEP reader using 20 individual wells with ring electrodes (b) to measure relative DEP force levels based on spatio-temporal variations in light scattering, as per example images shown in (d) for nDEP at 150 kHz, no DEP at 300 kHz and pDEP at 3 MHz field frequencies.

The DEP well device used to measure these responses is shown in Fig. 10b. Similarly, comparing the DEP response of MEF-Drp-KO cells (Fig. 10c(iii)) that exhibit a highly connected mitochondrial structure versus MEF-Mfn-KO cells (Fig. 10c(iv)) that exhibit a highly fragmented mitochondrial structure, it is apparent that the differences are in the 0.5-20 MHz region (verified by a t-test at each frequency to confirm separation at 95% significance level). It is noteworthy that DEP spectra in the 0.5-15 MHz frequency range are dominated by electrophysiology of the cell interior, wherein the mitochondria are situated. Hence, the pDEP levels for cells with a highly connected mitochondrial structure are consistently higher than those of cells with a highly fragmented mitochondrial structure, considering both cell types (HEK and MEFs) and independent of the method used to generate the respective mitochondrial phenotype (knockdown versus knockout). Figure 10d shows some sample images in the DEP-well device that are used to quantify varying DEP levels versus frequency, based on the spatio-temporal profiles of light scattering induced by DEP motion. Using a three-shell or four-layer dielectric model (Methods section), the respective spectra were fit to obtain dielectric properties for each subcellular region of interest (Table 1).

| Parameter (Unit) | shScramble (HEK) ($C_f$ = 25%) | shDRP (HEK) ($C_f$ = 95%) | Mfn-KO (MEF) ($C_f$ = 17%) | DRP-KO (MEF) ($C_f$ = 91%) |
|---|---|---|---|---|
| $\varepsilon_{mem}$ | 14 | 14 | 13 | 13 |
| $\sigma_{mem}$ (S/m) | $0.1 \times 10^{-6}$ | $0.1 \times 10^{-6}$ | $6 \times 10^{-5}$ | $8 \times 10^{-5}$ |
| $\varepsilon_{cytoplasm}$ | 60 | 60 | 60 | 65 |
| $\sigma_{cytoplasm}$ (S/m) | 0.30 | 0.52 | 0.35 | 0.56 |
| $\varepsilon_{nucEnvelope}$ | 25 | 25 | 25 | 25 |
| $\sigma_{nucEnvelope}$ (S/m) | $0.9 \times 10^{-3}$ | $3 \times 10^{-3}$ | $1.3 \times 10^{-3}$ | $3 \times 10^{-3}$ |
| $\varepsilon_{nucleus}$ | 60 | 60 | 60 | 60 |
| $\sigma_{nucleus}$ (S/m) | 1.4 | 1.4 | 1.5 | 1.4 |

Table 1: Fitted dielectric parameters to DEP spectra (details in SI)

Comparing the fitted dielectric parameters, it is apparent that an enhancement of mitochondrial connectivity increases conductivity of the cell interior, as reflected in that of the cytoplasm ($\sigma_{cyto}$) and nucleus envelope ($\sigma_{nucEnvelope}$). While Drp1 knockdown in HEK cells causes ~70% higher $\sigma_{cyto}$ levels and 3-fold higher $\sigma_{nucEnvelope}$ levels, Drp1 knockout in MEFs causes ~60% higher $\sigma_{cyto}$ levels and 3-fold higher $\sigma_{nucEnvelope}$ levels. Considering the two parameters, the influence of $\sigma_{cyto}$ would likely dominate over that of $\sigma_{nucEnvelope}$, since $\sigma_{cyto}$ is significantly higher (nearly 100-fold higher than $\sigma_{nucEnvelope}$). Hence, increasing $\sigma_{cyto}$ by 50-70% would substantially increase pDEP levels in the 0.5-15 MHz, whereas a three-fold rise in the substantially lower conductivity parameter ($\sigma_{nucEnvelope}$), causes only gradual alterations to slope of the DEP spectra in the 0.5-2 MHz region, especially at the chosen $\sigma_m$ of 0.15 S/m (>> $\sigma_{nucEnvelope}$).

**Single-cell quantification of pDEP trapping force levels**

The application of DEP towards frequency-selective isolation of cells based on their electrophysiology due to a particular mitochondrial structure requires that there be significant differences in their trapping force levels at optimal frequencies. While the DEP-well device measures the force spectra over a wide frequency range (0.1-20 MHz) by quantifying spatio-temporal alterations in light scattering due to the DEP motion of particles, it is an indirect method to measure DEP force on particle ensembles, without considering the efficacy of trapping forces for the purpose of DEP isolation at a particular flow rate. Hence, in order to quantify the differences in pDEP force levels between cells of differing mitochondrial structure, we use velocity tracking methods to directly measure force levels with single-cell sensitivity. Specifically, we choose an electrode-less DEP device configuration (Figure 11a) for these measurements, wherein the electric field is applied orthogonal to fluid flow.

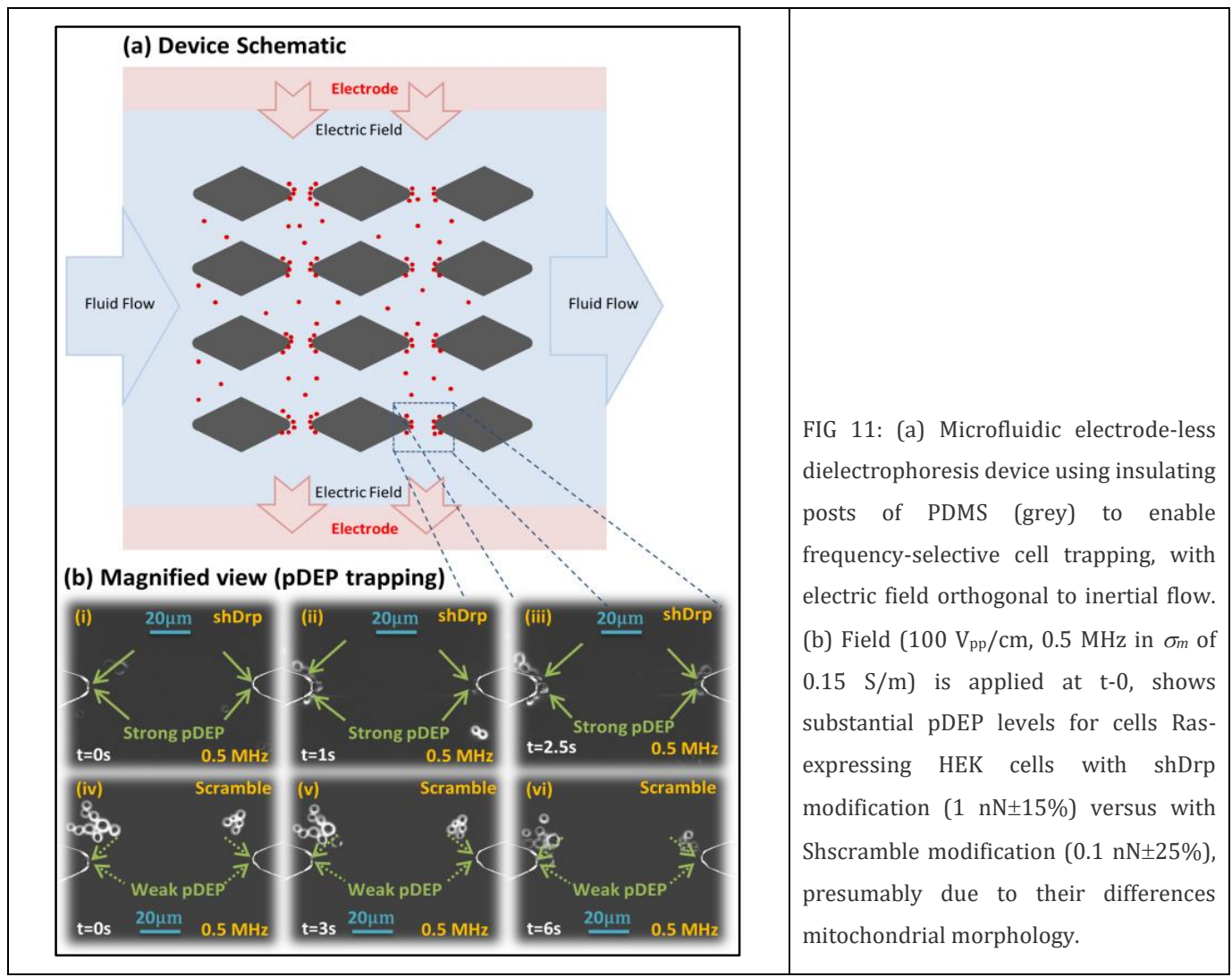

FIG 11: (a) Microfluidic electrode-less dielectrophoresis device using insulating posts of PDMS (grey) to enable frequency-selective cell trapping, with electric field orthogonal to inertial flow. (b) Field (100 $V_{pp}$/cm, 0.5 MHz in $\sigma_m$ of 0.15 S/m) is applied at t-0, shows substantial pDEP levels for cells Ras-expressing HEK cells with shDrp modification (1 nN±15%) versus with Shscramble modification (0.1 nN±25%), presumably due to their differences mitochondrial morphology.

This device is also significant to assess feasibility for selective cell isolation based on mitochondrial structure, since the spatial field non-uniformities created by constrictions due to insulating posts can enable frequency-selective isolation of cells at the constriction tips, based on magnitude of their pDEP levels versus orthogonal drag force due to flow. Specifically, we choose to compare HRas modified HEK cells expressing shDrp versus those expressing a scramble sequence to verify significant differences in pDEP for cells with a highly connected versus a highly fragmented mitochondrial structure. While significant differences in pDEP levels are obvious for the respective spectra over the 0.5-15 MHz range in Fig. 4a, we choose 0.5 MHz for these measurements, since power of the amplifier circuit that is used to generate the field over a wide spatial extent degrades at higher frequencies.

Based on the images in Figure 11b, rapid pDEP trapping is apparent in Figure 11b (i-iii) for shDrp-expressing cells with a highly connected mitochondrial structure exhibiting pDEP levels of 1 nN ± 15% based on ten independent single-cell measurements. The analogous measurement with HEK cells expressing oncogenic HRasG12V plus a scramble control, as per pDEP trapping images in Figure 11b (iv-vi), shows significantly lower pDEP levels that are measured at 0.1 nN ± 25% (based on ten independent single-cell measurements). These results based on direct pDEP measurements on single-cells validate the spectral distinctions of Figure 10.

Assuming a steady drop in pDEP force across constriction region of the device, a flow rate of ~1.35 μL/min is sufficient to ensure that the fluid velocity on the cells is significantly higher than the drag force due to $F_{pDEP}$ at the 0.1 nN level, while being lower than than the drag force due to $F_{pDEP}$ at the 1 nN level (see Supporting Information for device dimensions used in this calculation). Based on this flow rate, a starting concentration of $10^5$ cells/mL can be enriched for cells with the higher $F_{pDEP}$, at a rate of ~100 cells/min, to enable frequency-selective cell isolation at 0.5 MHz based on the cytoplasmic conductivity due to its mitochondrial features.

**Conclusions**

Alterations in connectivity of the intracellular mitochondrial network after select genetic modifications are shown to be quantified based on changes in the intracellular cytoplasmic conductivity, as determined by fitting dielectrophoretic spectra of the respective cells to a standard shell dielectric model. Specifically, direct inhibition of mitochondrial fission through shRNA-mediated knockdown of Drp1 increases the cytoplasmic conductivity by ~70% in human embryonic kidney cells and a full genetic knockout of Drp1 in mouse embryonic fibroblasts causes a 60% rise in cytoplasmic conductivity over cells deficient for mitochondrial fusion. Utilizing a frequency of 0.5 MHz, we demonstrate that human embryonic kidney cells with a highly connected mitochondrial network exhibit ~10-fold higher trapping forces under positive dielectrophoresis versus those with a highly fragmented mitochondrial network. Based on this, we envision a label-free platform for frequency-selective cell isolation to transform the discovery process, both for small molecule modulators of the mitochondrial dynamics machinery and for novel signaling pathways that regulate the machinery under a variety of physiological conditions [111, 112].

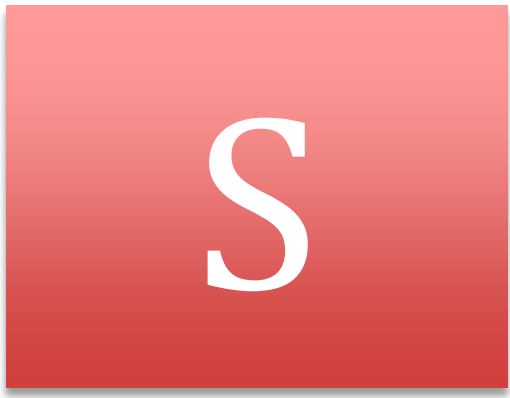
**Future work**

The work presented in this dissertation lays the foundations for developing point-of-care systems, which enables identifying subjects for optimal therapies based on genetic make-up or other molecular or cellular analysis. However it remains a lot to explore using these platforms. Nanoslit design, with optimal sensor placement was used to detect PSA versus sea of interfering HAMA to enable early detection of prostate cancer in the first chapter of this dissertation. However for future work, the same platform can be optimized and used to separate biomarkers of different diseases from their interfering species based on their different electrical double-layer induced polarization. This detection method can be further optimized by using the wave-sequencing methodology that we developed in ref 109.

On the mitochondrial morphology analysis in chapter two, we developed a platform to identify relationship between mitochondrial morphology and the genotype, as a proof of concept to show that morphology of cells can predict different properties of the cells. The same platform can be used to establish and identify relationships between morphology and number of physiological features important for tumor growth in a similar manner. These physiological features are straightforward to analyze and range from cell-level features to tumor-level features, as listed below:

(i)     Mitochondrial features can predict differences in metabolic profile of different tumor cell lines.
(ii)    Mitochondrial features present in a primary tumor can predict metastatic potential from xenograft models
(iii)   Mitochondrial heterogeneity can predict survival and metastatic potential
(iv)    Mitochondrial features can predict gene expression, mutation status, phospho-RTK signaling or cytokine expression
(v)     Mitochondrial network features can predict response to treatment in human patient derived xenografts.
(vi)    Accounting for inter-tumoral and intra-tumoral heterogeneity to predict therapies based on mitochondrial-shaping proteins

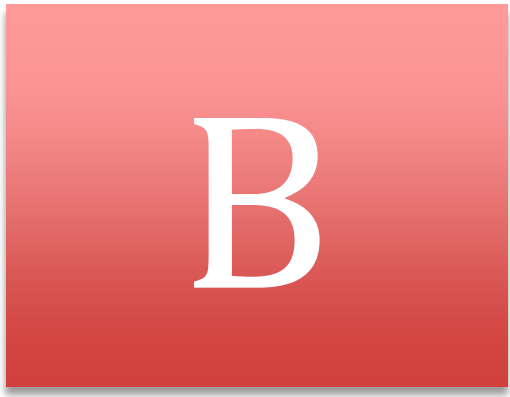

**Bibliography**

[1] J. S. Daniels and N. Pourmand, "Label-free impedance biosensors: Opportunities and challenges," (in English), *Electroanalysis,* vol. 19, no. 12, pp. 1239-1257, Jun 2007.

[2] V. Polaskova, A. Kapur, A. Khan, M. P. Molloy, and M. S. Baker, "High-abundance protein depletion: Comparison of methods for human plasma biomarker discovery," (in English), *Electrophoresis,* vol. 31, no. 3, pp. 471-482, Feb 2010.

[3] G. Q. Hu, Y. L. Gao, and D. Q. Li, "Modeling micropatterned antigen-antibody binding kinetics in a microfluidic chip," (in English), *Biosensors & Bioelectronics,* vol. 22, no. 7, pp. 1403-1409, Feb 15 2007.

[4] A. N. Hoofnagle and M. H. Wener, "The fundamental flaws of immunoassays and potential solutions using tandem mass spectrometry," *Journal of Immunological Methods,* vol. 347, no. 1–2, pp. 3-11, 8/15/ 2009.

[5] P. R. Nair and M. A. Alam, "Performance limits of nanobiosensors," (in English), *Applied Physics Letters,* vol. 88, no. 23, Jun 5 2006.

[6] M. Zimmermann, E. Delamarche, M. Wolf, and P. Hunziker, "Modeling and optimization of high-sensitivity, low-volume microfluidic-based surface immunoassays," (in English), *Biomedical Microdevices,* vol. 7, no. 2, pp. 99-110, Jun 2005.

[7] S. J. Kim, Y. A. Song, and J. Han, "Nanofluidic concentration devices for biomolecules utilizing ion concentration polarization: theory, fabrication, and applications," (in English), *Chemical Society Reviews,* vol. 39, no. 3, pp. 912-922, 2010.

[8] Y. C. Wang, A. L. Stevens, and J. Y. Han, "Million-fold preconcentration of proteins and peptides by nanofluidic filter," (in English), *Analytical Chemistry,* vol. 77, no. 14, pp. 4293-4299, Jul 15 2005.

[9] W. L. Hsu, D. J. E. Harvie, M. R. Davidson, H. Jeong, E. M. Goldys, and D. W. Inglis, "Concentration gradient focusing and separation in a silica nanofluidic channel with a non-uniform electroosmotic flow," (in English), *Lab on a Chip,* vol. 14, no. 18, pp. 3539-3549, 2014.

[10] D. W. Inglis, E. M. Goldys, and N. P. Calander, "Simultaneous Concentration and Separation of Proteins in a Nanochannel," (in English), *Angewandte Chemie-International Edition,* vol. 50, no. 33, pp. 7546-7550, 2011.

[11]    H. A. Pohl, "The motion and precipitation of suspensoids in divergent electric fields," *Applied Physics,* vol. 22, no. 7, pp. 869-871, 1951.

[12]    T. B. Jones, *Electromechanics of particles*. Cambridge ; New York: Cambridge University Press, 1995, p. 265 p.

[13]    R. Pethig, "Review article-dielectrophoresis: status of the theory, technology, and applications," *Biomicrofluidics,* vol. 4, no. 2, 2010.

[14]    H. Morgan and N. G. Green, *AC electrokinetics : colloids and nanoparticles* (Microtechnologies and microsystems series, no. 2). Baldock, Hertfordshire, England Philadelphia, Pa. Williston, VT: Research Studies Press ; Institute of Physics Pub. ; Distribution, North America, AIDC, 2003, pp. xvi, 324 p.

[15]    V. Chaurey, C. Polanco, C. F. Chou, and N. S. Swami, "Floating-electrode enhanced constriction dielectrophoresis for biomolecular trapping in physiological media of high conductivity," (in English), *Biomicrofluidics,* vol. 6, no. 1, Mar 2012.

[16]    N. Swami, C. F. Chou, V. Ramamurthy, and V. Chaurey, "Enhancing DNA hybridization kinetics through constriction-based dielectrophoresis," (in English), *Lab on a Chip,* vol. 9, no. 22, pp. 3212-3220, 2009.

[17]    N. S. Swami, C. F. Chou, and R. Terberueggen, "Two-potential electrochemical probe for study of DNA immobilization," *Langmuir,* vol. 21, no. 5, pp. 1937-41, Mar 1 2005.

[18]    K. T. Liao, M. Tsegaye, V. Chaurey, C. F. Chou, and N. S. Swami, "Nano-constriction device for rapid protein preconcentration in physiological media through a balance of electrokinetic forces," (in English), *Electrophoresis,* vol. 33, no. 13, pp. 1958-1966, Jul 2012.

[19]    K. T. Liao and C. F. Chou, "Nanoscale molecular traps and dams for ultrafast protein enrichment in high-conductivity buffers," *J Am Chem Soc,* vol. 134, no. 21, pp. 8742-5, May 30 2012.

[20]    B. J. Sanghavi, W. Varhue, J. L. Chavez, C. F. Chou, and N. S. Swami, "Electrokinetic Preconcentration and Detection of Neuropeptides at Patterned Graphene-Modified Electrodes in a Nanochannel," (in English), *Analytical Chemistry,* vol. 86, no. 9, pp. 4120-4125, May 6 2014.

[21] B. J. Sanghavi *et al.*, "Ultrafast immunoassays by coupling dielectrophoretic biomarker enrichment in nanoslit channel with electrochemical detection on graphene," *Lab on a Chip,* 10.1039/C5LC00840A 2015.

[22] B. G. Abdallah, S. Roy-Chowdhury, J. Coe, P. Fromme, and A. Ros, "High Throughput Protein Nanocrystal Fractionation in a Microfluidic Sorter," (in English), *Analytical Chemistry,* vol. 87, no. 8, pp. 4159-4167, Apr 21 2015.

[23] R. K. Anand, E. S. Johnson, and D. T. Chiu, "Negative Dielectrophoretic Capture and Repulsion of Single Cells at a Bipolar Electrode: The Impact of Faradaic Ion Enrichment and Depletion," (in English), *Journal of the American Chemical Society,* vol. 137, no. 2, pp. 776-783, Jan 21 2015.

[24] W. V. B. J. Sanghavi, A. Rohani, K. T. Liao, L. Bazydlo, C. F. Chou, N. S. Swami, "Ultrafast immunoassays by coupling dielectrophoretic biomarker enrichment on nano-slit molecular dam device with electrochemical detection," *Lab on a Chip,* 2015.

[25] N. S. S. a. C.-F. C. Kuo-Tang Laio, "Rapid monitoring of low abundance prostate specific antigen by protein nanoconstriction molecular dam," presented at the microTAS, Germany, 2013. Available:
http://www.rsc.org/images/loc/2013/PDFs/Papers/471_0719.pdf

[26] J. Schiffbauer, S. Park, and G. Yossifon, "Electrical Impedance Spectroscopy of Microchannel-Nanochannel Interface Devices," (in English), *Physical Review Letters,* vol. 110, no. 20, May 17 2013.

[27] D. G. Haywood, Z. D. Harms, and S. C. Jacobson, "Electroosmotic Flow in Nanofluidic Channels," (in English), *Analytical Chemistry,* vol. 86, no. 22, pp. 11174-11180, Nov 18 2014.

[28] S. S. Bahga, M. Bercovici, and J. G. Santiago, "Ionic strength effects on electrophoretic focusing and separations," (in English), *Electrophoresis,* vol. 31, no. 5, pp. 910-919, Mar 2010.

[29] J. Wu, K. Gerstandt, H. B. Zhang, J. Liu, and B. J. Hinds, "Electrophoretically induced aqueous flow through single-walled carbon nanotube membranes," (in English), *Nature Nanotechnology,* vol. 7, no. 2, pp. 133-139, Feb 2012.

[30] W. Thormann, C. X. Zhang, J. Caslavska, P. Gebauer, and R. A. Mosher, "Modeling of the impact of ionic strength on the electroosmotic flow in capillary electrophoresis with uniform and discontinuous buffer systems," (in English), *Analytical Chemistry,* vol. 70, no. 3, pp. 549-562, Feb 1 1998.

[31] A. Rohani, W. Varhue, Y. H. Su, and N. S. Swami, "Quantifying spatio-temporal dynamics of biomarker pre-concentration and depletion in microfluidic systems by intensity threshold analysis," (in English), *Biomicrofluidics,* vol. 8, no. 5, Sep 2014.

[32] I. Vlassiouk, S. Smirnov, and Z. Siwy, "Ionic selectivity of single nanochannels," (in English), *Nano Letters,* vol. 8, no. 7, pp. 1978-1985, Jul 2008.

[33] R. B. Schoch, J. Y. Han, and P. Renaud, "Transport phenomena in nanofluidics," (in English), *Reviews of Modern Physics,* vol. 80, no. 3, pp. 839-883, Jul-Sep 2008.

[34] H. Daiguji, P. D. Yang, and A. Majumdar, "Ion transport in nanofluidic channels," (in English), *Nano Letters,* vol. 4, no. 1, pp. 137-142, Jan 2004.

[35] B. J. Kirby, *Micro- and nanoscale fluid mechanics : transport in microfluidic devices*. New York: Cambridge University Press, 2010, pp. xxiii, 512 p.

[36] V. Chaurey, A. Rohani, Y. H. Su, K. T. Liao, C. F. Chou, and N. S. Swami, "Scaling down constriction-based (electrodeless) dielectrophoresis devices for trapping nanoscale bioparticles in physiological media of high-conductivity," *Electrophoresis,* vol. 34, no. 7, pp. 1097-104, Apr 2013.

[37] R. Y. Chein and B. G. Chung, "Numerical study of ionic current rectification through non-uniformly charged micro/nanochannel systems," (in English), *Journal of Applied Electrochemistry,* vol. 43, no. 12, pp. 1197-1206, Dec 2013.

[38] B. J. Sanghavi, O. S. Wolfbeis, T. Hirsch, and N. S. Swami, "Nanomaterial-based electrochemical sensing of neurological drugs and neurotransmitters," *Microchimica Acta,* 2014.

[39] R. Esfandyarpour, H. Esfandyarpour, M. Javanmard, J. S. Harris, and R. W. Davis, "Microneedle Biosensor: A Method for Direct Label-free Real Time Protein Detection," *Sens Actuators B Chem,* vol. 177, pp. 848-855, Feb 2013.

[40] R. Esfandyarpour, M. Javanmard, Z. Koochak, H. Esfandyarpour, J. S. Harris, and R. W. Davis, "Label-free electronic probing of nucleic acids and proteins at the nanoscale using the nanoneedle biosensor," (in English), *Biomicrofluidics,* vol. 7, no. 4, Jul 2013.

[41]	C. Yu, M. H. Davey, F. Svec, and J. M. Frechet, "Monolithic porous polymer for on-chip solid-phase extraction and preconcentration prepared by photoinitiated in situ polymerization within a microfluidic device," *Anal Chem,* vol. 73, no. 21, pp. 5088-96, Nov 1 2001.

[42]	S. Song and A. K. Singh, "On-chip sample preconcentration for integrated microfluidic analysis," *Anal Bioanal Chem,* vol. 384, no. 1, pp. 41-3, Jan 2006.

[43]	B. C. Giordano, D. S. Burgi, S. J. Hart, and A. Terray, "On-line sample pre-concentration in microfluidic devices: a review," *Anal Chim Acta,* vol. 718, pp. 11-24, Mar 9 2012.

[44]	C. C. Lin, J. L. Hsu, and G. B. Lee, "Sample preconcentration in microfluidic devices," (in English), *Microfluidics and Nanofluidics,* vol. 10, no. 3, pp. 481-511, Mar 2011.

[45]	J. Lichtenberg, N. F. de Rooij, and E. Verpoorte, "Sample pretreatment on microfabricated devices," *Talanta,* vol. 56, no. 2, pp. 233-66, Feb 11 2002.

[46]	J. R. Du, Y. J. Juang, J. T. Wu, and H. H. Wei, "Long-range and superfast trapping of DNA molecules in an ac electrokinetic funnel," *Biomicrofluidics,* vol. 2, no. 4, p. 44103, 2008.

[47]	M. Akbari, M. Bahrami, and D. Sinton, "Optothermal sample preconcentration and manipulation with temperature gradient focusing," (in English), *Microfluidics and Nanofluidics,* vol. 12, no. 1-4, pp. 221-228, Jan 2012.

[48]	K. D. Huang and R. J. Yang, "A nanochannel-based concentrator utilizing the concentration polarization effect," (in English), *Electrophoresis,* vol. 29, no. 24, pp. 4862-4870, Dec 2008.

[49]	D. Bottenus, T. Z. Jubery, Y. X. Ouyang, W. J. Dong, P. Dutta, and C. F. Ivory, "10 000-fold concentration increase of the biomarker cardiac troponin I in a reducing union microfluidic chip using cationic isotachophoresis," (in English), *Lab on a Chip,* vol. 11, no. 5, pp. 890-898, 2011.

[50]	D. Mampallil, D. Tiwari, D. van den Ende, and F. Mugele, "Sample preconcentration inside sessile droplets using electrowetting," *Biomicrofluidics,* vol. 7, no. 4, p. 44102, 2013.

[51] M. Kim and T. Kim, "Integration of nanoporous membranes into microfluidic devices: electrokinetic bio-sample pre-concentration," (in English), *Analyst,* vol. 138, no. 20, pp. 6007-6015, 2013.

[52] S. M. Kim, M. A. Burns, and E. F. Hasselbrink, "Electrokinetic protein preconcentration using a simple glass/poly(dimethylsiloxane) microfluidic chip," (in English), *Analytical Chemistry,* vol. 78, no. 14, pp. 4779-4785, Jul 15 2006.

[53] D. J. Bakewell, J. Bailey, and D. Holmes, "Advancing image quantification methods and tools for analysis of nanoparticle electrokinetics," (in English), *Aip Advances,* vol. 3, no. 10, Oct 2013.

[54] C. L. Asbury, A. H. Diercks, and G. van den Engh, "Trapping of DNA by dielectrophoresis," (in English), *Electrophoresis,* vol. 23, no. 16, pp. 2658-2666, Aug 2002.

[55] C. L. Asbury and G. van den Engh, "Trapping of DNA in nonuniform oscillating electric fields," (in English), *Biophysical Journal,* vol. 74, no. 2, pp. 1024-1030, Feb 1998.

[56] D. J. Bakewell and H. Morgan, "Measuring the frequency dependent polarizability of colloidal particles from dielectrophoretic collection data," (in English), *Ieee Transactions on Dielectrics and Electrical Insulation,* vol. 8, no. 3, pp. 566-571, Jun 2001.

[57] D. J. Bakewell and H. Morgan, "Quantifying dielectrophoretic collections of sub-micron particles on microelectrodes," (in English), *Measurement Science & Technology,* vol. 15, no. 1, pp. 254-266, Jan 2004.

[58] D. J. Bakewell and H. Morgan, "Dielectrophoresis of DNA: Time- and frequency-dependent collections on microelectrodes (vol 5, pg 1, 2006)," (in English), *Ieee Transactions on Nanobioscience,* vol. 5, no. 2, pp. 139-146, Jun 2006.

[59] V. Farmehini, A. Rohani, Y.-H. Su, and N. Swami, "Wide bandwidth power amplifier for frequency-selective insulator-based dielectrophoresis," *Lab Chip,* 2014.

[60] L. J. Kricka, "Human anti-animal antibody interferences in immunological assays," (in English), *Clinical Chemistry,* vol. 45, no. 7, pp. 942-956, Jul 1999.

[61] S. Park, F. H. Wians Jr, and J. A. Cadeddu, "Spurious prostate-specific antigen (PSA) recurrence after radical prostatectomy: Interference by human antimouse

heterophile antibodies," *International Journal of Urology,* vol. 14, no. 3, pp. 251-253, 2007.

[62] C. Poyet, D. Hof, T. Sulser, and M. Müntener, "Artificial Prostate-Specific Antigen Persistence After Radical Prostatectomy," *Journal of Clinical Oncology,* vol. 30, no. 5, pp. e62-e63, February 10, 2012 2012.

[63] L. G. Gomella *et al.*, "Screening for prostate cancer: the current evidence and guidelines controversy," (in English), *Canadian Journal of Urology,* vol. 18, no. 5, pp. 5875-5883, Oct 2011.

[64] S. S. Levinson and J. J. Miller, "Towards a better understanding of heterophile (and the like) antibody interference with modem immunoassays," (in English), *Clinica Chimica Acta,* vol. 325, no. 1-2, pp. 1-15, Nov 2002.

[65] I. Ermolina and H. Morgan, "The electrokinetic properties of latex particles: comparison of electrophoresis and dielectrophoresis," *Journal of Colloid and Interface Science,* vol. 285, no. 1, pp. 419-428, 5/1/ 2005.

[66] A. Rohani, W. Varhue, K. T. Liao, C. F. Chou, and N. S. Swami, "Nanoslit design for ion conductivity gradient enhanced dielectrophoresis for ultrafast biomarker enrichment in physiological media," *Biomicrofluidics,* vol. 10, no. 3, p. 033109, May 2016.

[67] B. J. Sanghavi *et al.*, "Aptamer-functionalized nanoparticles for surface immobilization-free electrochemical detection of cortisol in a microfluidic device," *Biosens Bioelectron,* vol. 78, pp. 244-52, Apr 15 2016.

[68] A. Rohani, B. J. Sanghavi, A. Salahi, K. T. Liao, C. F. Chou, and N. S. Swami, "Frequency-selective electrokinetic enrichment of biomolecules in physiological media based on electrical double-layer polarization," (in English), *Nanoscale,* vol. 9, no. 33, pp. 12124-12131, Sep 7 2017.

[69] M. Corrado, L. Scorrano, and S. Campello, "Mitochondrial Dynamics in Cancer and Neurodegenerative and Neuroinflammatory Diseases," *International Journal of Cell Biology,* vol. 2012, p. 13, 2012, Art. no. 729290.

[70] S. Grandemange, S. Herzig, and J.-C. Martinou, "Mitochondrial dynamics and cancer," *Seminars in Cancer Biology,* vol. 19, no. 1, pp. 50-56, 2// 2009.

[71] B. Westermann, "Mitochondrial fusion and fission in cell life and death," *Nat Rev Mol Cell Biol,* 10.1038/nrm3013 vol. 11, no. 12, pp. 872-884, 12//print 2010.

[72] D. Hanahan and Robert A. Weinberg, "Hallmarks of Cancer: The Next Generation," *Cell,* vol. 144, no. 5, pp. 646-674.

[73] Jennifer A. Kashatus *et al.*, "Erk2 Phosphorylation of Drp1 Promotes Mitochondrial Fission and MAPK-Driven Tumor Growth," *Molecular Cell,* vol. 57, no. 3, pp. 537-551.

[74] E. E. Griffin, S. A. Detmer, and D. C. Chan, "Molecular mechanism of mitochondrial membrane fusion," *Biochimica et Biophysica Acta (BBA)-Molecular Cell Research,* vol. 1763, no. 5, pp. 482-489, 2006.

[75] J. A. Kashatus *et al.*, "Erk2 Phosphorylation of Drp1 Promotes Mitochondrial Fission and MAPK-Driven Tumor Growth," (in English), *Molecular Cell,* vol. 57, no. 3, pp. 537-551, Feb 5 2015.

[76] Q. Xie *et al.*, "Mitochondrial control by DRP1 in brain tumor initiating cells," (in English), *Nature Neuroscience,* vol. 18, no. 4, pp. 501-510, Apr 2015.

[77] A. Chevrollier *et al.*, "Standardized mitochondrial analysis gives new insights into mitochondrial dynamics and OPA1 function," *The International Journal of Biochemistry & Cell Biology,* vol. 44, no. 6, pp. 980-988, 6// 2012.

[78] J.-Y. Peng *et al.*, "Automatic Morphological Subtyping Reveals New Roles of Caspases in Mitochondrial Dynamics," *PLoS Comput Biol,* vol. 7, no. 10, p. e1002212, 2011.

[79] Y. Reis *et al.*, "Multi-Parametric Analysis and Modeling of Relationships between Mitochondrial Morphology and Apoptosis," *PLoS ONE,* vol. 7, no. 1, p. e28694, 2012.

[80] E. Lihavainen, J. Mäkelä, J. N. Spelbrink, and A. S. Ribeiro, "Mytoe: automatic analysis of mitochondrial dynamics," *Bioinformatics,* vol. 28, no. 7, pp. 1050-1051, April 1, 2012 2012.

[81] A. P. Leonard *et al.*, "Quantitative analysis of mitochondrial morphology and membrane potential in living cells using high-content imaging, machine learning, and morphological binning," *Biochimica et Biophysica Acta (BBA) - Molecular Cell Research,* vol. 1853, no. 2, pp. 348-360, 2// 2015.

[82] W. J. H. Koopman, H.-J. Visch, J. A. M. Smeitink, and P. H. G. M. Willems, "Simultaneous quantitative measurement and automated analysis of mitochondrial morphology,

mass, potential, and motility in living human skin fibroblasts," *Cytometry Part A,* vol. 69A, no. 1, pp. 1-12, 2006.

[83]   J. Wakabayashi *et al.*, "The dynamin-related GTPase Drp1 is required for embryonic and brain development in mice," (in English), *Journal of Cell Biology,* vol. 186, no. 6, pp. 805-816, Sep 21 2009.

[84]   J. Jonkers, R. Meuwissen, H. van der Gulden, H. Peterse, M. van der Valk, and A. Berns, "Synergistic tumor suppressor activity of BRCA2 and p53 in a conditional mouse model for breast cancer," (in English), *Nature Genetics,* vol. 29, no. 4, pp. 418-425, Dec 2001.

[85]   M. Karbowski *et al.*, "Spatial and temporal association of Bax with mitochondrial fission sites, Drp1, and Mfn2 during apoptosis," *The Journal of cell biology,* vol. 159, no. 6, pp. 931-938, 2002.

[86]   P. Soille, *Morphological image analysis : principles and applications*, 2nd ed. Berlin ; New York: Springer, 2003, pp. xvi, 391 p.

[87]   A. Kasahara and L. Scorrano, "Mitochondria: from cell death executioners to regulators of cell differentiation," (in English), *Trends in Cell Biology,* vol. 24, no. 12, pp. 761-770, Dec 2014.

[88]   H. C. Chen and D. C. Chan, "Emerging functions of mammalian mitochondrial fusion and fission," (in English), *Human Molecular Genetics,* vol. 14, pp. R283-R289, Oct 15 2005.

[89]   A. Ferree and O. Shirihai, "Mitochondrial Dynamics: The Intersection of Form and Function," (in English), *Mitochondrial Oxidative Phosphorylation: Nuclear-Encoded Genes, Enzyme Regulation, and Pathophysiology,* vol. 748, pp. 13-40, 2012.

[90]   D. Hanahan and R. A. Weinberg, "Hallmarks of Cancer: The Next Generation," (in English), *Cell,* vol. 144, no. 5, pp. 646-674, Mar 4 2011.

[91]   S. Hoppins and J. Nunnari, "Mitochondrial Dynamics and Apoptosis-the ER Connection," (in English), *Science,* vol. 337, no. 6098, pp. 1052-1054, Aug 31 2012.

[92]   L. Scorrano, "Keeping mitochondria in shape: a matter of life and death," (in English), *European Journal of Clinical Investigation,* vol. 43, no. 8, pp. 886-893, Aug 2013.

[93]   B. Westermann, "Mitochondrial fusion and fission in cell life and death," (in English), *Nature Reviews Molecular Cell Biology,* vol. 11, no. 12, pp. 872-884, Dec 2010.

[94]   B. Westermann, "Bioenergetic role of mitochondrial fusion and fission," *Biochimica et Biophysica Acta (BBA)-Bioenergetics,* vol. 1817, no. 10, pp. 1833-1838, 2012.

[95]   L. M. Ferreira, "Cancer metabolism: the Warburg effect today," *Experimental and molecular pathology,* vol. 89, no. 3, pp. 372-380, 2010.

[96]   M. G. Vander Heiden, L. C. Cantley, and C. B. Thompson, "Understanding the Warburg effect: the metabolic requirements of cell proliferation," *science,* vol. 324, no. 5930, pp. 1029-1033, 2009.

[97]   M. N. Serasinghe *et al.*, "Mitochondrial Division Is Requisite to RAS-Induced Transformation and Targeted by Oncogenic MAPK Pathway Inhibitors," (in English), *Molecular Cell,* vol. 57, no. 3, pp. 521-536, Feb 5 2015.

[98]   H. A. Pohl, "Dielectrophoresis : the behavior of neutral matter in nonuniform electric fields," ed. Cambridge ;: Cambridge University Press, 1978.

[99]   Y. H. Su *et al.*, "Quantitative dielectrophoretic tracking for characterization and separation of persistent subpopulations of Cryptosporidium parvum," (in English), *Analyst,* vol. 139, no. 1, pp. 66-73, 2014.

[100]  Y.-H. Su, C. A. Warren, R. L. Guerrant, and N. S. Swami, "Dielectrophoretic Monitoring and Interstrain Separation of Intact Clostridium difficile Based on Their S(Surface)-Layers," *Analytical Chemistry,* vol. 86, no. 21, pp. 10855-10863, 2014/11/04 2014.

[101]  Z. R. Gagnon, "Cellular dielectrophoresis: Applications to the characterization, manipulation, separation and patterning of cells," (in English), *Electrophoresis,* vol. 32, no. 18, pp. 2466-2487, Sep 2011.

[102]  L. M. Broche, K. F. Hoettges, S. L. Ogin, G. E. N. Kass, and M. P. Hughes, "Rapid, automated measurement of dielectrophoretic forces using DEP-activated microwells," (in English), *Electrophoresis,* vol. 32, no. 17, pp. 2393-2399, Sep 2011.

[103]  T. Young, F. Mei, J. Liu, R. C. Bast, A. Kurosky, and X. Cheng, "Proteomics analysis of H-RAS-mediated oncogenic transformation in a genetically defined human ovarian cancer model," *Oncogene,* vol. 24, no. 40, pp. 6174-6184, 2005.

[104]  H. C. Chen, S. A. Detmer, A. J. Ewald, E. E. Griffin, S. E. Fraser, and D. C. Chan, "Mitofusins Mfn1 and Mfn2 coordinately regulate mitochondrial fusion and are

essential for embryonic development," (in English), *Journal of Cell Biology,* vol. 160, no. 2, pp. 189-200, Jan 20 2003.

[105] E. Silva Ramos, N. G. Larsson, and A. Mourier, "Bioenergetic roles of mitochondrial fusion," *Biochim Biophys Acta,* vol. 1857, no. 8, pp. 1277-83, Aug 2016.

[106] S. H. Ali and J. A. DeCaprio, "Cellular transformation by SV40 large T antigen: interaction with host proteins," in *Seminars in cancer biology*, 2001, vol. 11, no. 1, pp. 15-22: Elsevier.

[107] M. Castellarnau, A. Errachid, C. Madrid, A. Juarez, and J. Samitier, "Dielectrophoresis as a tool to characterize and differentiate isogenic mutants of Escherichia coli," *Biophysical journal,* vol. 91, no. 10, pp. 3937-3945, 2006.

[108] Y.-H. Su, A. Rohani, C. A. Warren, and N. S. Swami, "Tracking Inhibitory Alterations during Interstrain Clostridium difficile Interactions by Monitoring Cell Envelope Capacitance," *ACS Infectious Diseases,* vol. 2, no. 8, pp. 544-551, 2016.

[109] P. R. Gascoyne, S. Shim, J. Noshari, F. F. Becker, and K. Stemke - Hale, "Correlations between the dielectric properties and exterior morphology of cells revealed by dielectrophoretic field - flow fractionation," *Electrophoresis,* vol. 34, no. 7, pp. 1042-1050, 2013.

[110] A. Rohani, W. Varhue, Y. H. Su, and N. S. Swami, "Electrical tweezer for highly parallelized electrorotation measurements over a wide frequency bandwidth," *Electrophoresis,* vol. 35, no. 12-13, pp. 1795-1802, 2014.

[111] A. Rohani, J. H. Moore, J. A. Kashatus, H. Sesaki, D. F. Kashatus, and N. S. Swami, "Label-Free Quantification of Intracellular Mitochondrial Dynamics Using Dielectrophoresis," *Anal Chem,* vol. 89, no. 11, pp. 5757-5764, Jun 06 2017.

[112] R. E. Fernandez, A. Rohani, V. Farmehini, and N. S. Swami, "Review: Microbial analysis in dielectrophoretic microfluidic systems," *Anal Chim Acta,* vol. 966, pp. 11-33, May 08 2017.